\documentclass[a4paper]{article}
\usepackage{graphicx}
\usepackage{multicol}
\usepackage{epsfig}
\usepackage{float}

\newenvironment{changemargin}[2]{\begin{list}{}{%
\setlength{\topsep}{0pt}%
\setlength{\leftmargin}{0pt}%
\setlength{\rightmargin}{0pt}%
\setlength{\listparindent}{\parindent}%
\setlength{\itemindent}{\parindent}%
\setlength{\parsep}{0pt plus 1pt}%
\addtolength{\leftmargin}{#1}%
\addtolength{\rightmargin}{#2}%
}\item }{\end{list}}

\title{
\begin{flushright}
\small  AMS note 2005-12-01\\
\small LAPP-EXP-2005-21
\end{flushright}
\vspace{40pt} {AMS02 Ecal $\gamma$ trigger performance measured at the\\October 2004 CERN test beam\\} \vspace{40pt} {\normalsize
Pierre Brun\footnotemark[1] \footnotemark[2]
 and
Sylvie Rosier-Lees\footnotemark[1]}
 \footnotetext[1]{Laboratoire d'Annecy-le-vieux de Physique des Particules, CNRS/IN2P3/Univ. de Savoie}
 \footnotetext[2]{brun@lapp.in2p3.fr}
\vspace{80pt}
\begin{abstract}
\begin{changemargin}{0.5cm}{0.5cm}
\textit{Test beam data collected in October 2004 at CERN PS to validate the AMS 02 Ecal Intermediate Board (EIB) are analyzed. After
describing the experimental setup and the event samples, results concerning noise measurement, trigger efficiency and
threshold accuracy are presented. They demonstrate that the EIB fulfils the physics requirements. Therefore the analog part
of the trigger is validated, and hardware choices are also made towards the final device.}
\end{changemargin}
\end{abstract}
}
\date{}

\begin{document}

\maketitle

\newpage

\section{Goals of the test beam}

\subsection{Introduction}
This analysis concerns data taken between September $16^{th}$ and October $7^{th}$ 2004 at CERN. The test beam involved both
Tracker and Ecal subgroups, and partly aimed to combine energy measurement with the
Ecal and tracking with some silicon ladders in a magnetic field.\\
Once installed on the ISS, AMS02 main trigger signal will be provided by the Time Of Flight detectors, which sign the passage
of charged particles. In order to perform $\gamma$ ray astronomy with AMS02, methods based on the conversion of photons in
the upper detector, requiring the $e^{\pm}$ pair rigidity measurement in the Tracker have been made up. In addition to this,
the Ecal working group developed a specific trigger only involving the calorimeter, which we call standalone $\gamma$ trigger
(\cite{note0}, \cite{note1}, \cite{note2}). Its signal has to be fast in order to be able to shortcut the main trigger. This
method gives a complementary $\gamma$ detection method, providing a very good energy resolution and sensitivity in the
GeV-TeV \cite{loic}. This test beam was the opportunity to validate some parts of the trigger chain, the Ecal Intermediate
Board (EIB) prototype and to test the performance of the standalone $\gamma$ trigger: stability, effective thresholds, noise
level and efficiency.

\subsection{Ecal instrumentation}
AMS02 Ecal consists of alternated planes of lead and scintillating fibers. When a particle crosses the detector,
photomultiplier tubes (PMT) collect light from fibers, each PMT having 4 distinct pixels.
\begin{figure}[H]
\centering
\includegraphics[width=10cm]{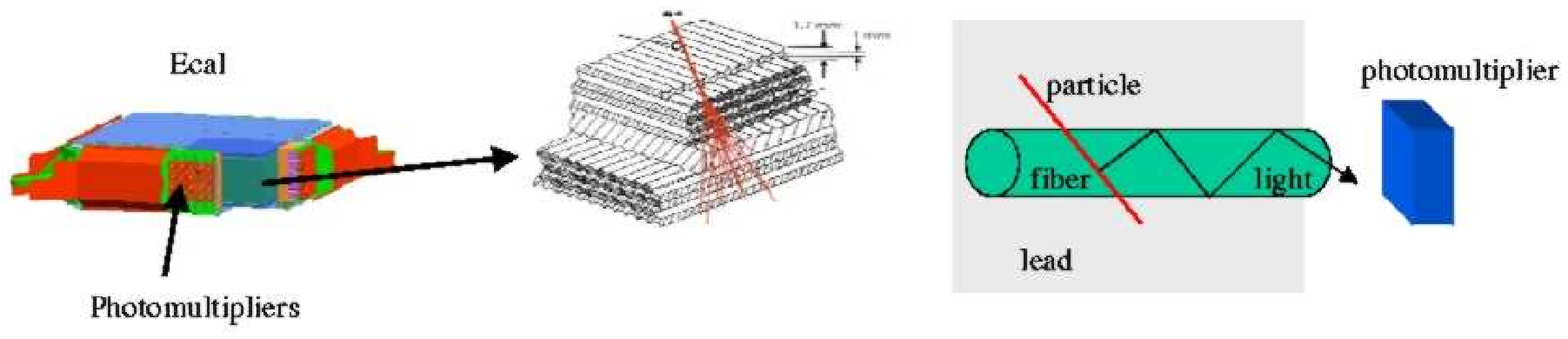}
\includegraphics[width=4cm]{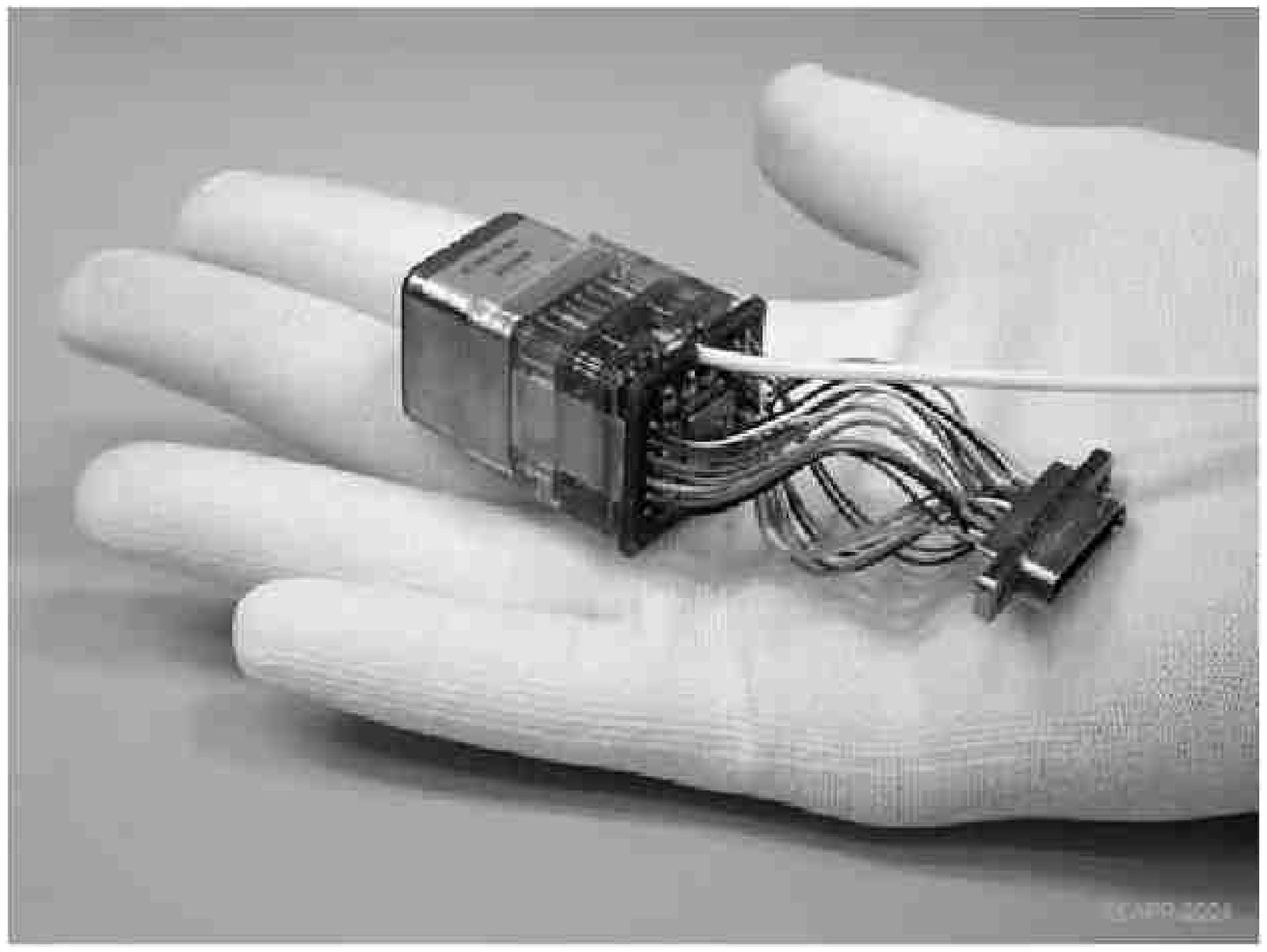}
\caption{Shower light collection in the Ecal}
\label{lightcol}
\end{figure}
For each PMT, 9 wires provide digitized signals from front end electronics (EFE) : 2 from each pixel (low gain / high gain)
and 1 dynode. In addition, the analog dynode signal which is not shaped and digitized (then faster) is amplified, compared to
a threshold and used for the trigger.
\begin{figure}[H]
\centering
\includegraphics[width=12cm]{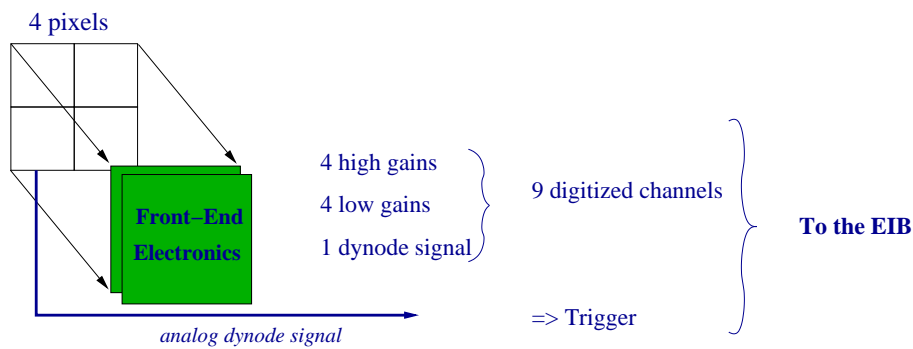}
\caption{Ecal PMT readout}
\label{readout}
\end{figure}
By recording data from both the EFE and the comparator response, we aim to measure the performance of this trigger system. We
expect the trigger to be efficient for a 100 MeV local energy deposition, it has been previously measured \cite{loic} that it
corresponds to a 2.5 mV dynode signal. With the 3 V supply used, an amplification by a factor of 10 has been chosen. 2 types
of amplifiers  have been tested and the final choice is to be one of this test beam results. Next section give a complete
description of the methods and measurements.

\section{Experimental conditions}

\subsection{Beam and detectors}

\begin{figure}[H]
\centering
\includegraphics[width=6cm]{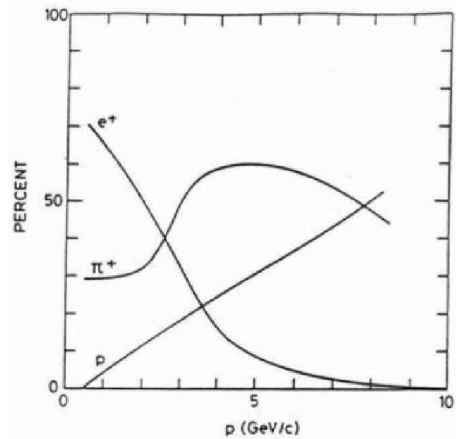}
\caption{Beam composition}
\label{faisceau}
\end{figure}

CERN's Proton Synchrotron (PS) provided electrons, pions, kaons and muons at 3 energies (3, 5 and 7 GeV depending on the
period). This test beam took place at T7 line East hall. Figure \ref{faisceau}  show the relative T7 line beam composition
versus the energy.

\begin{figure}[H]
\centering
\includegraphics[width=13cm]{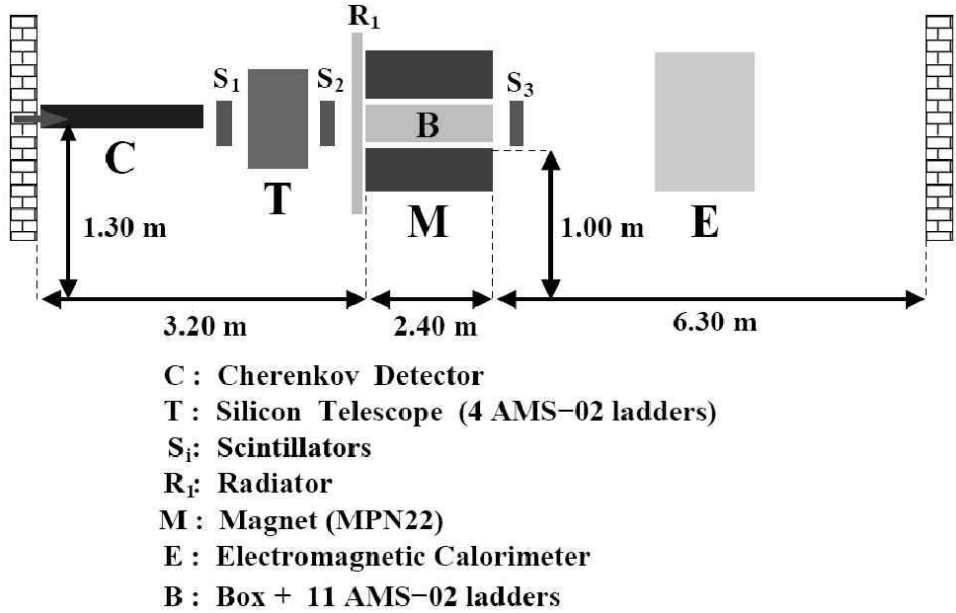}
\caption{Schematic representation of the experimental zone \cite{divic}} \label{zone}
\end{figure}

The experimental zone is represented on figure \ref{zone}, it contained the Ecal (E), 2 Silicon telescopes (T and B), a
magnet (M), 3 scintillators  planes ($S_{i}$) and a 2 \v{C}erenkov counter (C, one of them was outside the zone).

Two \v{C}erenkov detectors were placed before the magnet and used for trigger purpose in complement to 3 scintillators (the
trigger logic is described below). The Tracker telescope was dipped into the magnetic field provided by the magnet. The Ecal
was placed outside the magnetic field at the end of the chain. An additional silicon ladder allowed to determine the beam
position just before it enters the Ecal. It was possible to place a Tungsten radiator before the magnet in order to enhance
bremsstrahlung from electrons, this was used by the Tracker working group and is not relevant for our study.

\subsection{General triggers}

A main trigger drove both Ecal and Tracker data acquisition (\cite{dani}, \cite{divic}). The latter was triggered when the 3
scintillators gave signals in coincidence. Then information provided by the 2 \v{C}erenkov detectors allowed to distinguish
electrons/muons and charged pions:
\begin{itemize}
\item{
Beam = $S_{1}.S_{2}.S_{3}$
}
\end{itemize}
and, depending on the chosen type of particle:
\begin{itemize}
\item{
  Trigger = $Beam.\check{C}$
}
\item[]{or}
\item{
  Trigger = $Beam$
}
\end{itemize}

\begin{figure}[H]
\centering
\includegraphics[width=13cm]{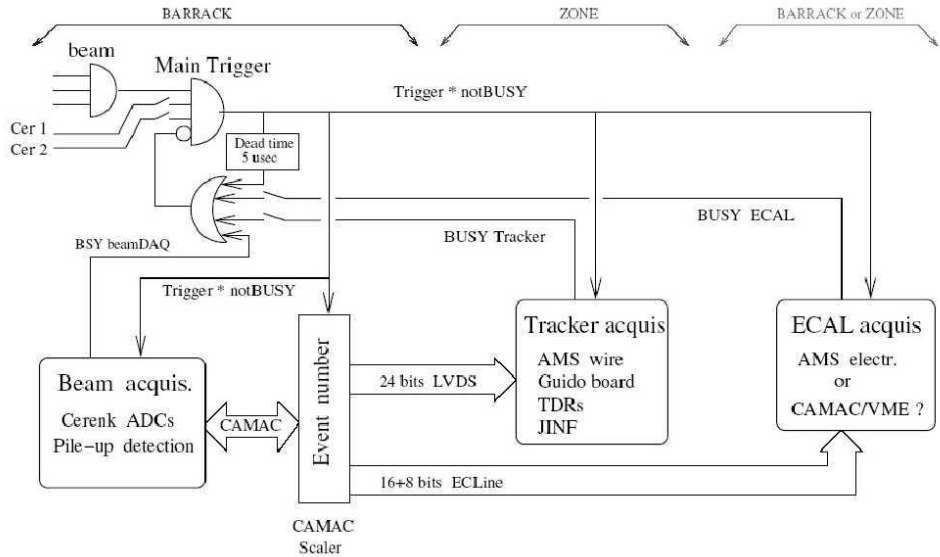}
\caption{Main trigger logic \cite{divic}} \label{trigger}
\end{figure}

It was a key issue that both Ecal and Tracker have a common event number for combined studies. So the event number was
provided by the main trigger logic described on figure \ref{trigger}.

This trigger configuration was the basic one and other Ecal
acquisition logics have been tested in order to study the $\gamma$
trigger. The former are detailed
in next sections.\\

\subsection{Ecal setup}
\begin{figure}[H]
\centering
\includegraphics[width=6cm]{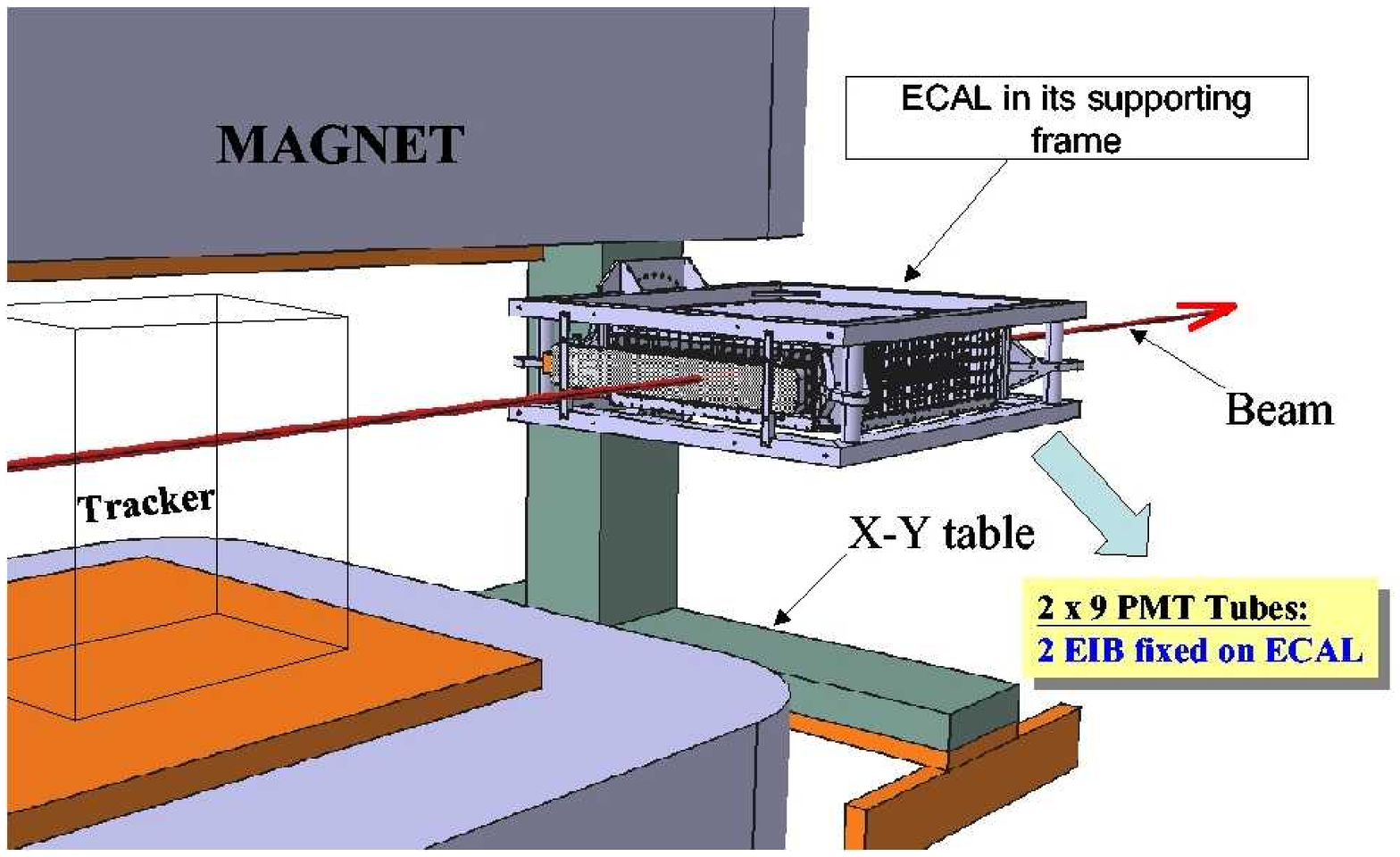}
\includegraphics[width=8cm]{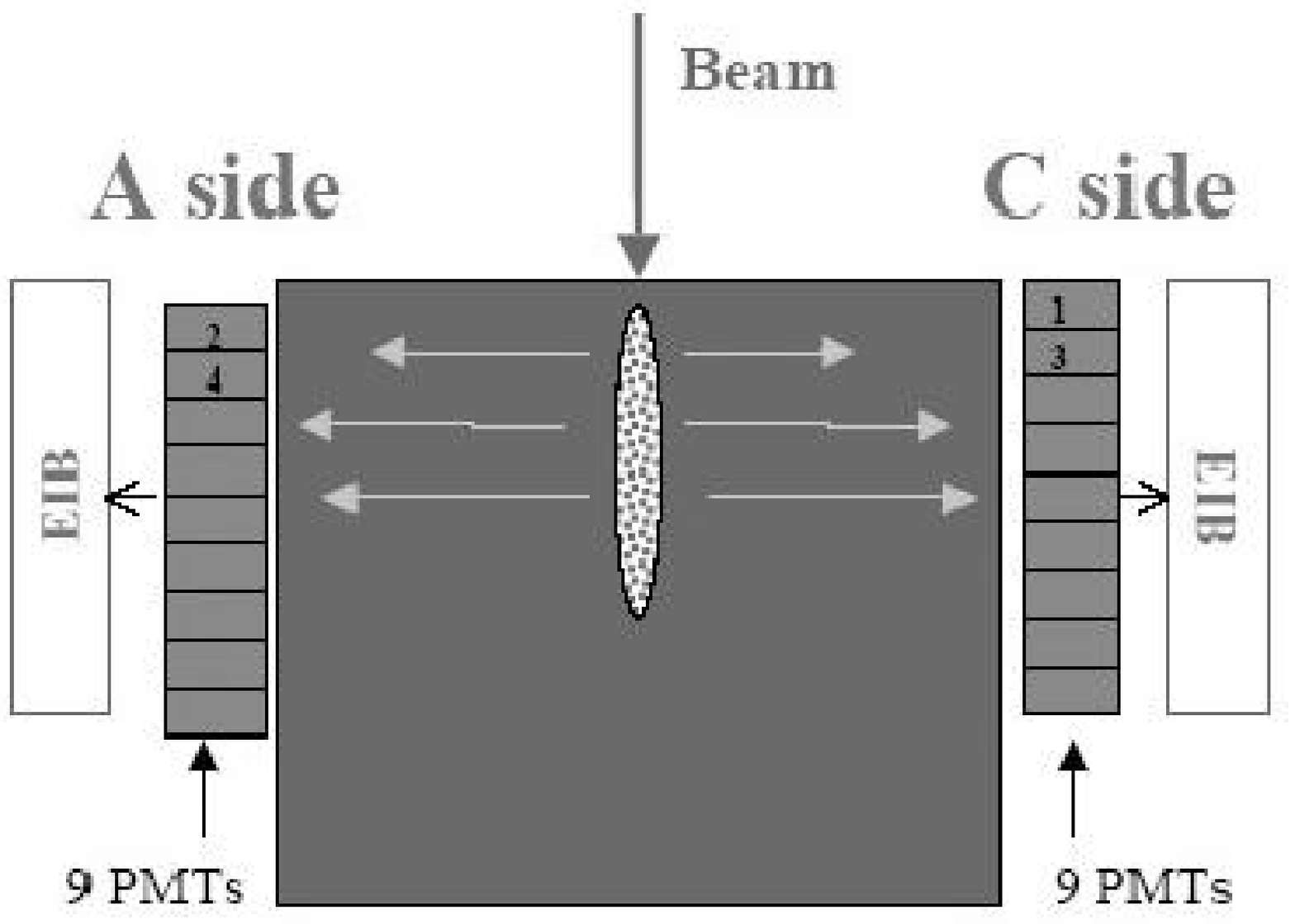}
\caption{Relative beam Ecal positions (left, \cite{franck}) and PMT numbering} \label{franck}
\end{figure}

The Ecal equipment consisted in 2 EIBs and therefore 18 PMTs (9 per EIB). In order to be sensitive to most of the shower, the
Ecal was placed in horizontal position, the beam entering in it by its side (see figure \ref{franck}). In this configuration
80\% of the energy was measured by the Ecal. An additional silicon ladder was placed on the Ecal side to give an accurate
determination of the beam entering point.

\subsection{Event statistics}

\begin{figure}[H]
  \centering
  \includegraphics[width=8.2cm]{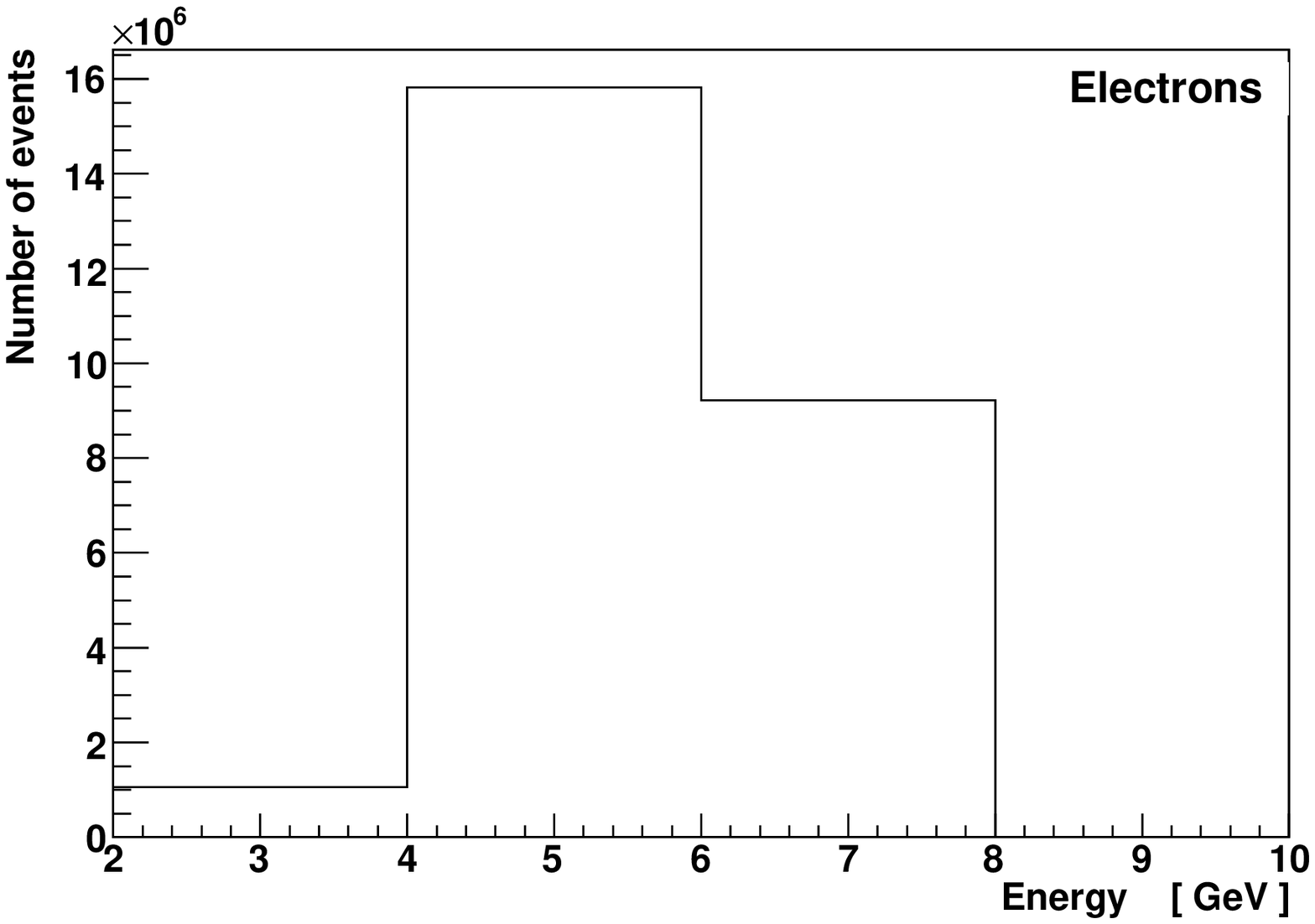}
  \includegraphics[width=8.2cm]{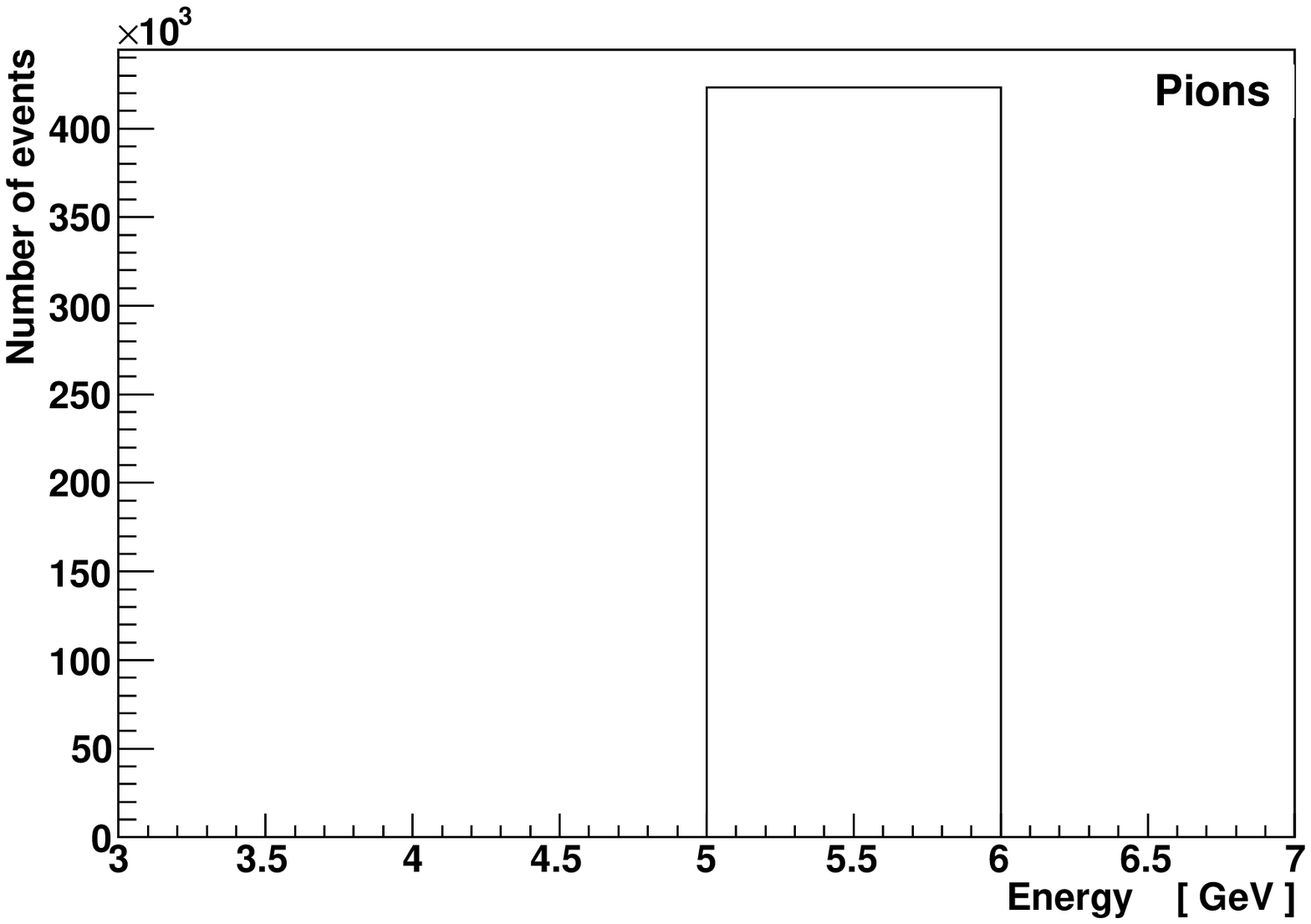}
  \includegraphics[width=8.2cm]{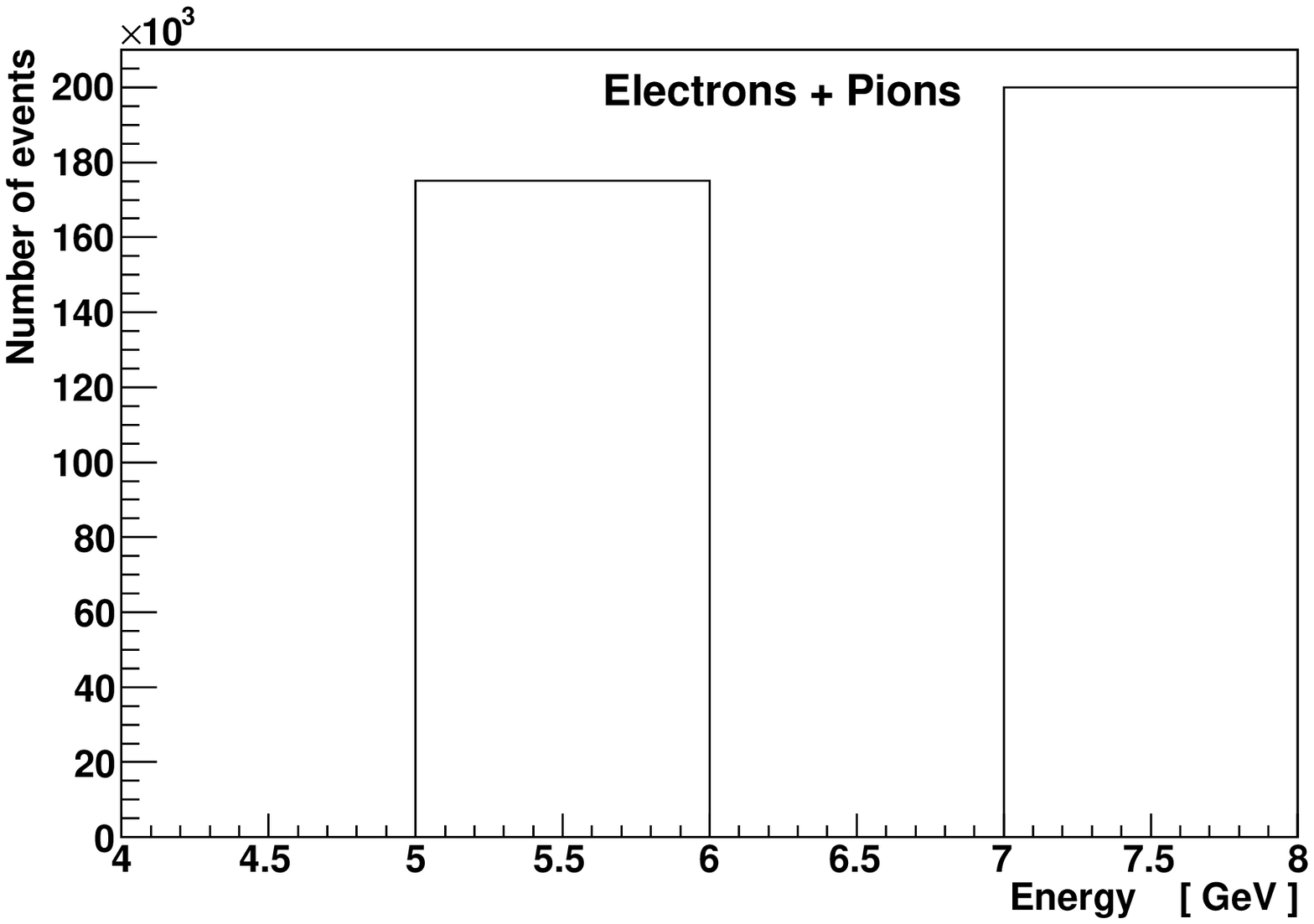}
  \includegraphics[width=8.2cm]{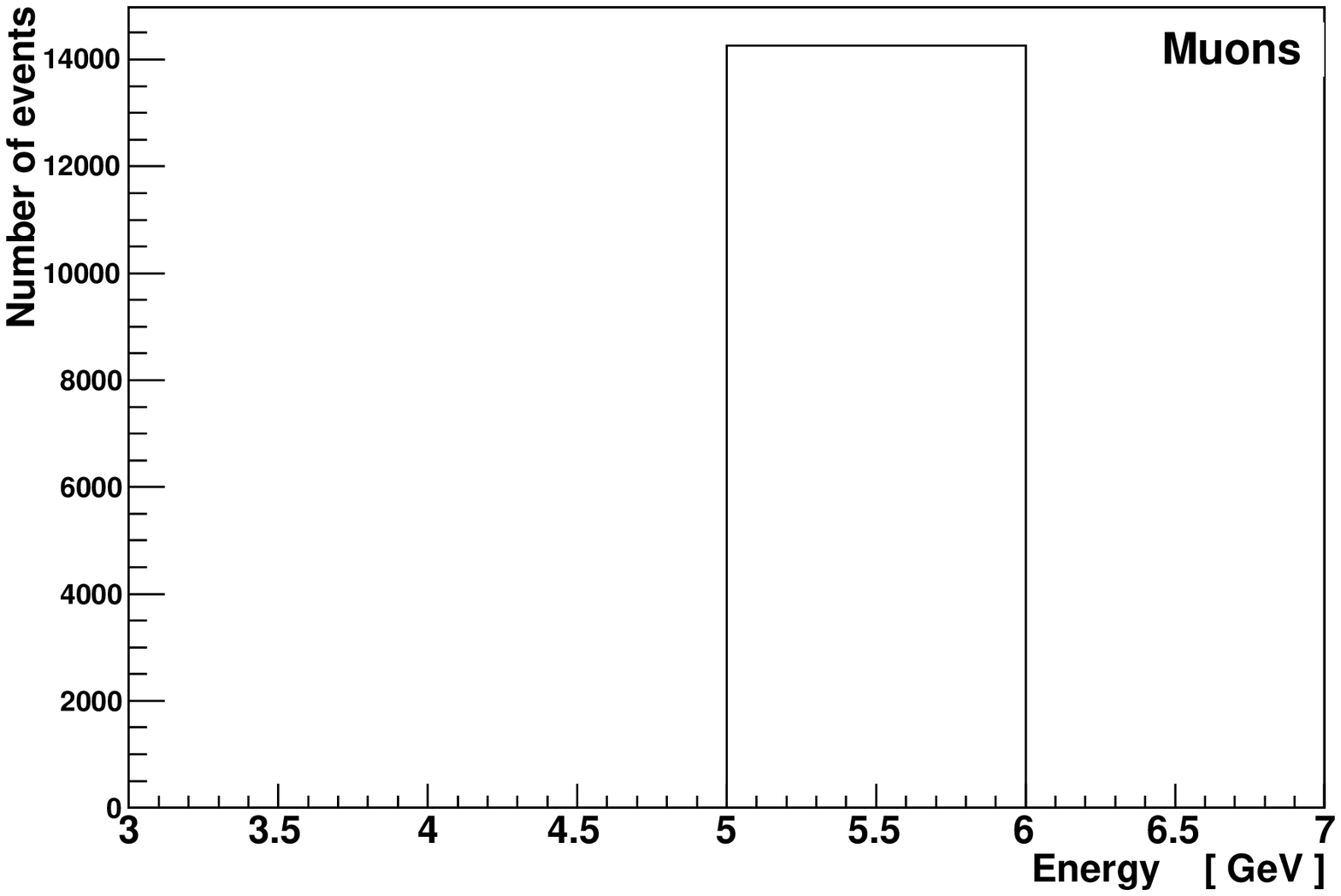}
  \caption{Beam energy for each type of particle}
  \label{energy}
\end{figure}

A large amount of data has been recorded, with about 700 runs and total of $1.7\:10^{7}$ events. Next table shows the number
of events for each type of particles, and figure \ref{energy} shows their energy repartition.

\begin{figure}[H]
  \centering
  \begin{tabular}{c|c|c|c|c}
    Particle type(s) : & electrons & pions & electrons + pions & muons \\
    \hline
    Number of events : & $1.6\:10^{7}$ & $4.2\:10^{5}$ & $3.7\:10^{5}$ & $1.4\:10^{4}$\\
  \end{tabular}
  \caption{Number of events for each type of particle}
\end{figure}

Other parameters differentiate runs, like whether the magnet or
the radiators are in use or not. Nevertheless these different
configurations do not have any implication for the study presented
here and they will no more be mentioned.

\subsection{Different analysis aspects}

This test beam had two main goals from the Ecal point of view:
\begin{itemize}
  \item{To validate the bus part, i.e. what concerns the data and logic command}
  \item{To test and validate the analog part of the $\gamma$ trigger}
\end{itemize}
In this context, the following analysis is divided in two parts.
In the first, we analyze the bus part with pedestal measurements,
time stability of the electronics and we show that data
acquisition do not present any problem.

\section{Analysis part 1 : data and logic command}

\subsection{Conditions stability}
As shown in the following, data taking conditions were very good, only a few problem occurred. On the C side, no coherent
data were collected from PMT C-9, due to a problem on the readout board. This has very few implications on our analysis since
it is the most far from the beam entering point, and therefore would have had very few signal. A more important problem
occurred on the A side, since the EIB has been changed because of some connection defaults. This problem will be detailed
below.

\subsection{Pedestal measurement}

\subsubsection{Test beam measurements}
During he whole test beam, pedestals have been regularly measured. It allows to have a good calibration at any time and to
follow the global conditions evolution. As an illustration, figure \ref{ped0} is the signal distribution for an anode channel
on PMT C2 recorded during a typical pedestal run (which contain 1000 events).

\begin{figure}[H]
\centering
\includegraphics[width=8.2cm]{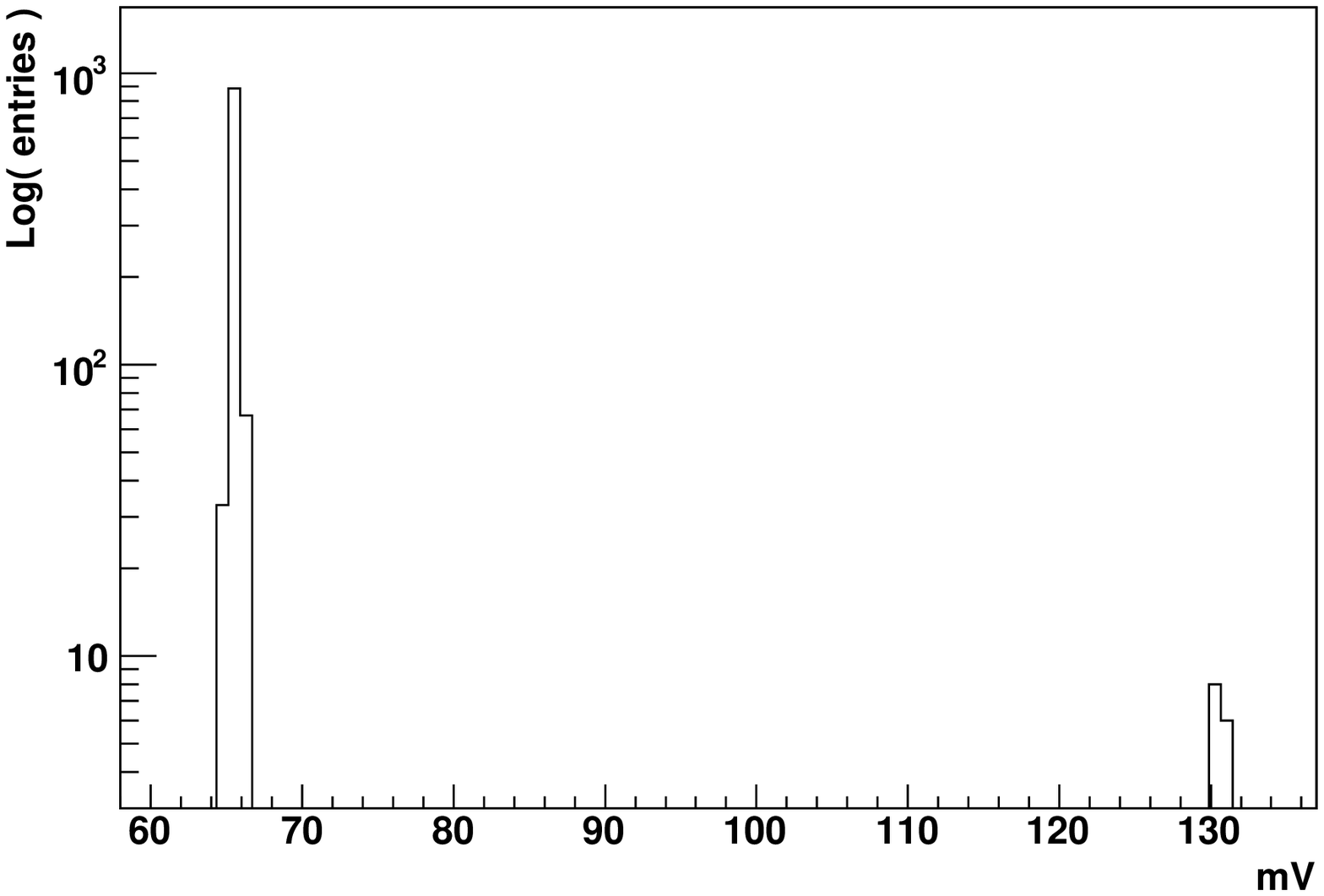}
\includegraphics[width=8.2cm]{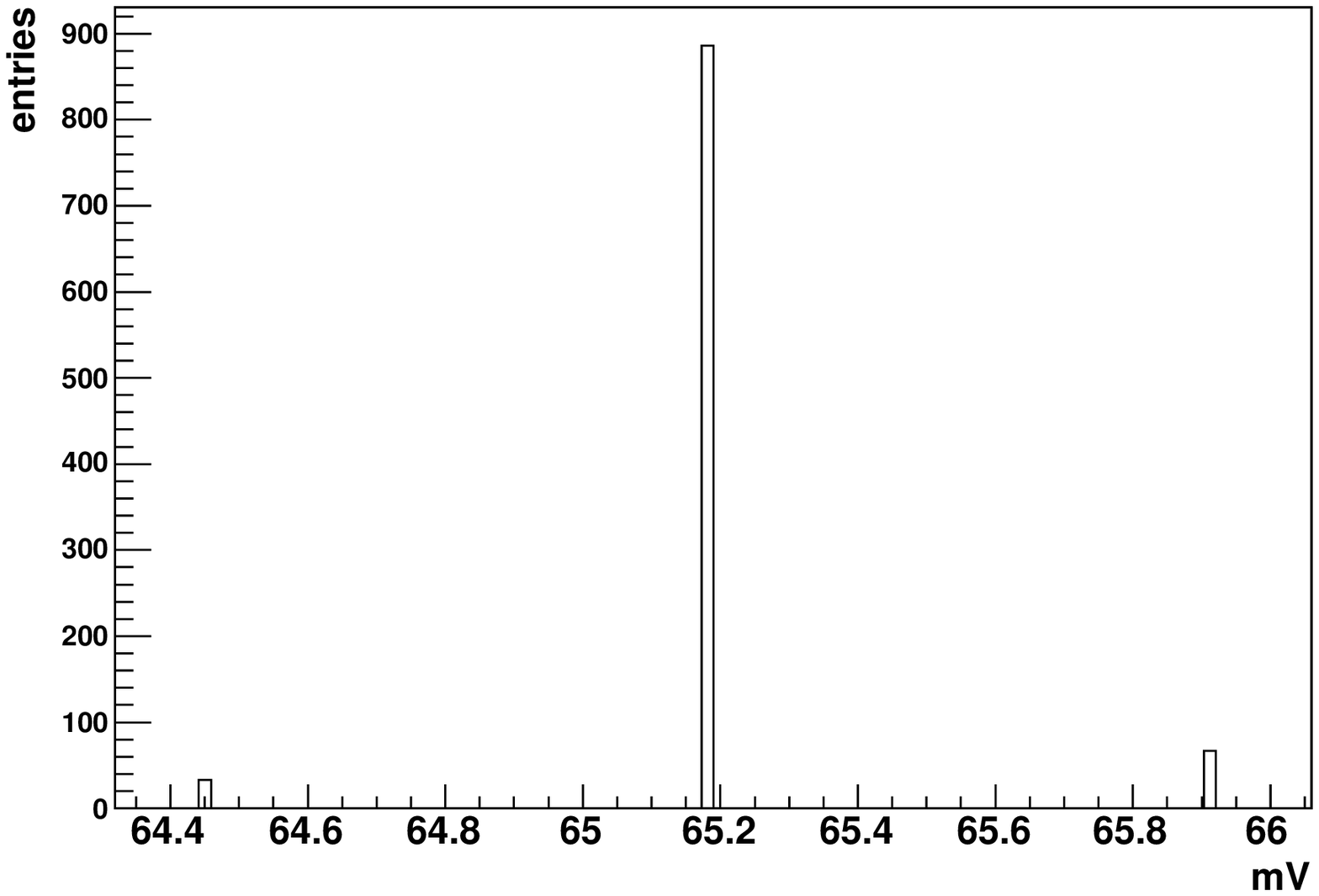}
\caption{Signal distribution for an anode channel on PMT C2 during a pedestal run}
\label{ped0}
\end{figure}

On the left plot we see that a few events are far above pedestal (which is here of about 66 mV). These are probably due to
the passage of a particle in the detector during the run (the beam was not stopped during pedestal runs). On the right, we
can see the same figure zoomed around the peak. In the following, each time we present a pedestal value, it comes from a
gaussian fit to this kind of distribution, after removing values far above the main peak. Note that an ADC channel is 0.73 mV
and the standard deviation from the fits are always much lower than one ADC channel.

\subsubsection{Stability for the whole test beam duration}

\begin{figure}[H]
\centering
\includegraphics[width=8.2cm]{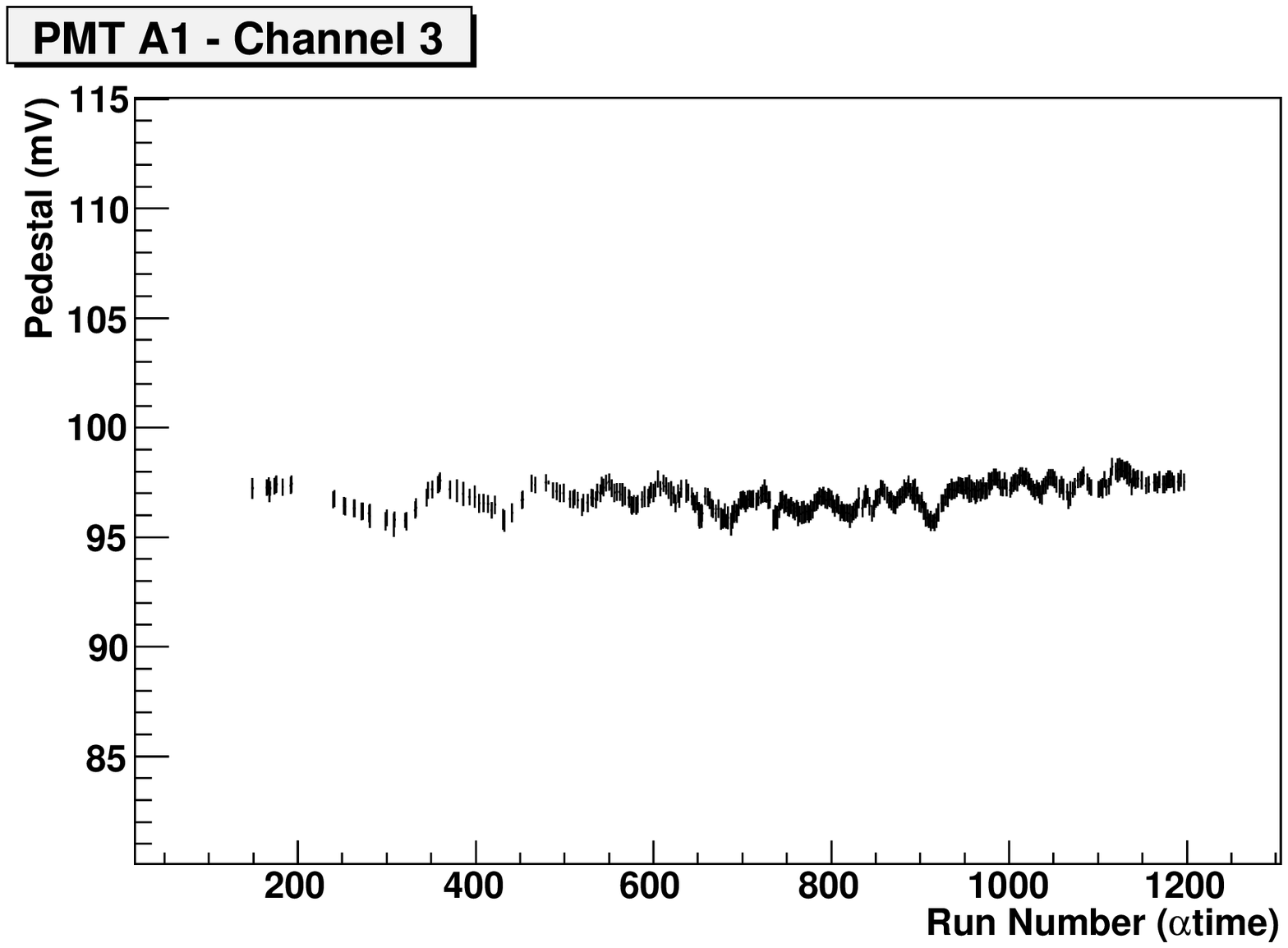}
\includegraphics[width=8.2cm]{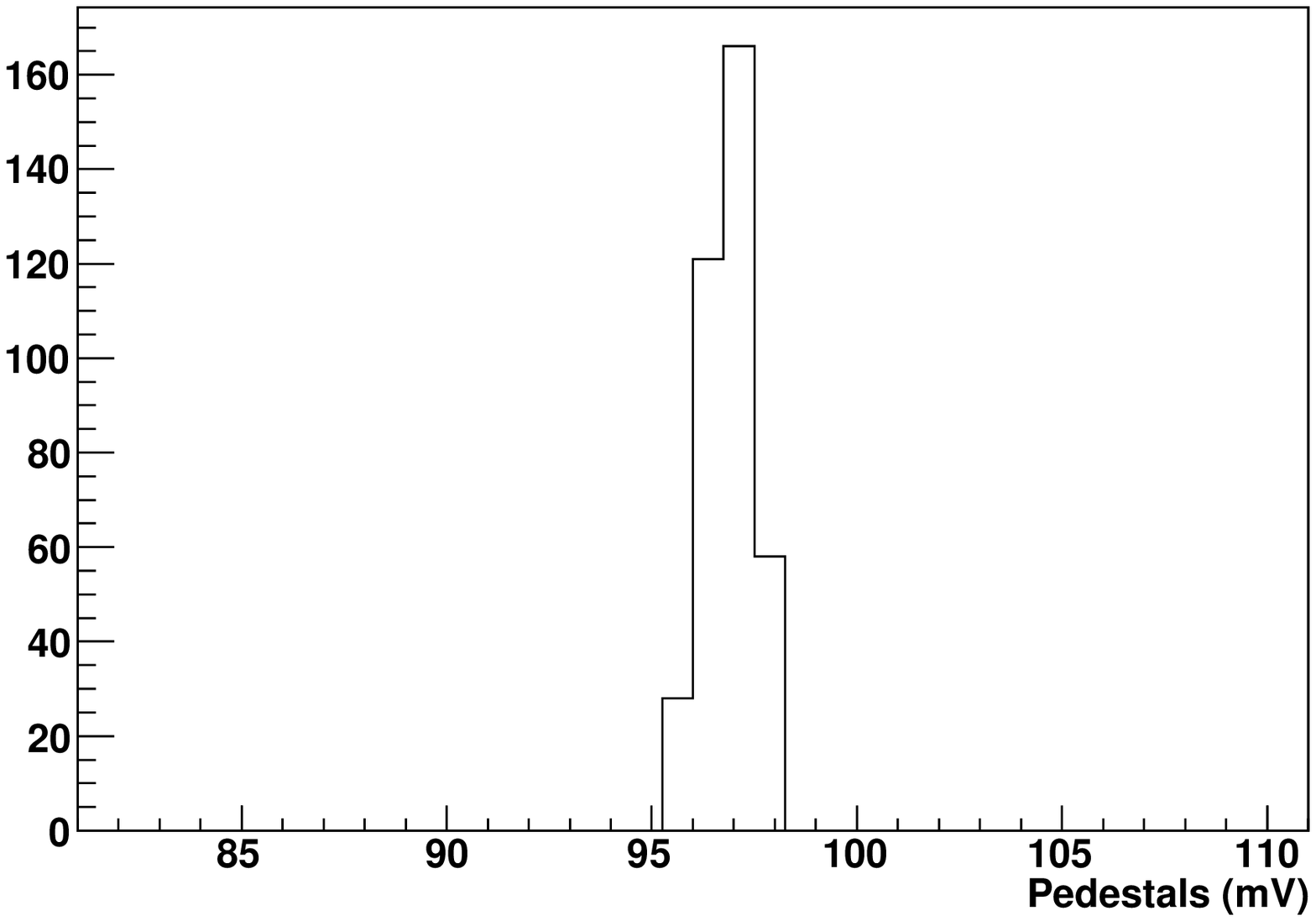}
\includegraphics[width=8.2cm]{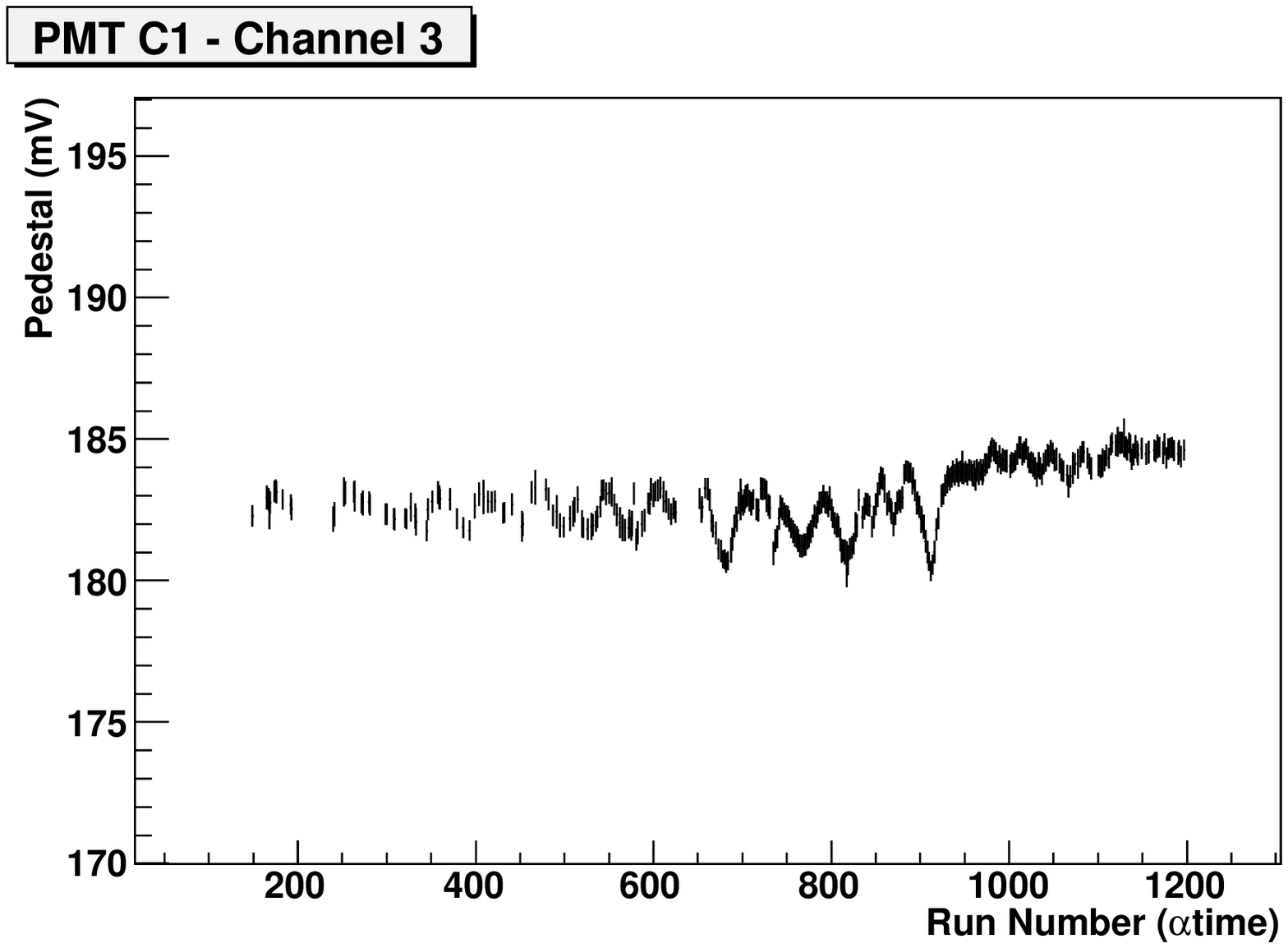}
\includegraphics[width=8.2cm]{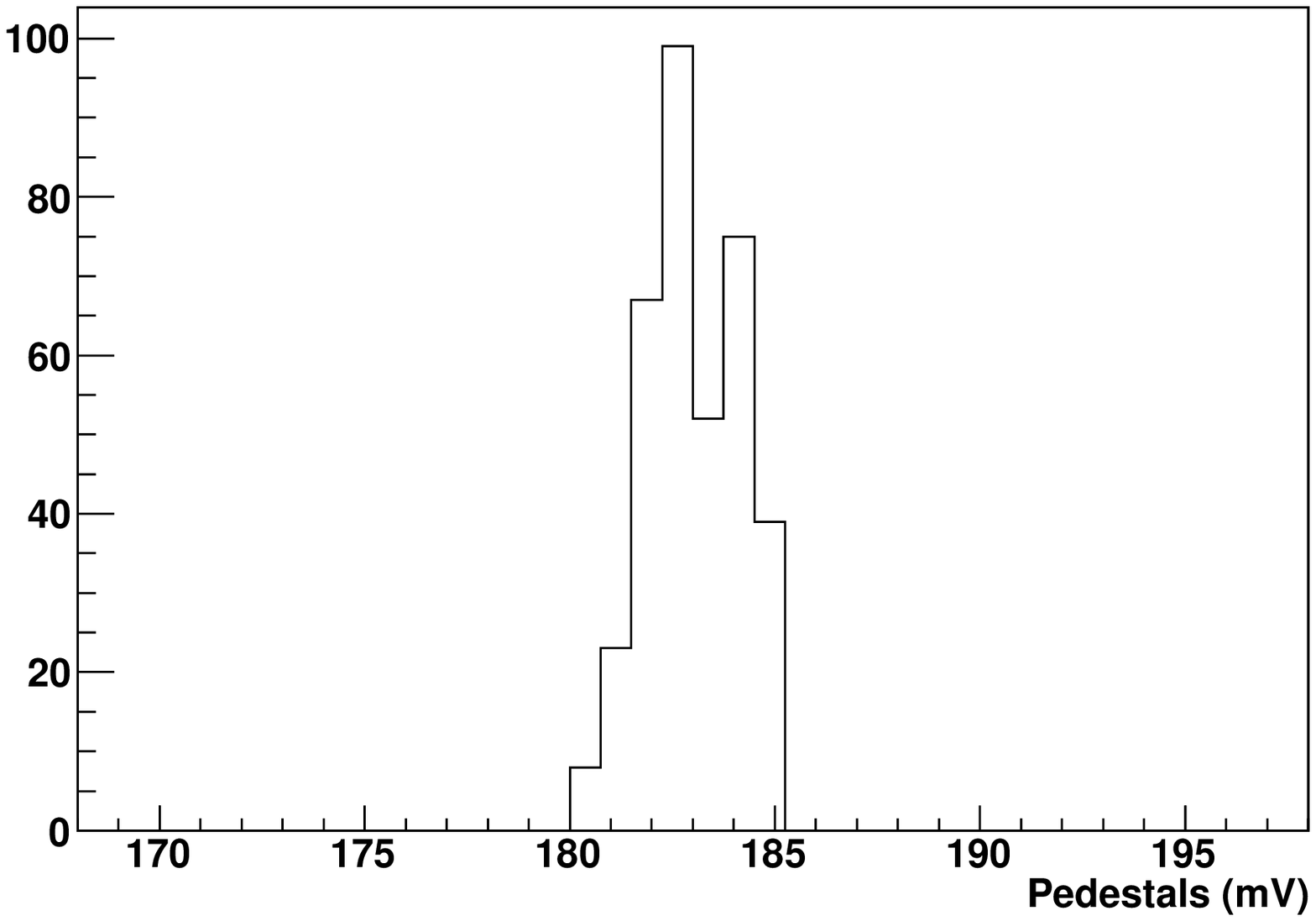}
\caption{Typical pedestal evolution curves (on the left) and distribution (on the right)
  for 2 low gain anode channels on PMT A1 (top) and C1 (bottom)}
\label{pedevol}
\end{figure}

Pedestal runs have been taken all the way long, in general each time a physical run ended. The purpose of these measurements
are to subtract an up-to-date pedestal value to the data during the analysis. It also permits to follow the general evolution
of the electronics. For each pedestal run, a distribution similar to figure \ref{ped0} is fitted. Then we obtain the behavior
of the pedestal evolution versus time, as shown on figure \ref{pedevol} (here for one PMT channel on each side). This figure
shows on the left hand side the evolution versus time, and on the right hand side the pedestal distribution for the whole
test beam duration. Similar plots are obtained for each channel and each PMT. The tables of figure \ref{pedgen} and
\ref{pedgen2} contain a summary of all these results. For a given channel and a given PMT, the ``mean'' value is the mean
pedestal value for the whole test beam and ``rms'' is the root mean square of this distribution.

\begin{figure}[H]
\centering
\begin{tabular}{c|c *9{||c}}
\multicolumn{2}{c|}{ PMT A$\rightarrow$} & 1 & 2 & 3 & 4 & 5 & 6 & 7 & 8 & 9\\
\multicolumn{2}{c|}{ $\downarrow$ Channels}   &   &   &   &   &   &   &   &   &  \\
\hline \hline
1 & mean &131.1 & 233.3& 164.6& 72.5 & 83.6&189.6 &134.4 &146.7 & 67.8\\
\cline{2-11}
 & rms &0.46 &0.90 &0.79 & 5.62& 0.75&0.67 &1.10 &0.78 & 0.67 \\
\hline \hline
2 & mean & 103.7&63.42& 194.7 &160.4 &124.5 &103.3 & 115.8 &53.32  & 69.5\\
\cline{2-11}
 & rms & 0.70& 0.75&0.80 &67.1 & 0.66& 0.84& 1.02&1.15& 0.33\\
\hline \hline
3 & mean &96.9 &166.4 &149.3 &89.4 &124.1 &121..9 &119.8  &127.8 & 133.2\\
\cline{2-11}
 & rms & 0.57&0.80&0.71 & 5.31& 0.69& 0.74&1.02 & 0.84 & 0.32\\
\hline \hline
4 & mean & 171.4 & 218.5&130.6 & 73.06& 145.4& 114.4&115.7 &127.2 & 117.4\\
\cline{2-11}
 & rms &0.52 &0.90&0.74 &5.48 &0.80 &0.49 & 1.13 & 0.67 & 0.39\\
\hline \hline
5 & mean & 147.9 &203.4&174.1 & 157.8& 114 & 96.7  &144.3 &184.2 & 38.4\\
\cline{2-11}
 & rms &0.74 &0.91&1.12 &68 & 0.53& 0.61& 1.08 &0.55 & 0.54\\
\hline \hline
6 & mean & 174.4 &118.6&58.71 & 260.7& 125.6& 228.7  & 164.4&163.8 & 141.7\\
\cline{2-11}
 & rms &0.69 &0.87&0.76 &79.2 & 125.6& 228.7& 164.4 &163.8 & 0.50\\
\hline \hline
7 & mean & 134.1 &160.2&177.4 &124.1 &138.5 &181.1   & 220.9& 87.4& 164.2\\
\cline{2-11}
 & rms & 0.58&0.74&0.66 &5.25 &0.50 &0.54 &1.10  &0.91 &0.38\\
\hline \hline
8 & mean & 136 &97.9&178.5 & 171& 130.3& 205.2&174.7&172.4 &92.18\\
\cline{2-11}
 & rms &0.21 &1.01& 0.68 & 70.98& 0.70& 0.62& 1.05  & 0.73&0.41\\
\hline \hline
9 & mean &451.8  &328.1 &470.8 & 483.8&459.1 &424.9  &507.3 & 393.8&395.8\\
\cline{2-11}
 & rms &3.34 & 32.6&9.68 &8.97 &5.9 &2.89 & 7.9  & 0.46&10.26\\
\hline\hline\hline
9 & mean & 326.8 & 274.6 & 332.9 & 359.3 & 343 & 314.4 & 361.9 & 325.7 & 412.6 \\
\cline{2-11}
EIB 1 & rms & 0.35 & 0.42 & 0.24 & 19.02 & 0.25 & 0.34 & 0.36 & 0.47 & 0.62 \\
\hline\hline
9 & mean & 332.1 & 348 & 332.9 & 355.35 & 333.3 & 309.5 & 374.7 & 288.1 & 389.4 \\
\cline{2-11}
EIB 2 & rms & 0.48 & 0.36 & 0.26 & 0.24 & 0.48 & 0.06 & 0.19 & 0.44 & 0.25 \\
\end{tabular}
\caption{Mean pedestal values and RMS for all A side PMTs and all channels during all test beam runs} \label{pedgen}
\end{figure}

\begin{figure}[H]
\centering
\begin{tabular}{c|c *9{||c}}
\multicolumn{2}{c|}{ PMT C$\rightarrow$} & 1 & 2 & 3 & 4 & 5 & 6 & 7 & 8 & 9\\
\multicolumn{2}{c|}{ $\downarrow$ Channels}   &   &   &   &   &   &   &   &   &  \\
\hline \hline
1 & mean & 140.1 & 65.52 & 101.3 & 76.77 & 159.2 & 92.0 & 205.6 & 143.4 & - \\
\cline{2-11}
 & rms & 0.96 & 1.05 & 0.48 & 0.74 & 0.50 & 0.76 & 0.57 & 0.93 & - \\
\hline \hline
2 & mean & 169.6 & 95.15 & 98.54 & 121.9 & 51 & 95.71 & 40.85 & 192 & - \\
\cline{2-11}
 & rms & 0.87 & 0.80 & 0.75 & 0.60 & 0.67 & 0.50 & 1.01 & 0.58 & - \\
\hline \hline
3 & mean & 183 & 91.13 & 180 & 131.7 & 90.0 & 150.1 & 131 & 188.5 & - \\
\cline{2-11}
 & rms & 1.09 & 0.95 & 0.63 & 0.73 & 0.81 & 0.87 & 0.80 & 0.81 & - \\
\hline \hline
4 & mean & 154.1 & 97.34 & 128.5 & 82.76 & 136.1 & 115.8 & 179.6 & 95.26 & - \\
\cline{2-11}
 & rms & 1.02 & 0.79 & 0.45 & 0.82 & 0.99 & 0.72 & 0.43 & 0.63 & - \\
\hline \hline
5 & mean & 172.4 & 82.14 & 232.7 & 112.6 & 56.22 & 107 & 176.2 & 92.7 & - \\
\cline{2-11}
 & rms & 0.99 & 0.92 & 0.42 & 0.94 & 0.74 & 0.78 & 0.74 & 0.98 & - \\
\hline \hline
6 & mean &  150.5 & 166.7 & 149.8 & 120.4 & 206.1 & 87.94 & 89.97 & 141.8 & - \\
\cline{2-11}
 & rms & 1.09 & 0.74 & 0.58 & 0.62 & 0.65 & 0.86 & 0.78 & 0.84 & - \\
\hline \hline
7 & mean & 121.9 & 230.6 & 77.47 & 126.9 & 157.8 & 157.3 & 98.57 & 82.54 & - \\
\cline{2-11}
 & rms & 0.67 & 0.54 & 0.79 & 0.65 & 0.47 & 0.63 & 0.86 & 0.90 & - \\
\hline \hline
8 & mean & 120.8 & 129.7 & 189 & 267.2 & 126.1 & 115.2 & 98.41 & 159.3 & -\\
\cline{2-11}
 & rms & 0.78 & 0.88 & 0.71 & 0.66 & 0.60 & 0.79 & 1.04 & 0.96 & - \\
\hline \hline
9 & mean & 573.5 & 585.9 & 584.3 & 604.7 & 598.9 & 615.4 & 574.4 & 599.8 & - \\
\cline{2-11}
 & rms & 0.55 & 0.40 & 0.43 & 0.49 & 0.39 & 0.31 & 0.21 & 0.40 & - \\
\end{tabular}
\caption{Mean pedestal values and RMS for all C side PMTs and all channels during all test beam runs} \label{pedgen2}
\end{figure}

\begin{figure}[H]
\centering
\includegraphics[width=18cm]{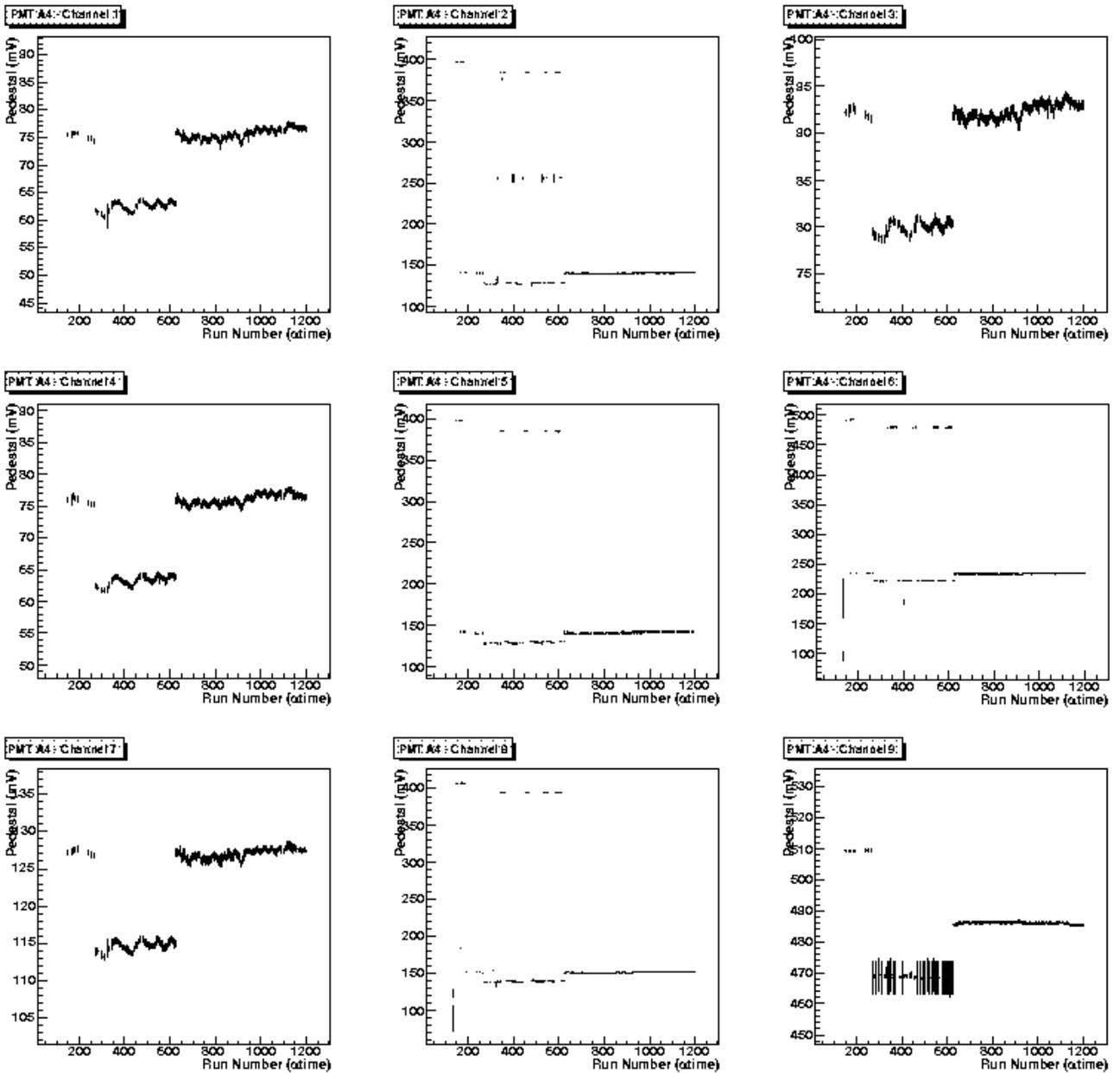}
\caption{PMT A4 pedestal evolution: this illustrates that after the change of EIB
  (around run number 600), the PMT A4 channels pedestals are very stable}
\label{A4dyn}
\end{figure}

Tables from figures \ref{pedgen} and \ref{pedgen2} show that C side PMTs have extremely stable pedestals, with an overall
standard deviation which is often below 1 ADC channel (= 0.73 mV) and always below 1 mV. On A side the situation is the same
with a few exceptions detailed below. This stability is a first good result because it ensures good data taking condition and
is a first indication that the bus part of the device works correctly. Some figures on the A side appear to be quite bad,
like the fluctuations of the A4 pedestal value or the variability of all the A side dynodes. These problems are related to
the change of EIB on the A side. Before it was changed, some fluctuations were observed in particular on PMT A4. After the
change the stability was back. Figure \ref{A4dyn} at the end of this note shows PMT A4 pedestal evolution for all its
channels. One can see clearly there that after the EIB change, their value become stable again. To illustrate this, the
dynode pedestal values are written in table \ref{pedgen} for the first EIB and the second one separately.

\subsection{Comparison with previous measurements}
The 18 PMTs used in this test beam had their pedestals previously measured at LAPP on test bench. The first check we did was
to see if we measured the same pedestals once installed in the experimental zone with the whole acquisition apparatus (in
particular the EIB plugged on). Figure \ref{ped0} shows the correlation between the anode pedestals previously measured at
LAPP and during the test beam.

\begin{figure}[H]
\centering
\includegraphics[width=8.2cm]{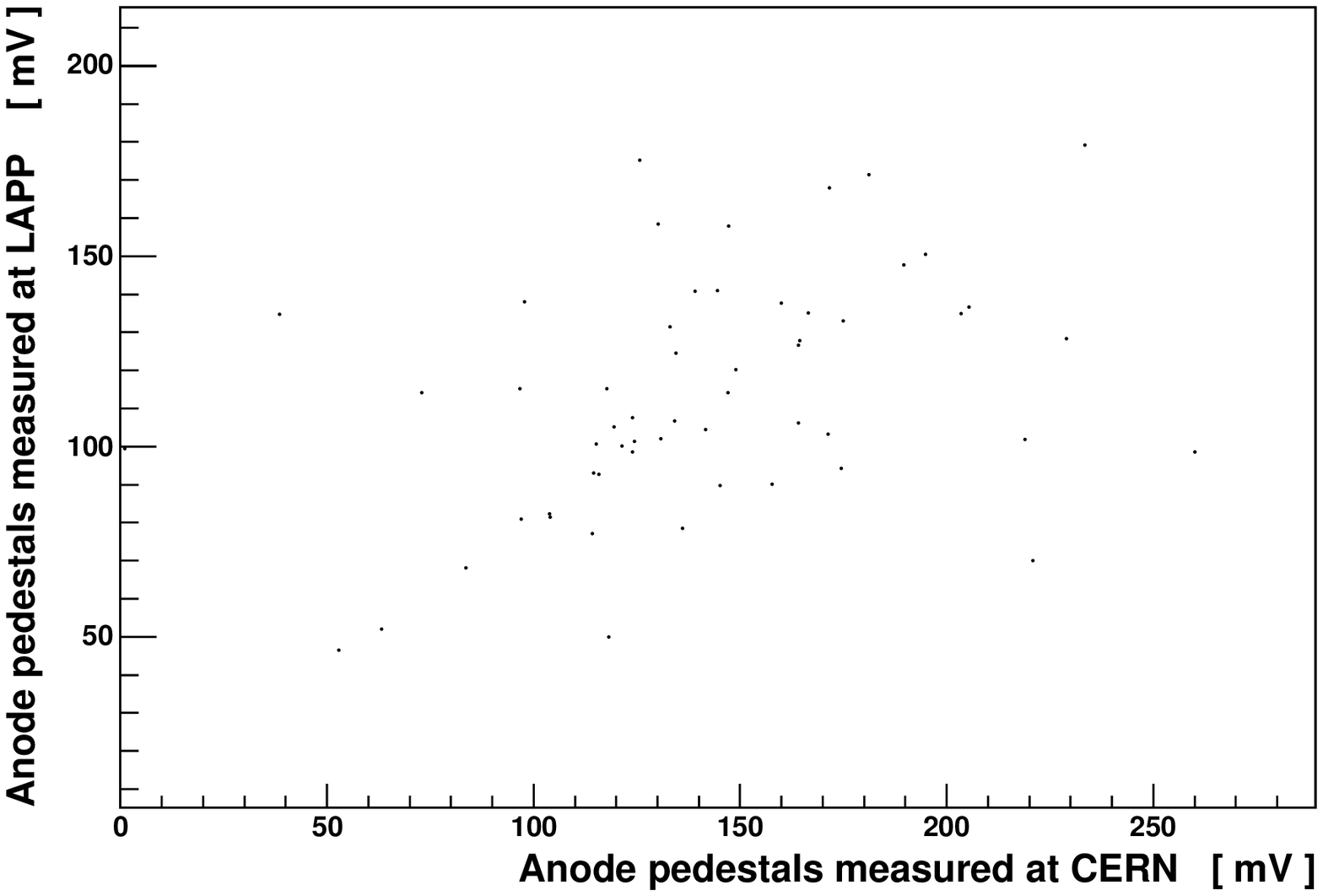}
\includegraphics[width=8.2cm]{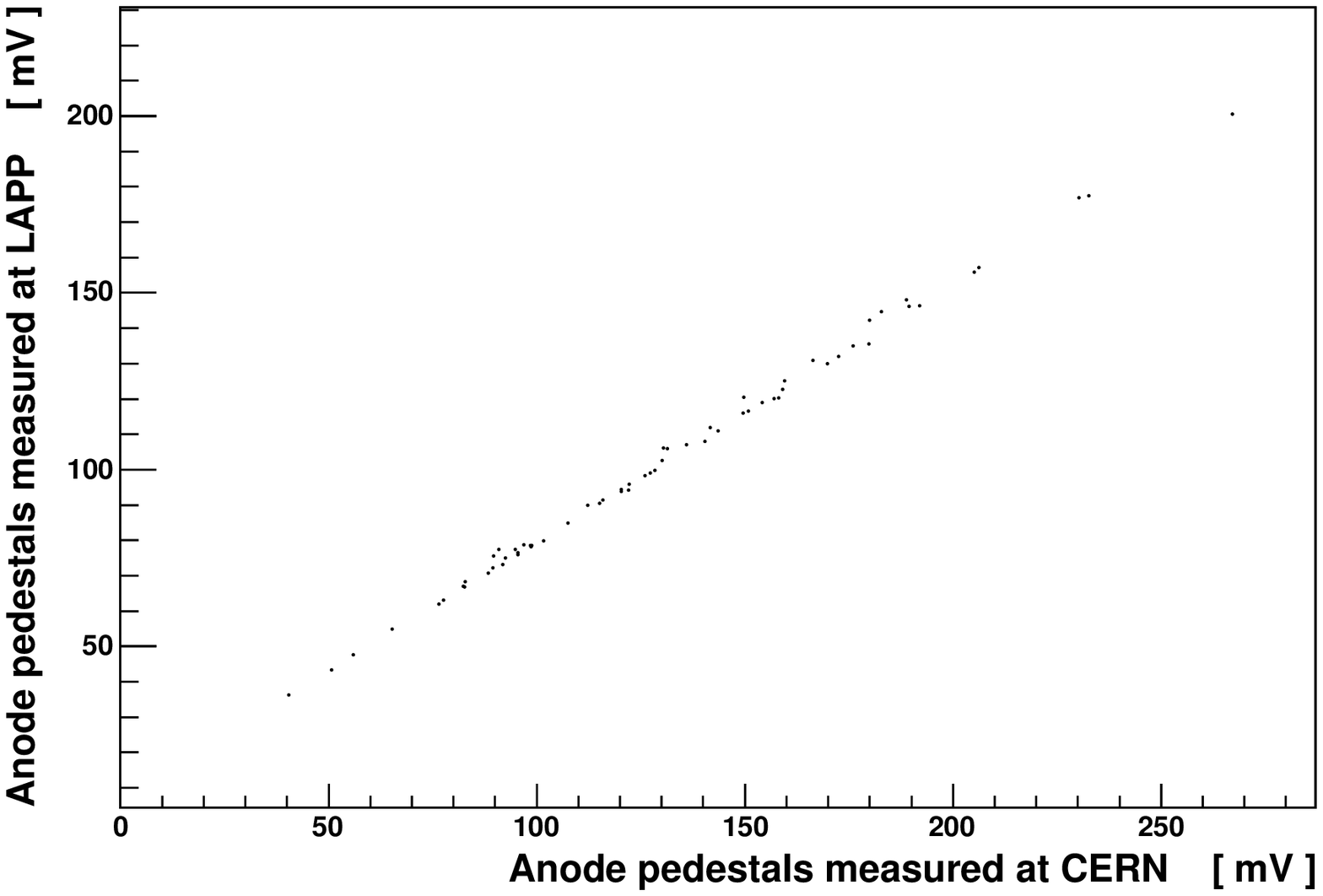}
\caption{Comparison of anode pedestal measurement at LAPP and during this
  test beam at CERN (left: A side, right: C side)}
\label{ped0}
\end{figure}

\begin{figure}[H]
\centering
\includegraphics[width=8.2cm]{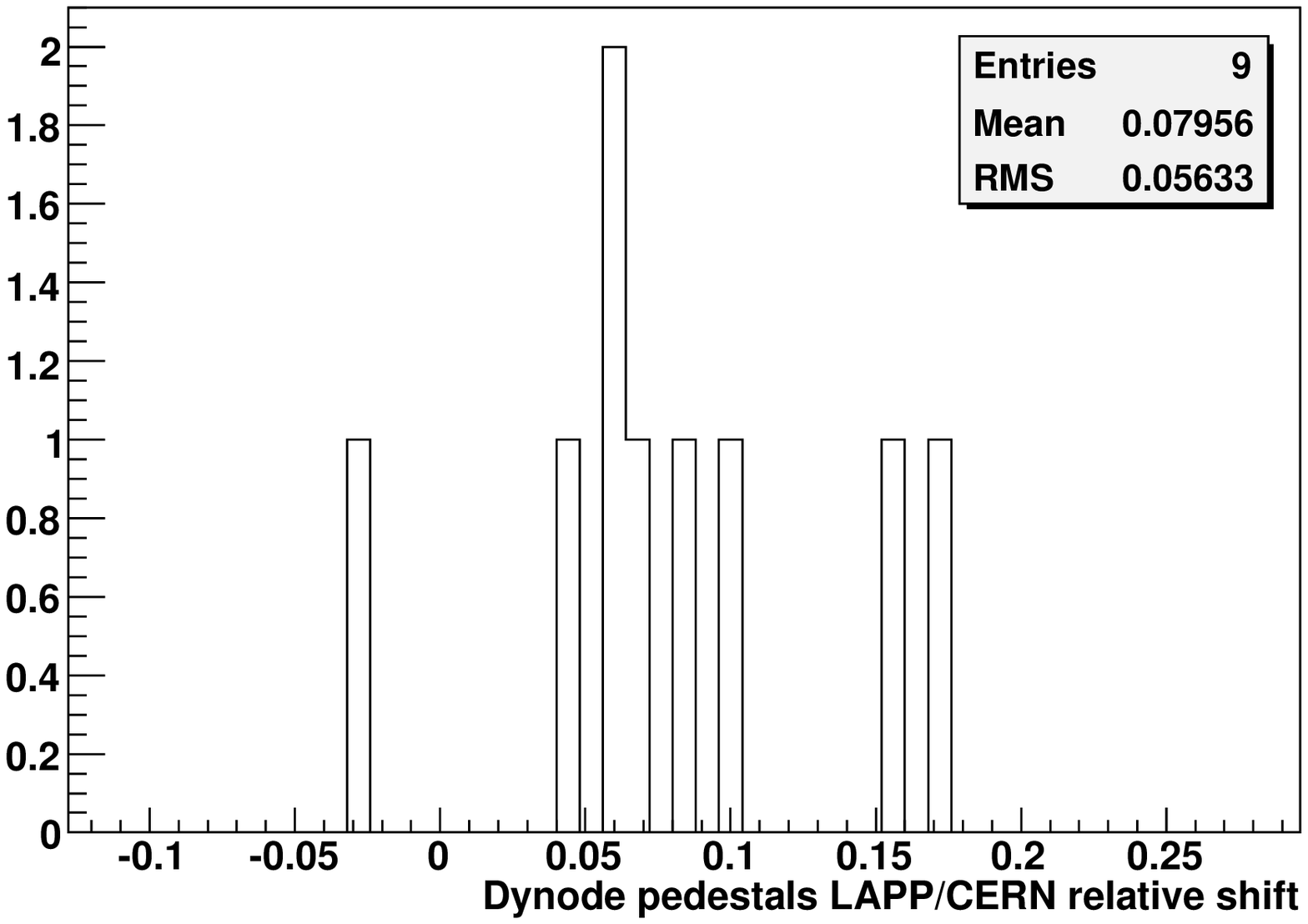}
\includegraphics[width=8.2cm]{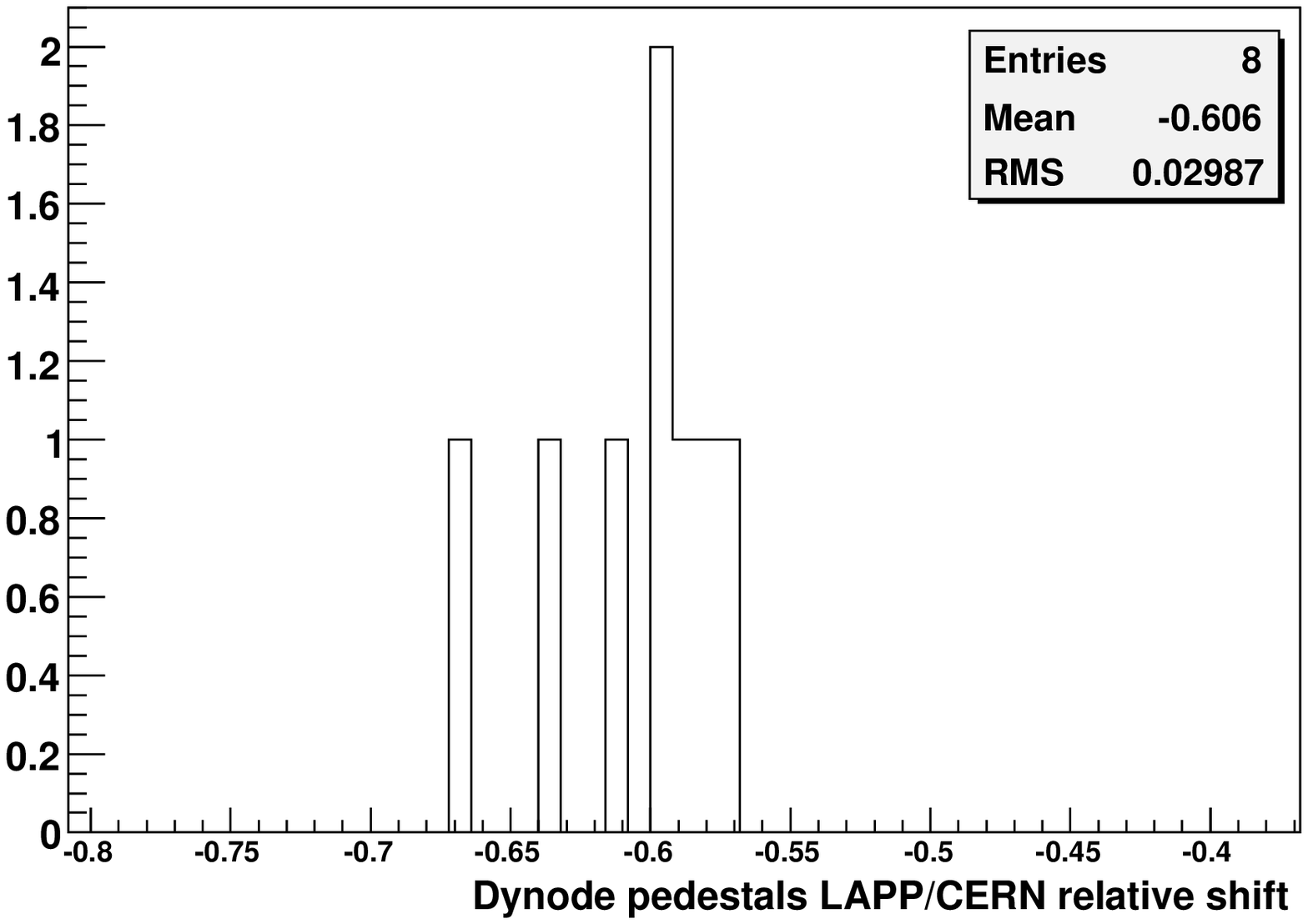}
\caption{Relative dynode pedestal shifts between previous LAPP measurements and
  this test beam measurements (left: A side, right: C side)}
\label{pedshift}
\end{figure}

As we can see the A side pedestals are shifted, unlike the C side ones which are very correlated. Concerning the dynode, the
different amplifiers have different effects on it. The amplifiers are connected to the dynode and could modify the pedestal
value. The two types of amplifier have different feedback loops so we expect having a positive dynode pedestal shift on the A
side and a negative one on the other side. Figure \ref{pedshift} shows the relative dynode pedestal shifts for all PMTs.

\subsection{Pedestal fluctuations}

\begin{figure}[H]
\centering
\includegraphics[width=8.2cm]{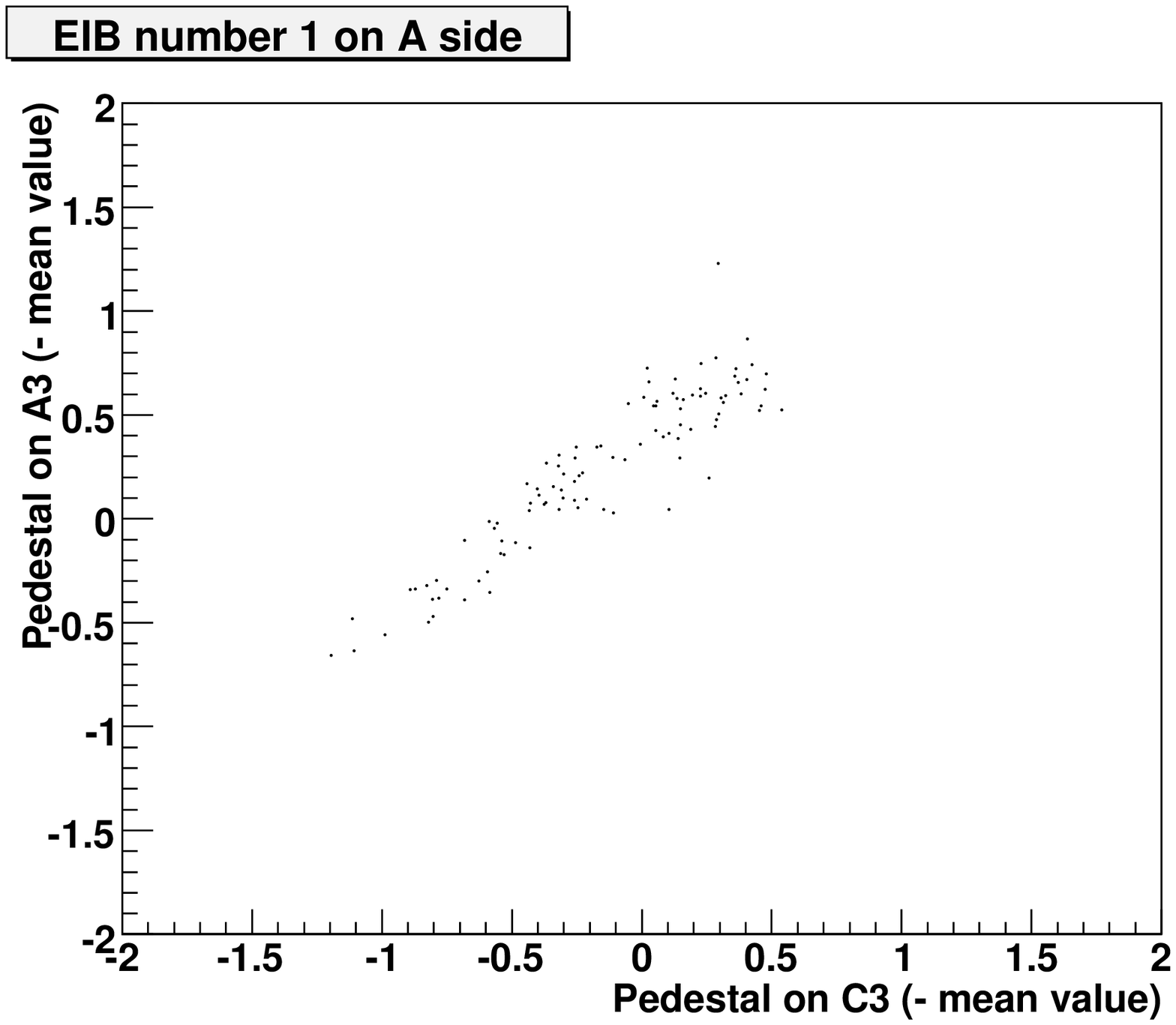}
\includegraphics[width=8.2cm]{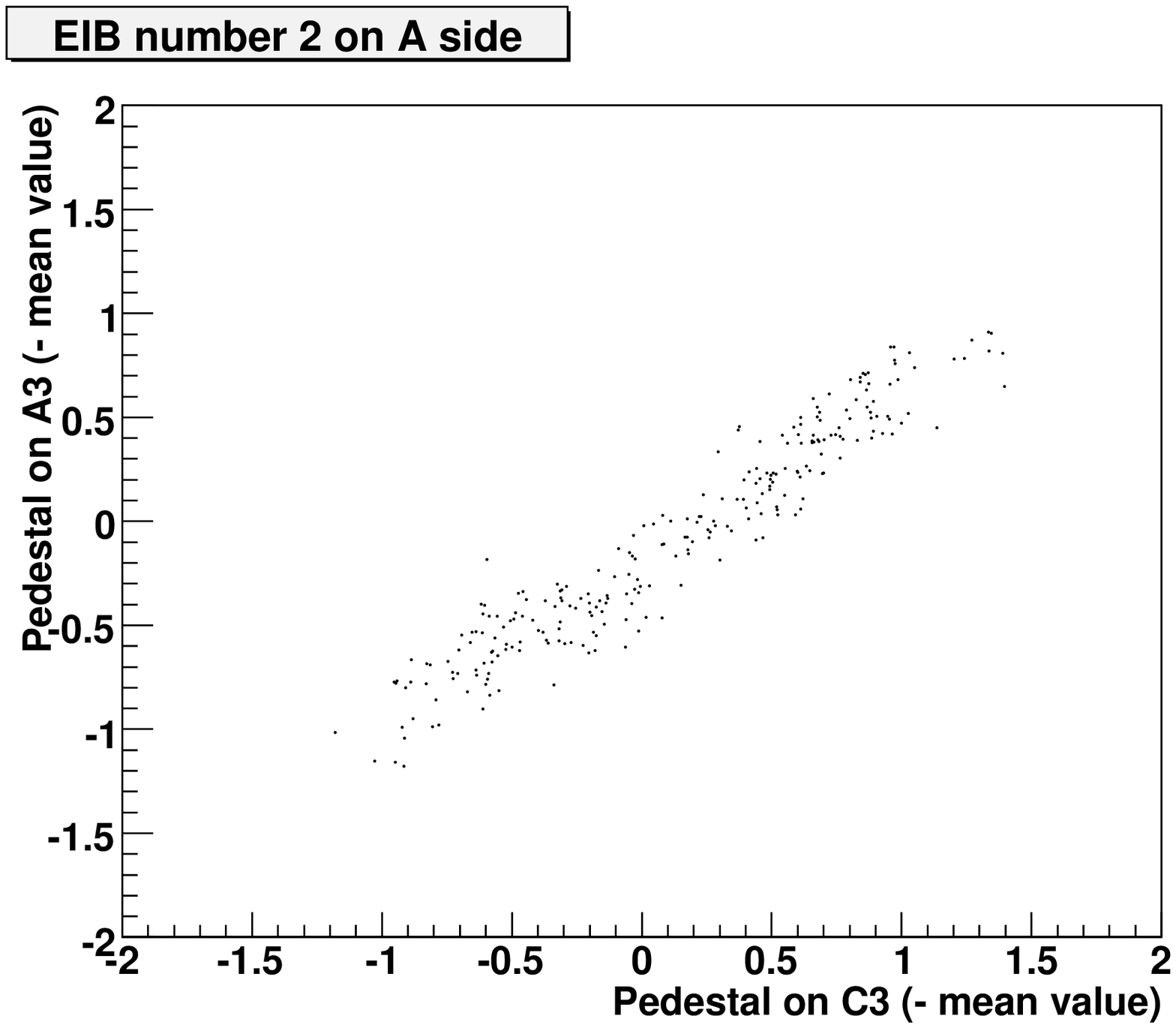}
\caption{Pedestal fluctuations correlation between both sides}
\label{correl}
\end{figure}

One can wonder if the low fluctuations observed arise from the electronics or has an external cause. In this part we show
that these fluctuations are not intrinsic to the electronics. We can see that by looking if fluctuations from one PMT to
another are correlated. The pedestal fluctuations are in fact correlated one to another, as shown on figure \ref{correl}.
This figure shows the fluctuation of an A side channel pedestal around its mean value versus the same thing on the C side.
This tends to demonstrate that the fluctuations are due to external conditions. Pedestal fluctuation correlations are also
observed between PMTs from the same side but this result is more constraining for the origin of fluctuations. In particular,
A and C side fluctuation are correlated whereas they have nothing in common except the low voltage power supply and the room
conditions like temperature for instance. A thermal vacuum test bench in Annecy allowed to measure these fluctuations
\cite{videtherm}. We conclude that the pedestal fluctuations are not due to the electronics but to external factors.

\subsection{Data}

\begin{figure}
  \centering
  \includegraphics[width=3cm]{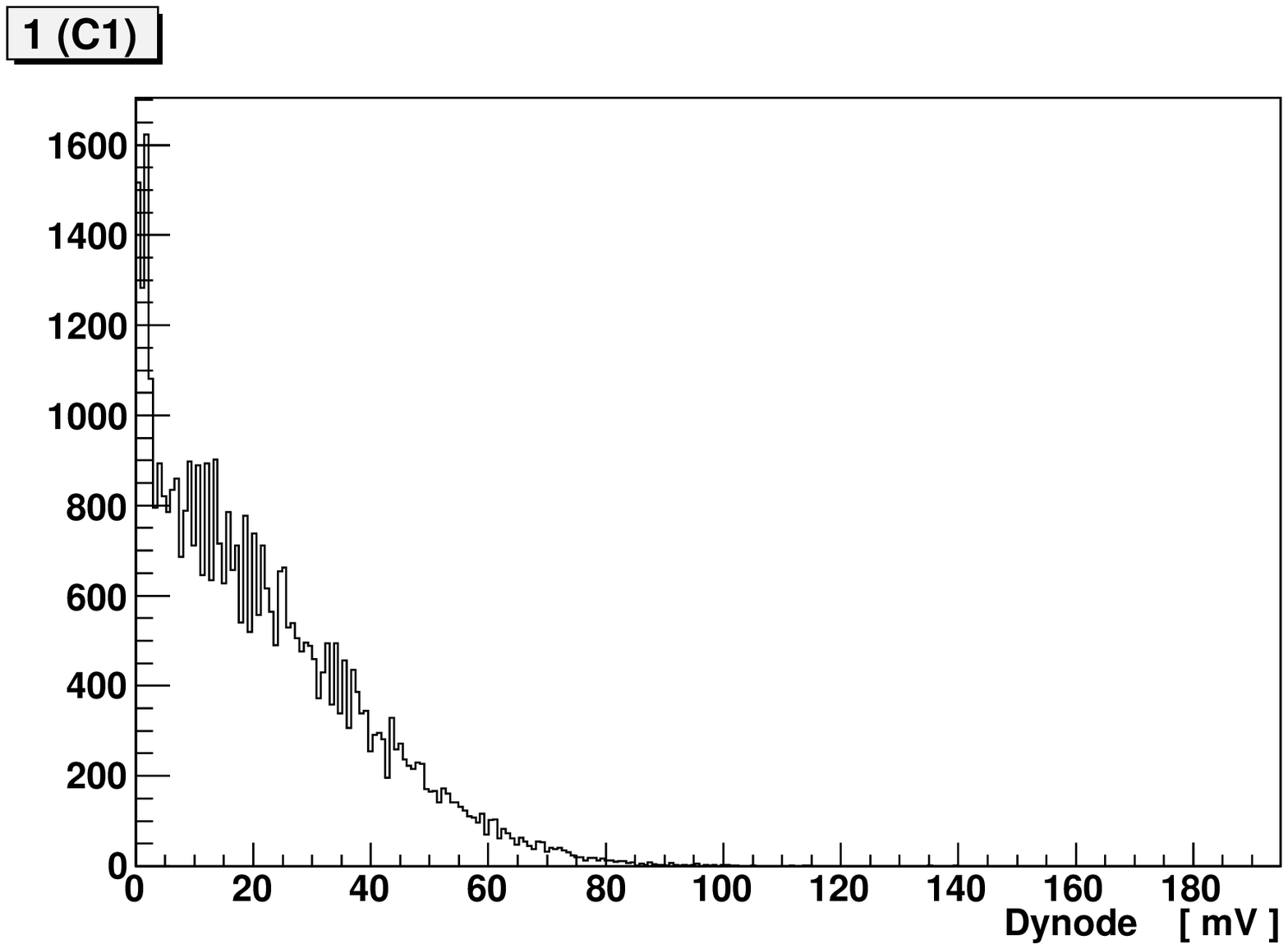}
  \includegraphics[width=3cm]{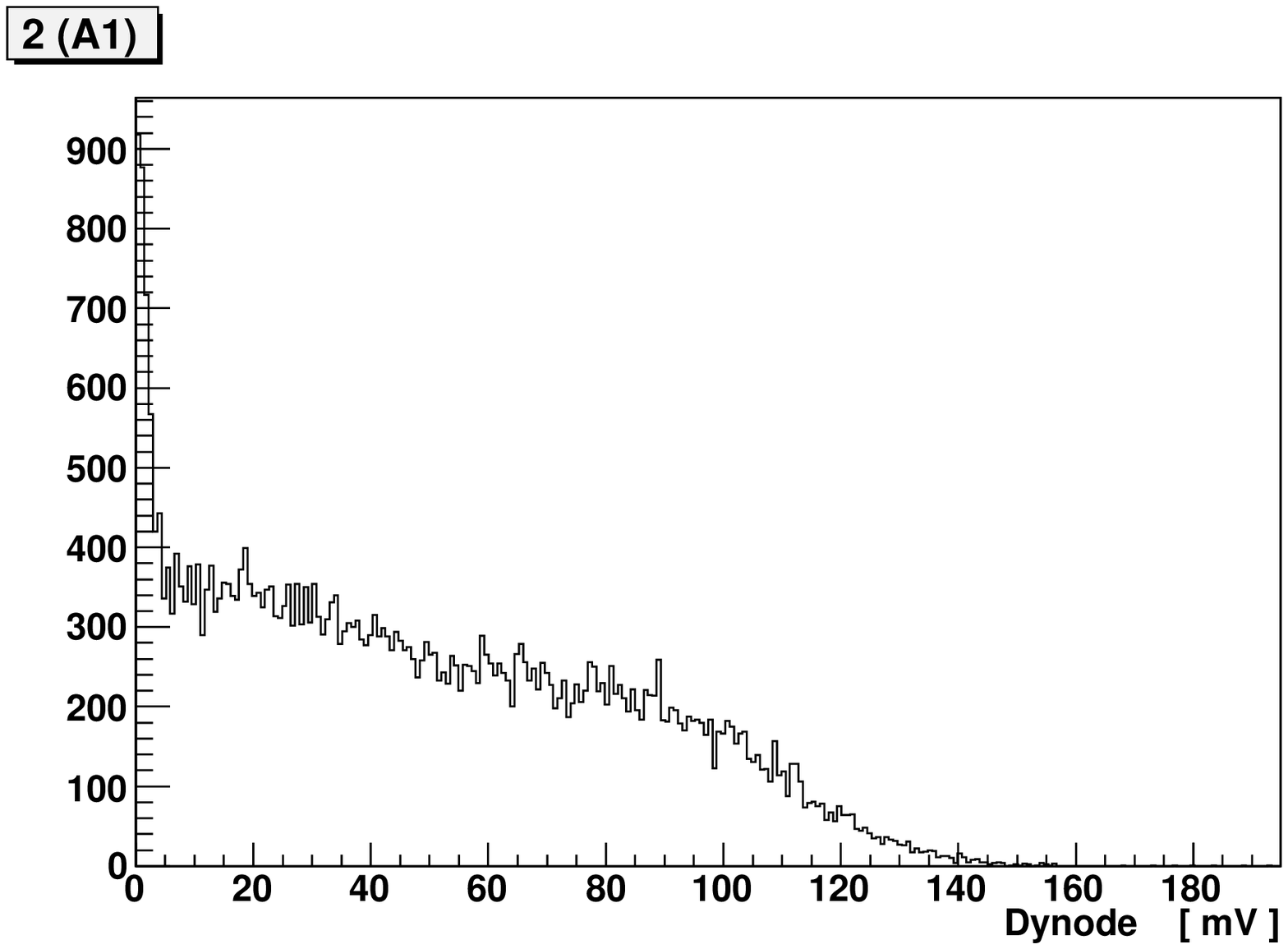}\\
  \includegraphics[width=3cm]{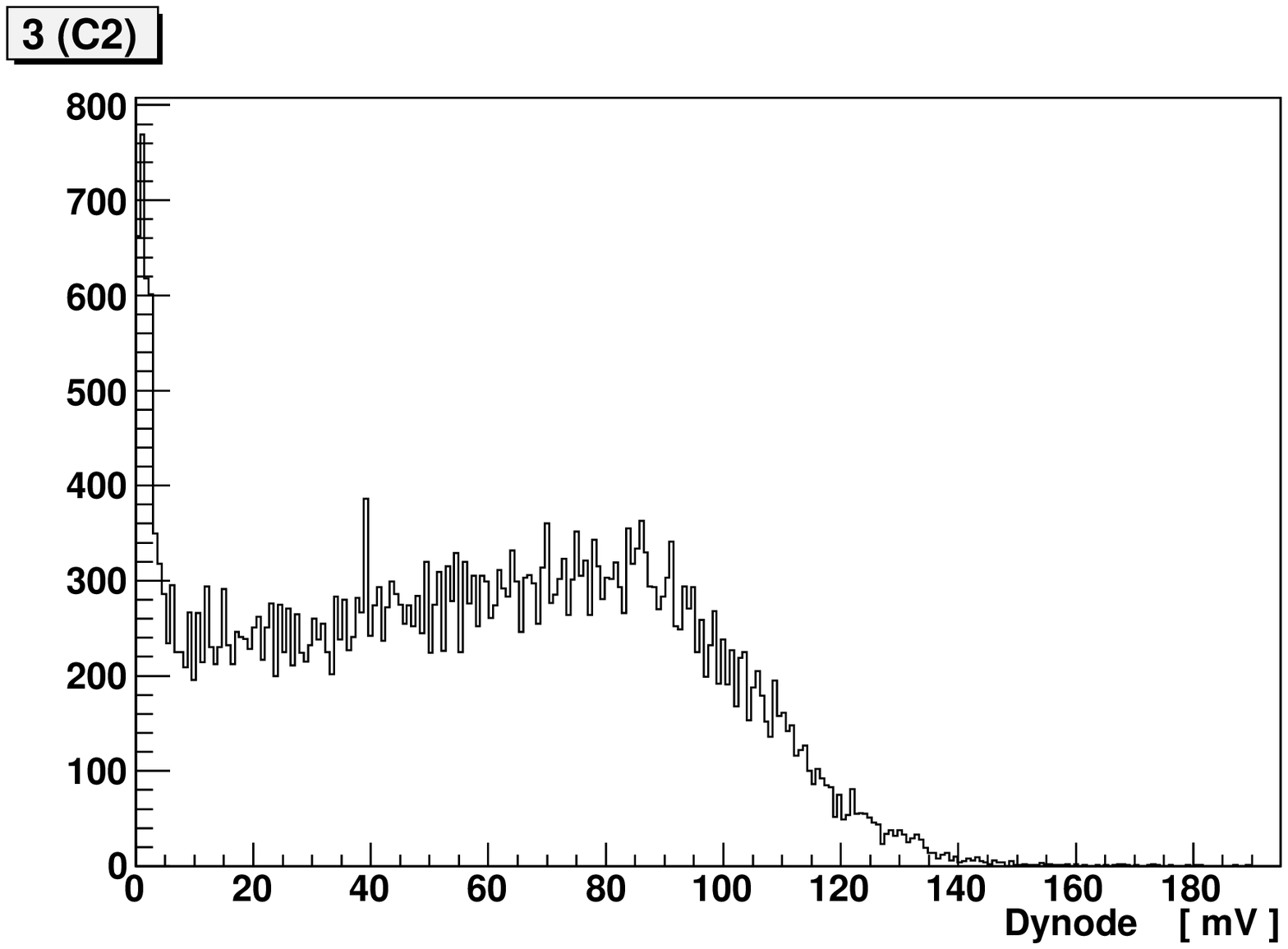}
  \includegraphics[width=3cm]{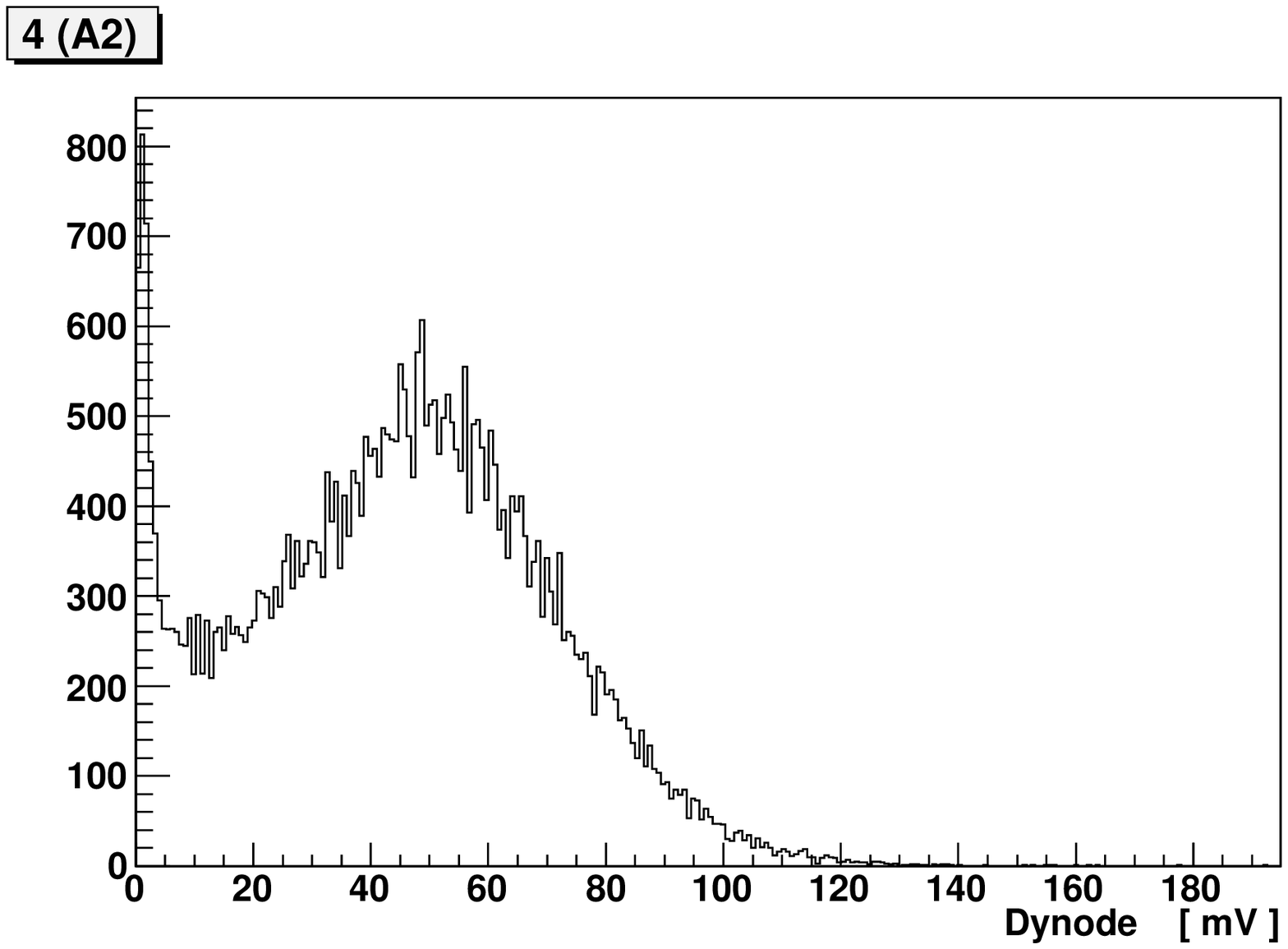}\\
  \includegraphics[width=3cm]{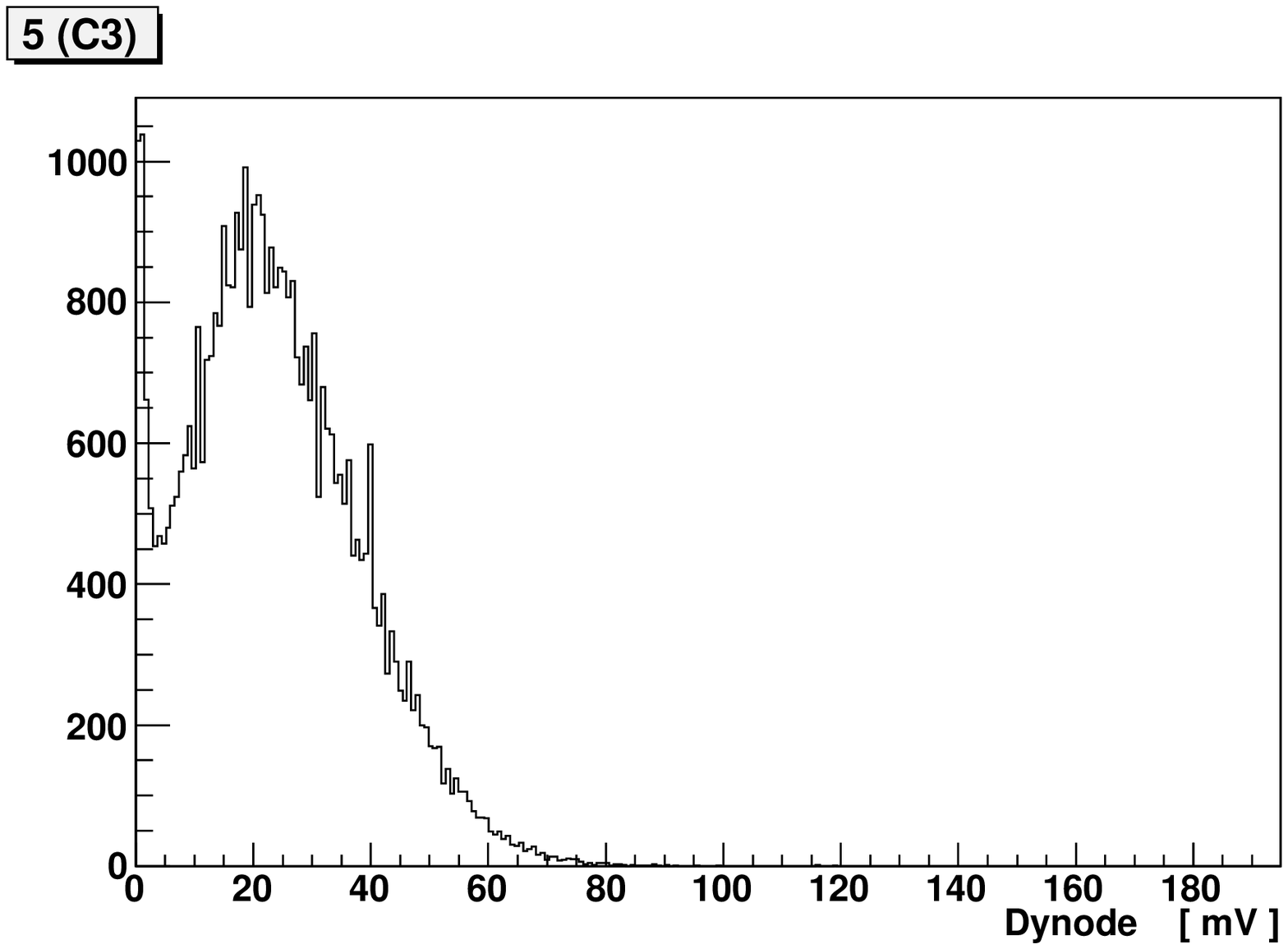}
  \includegraphics[width=3cm]{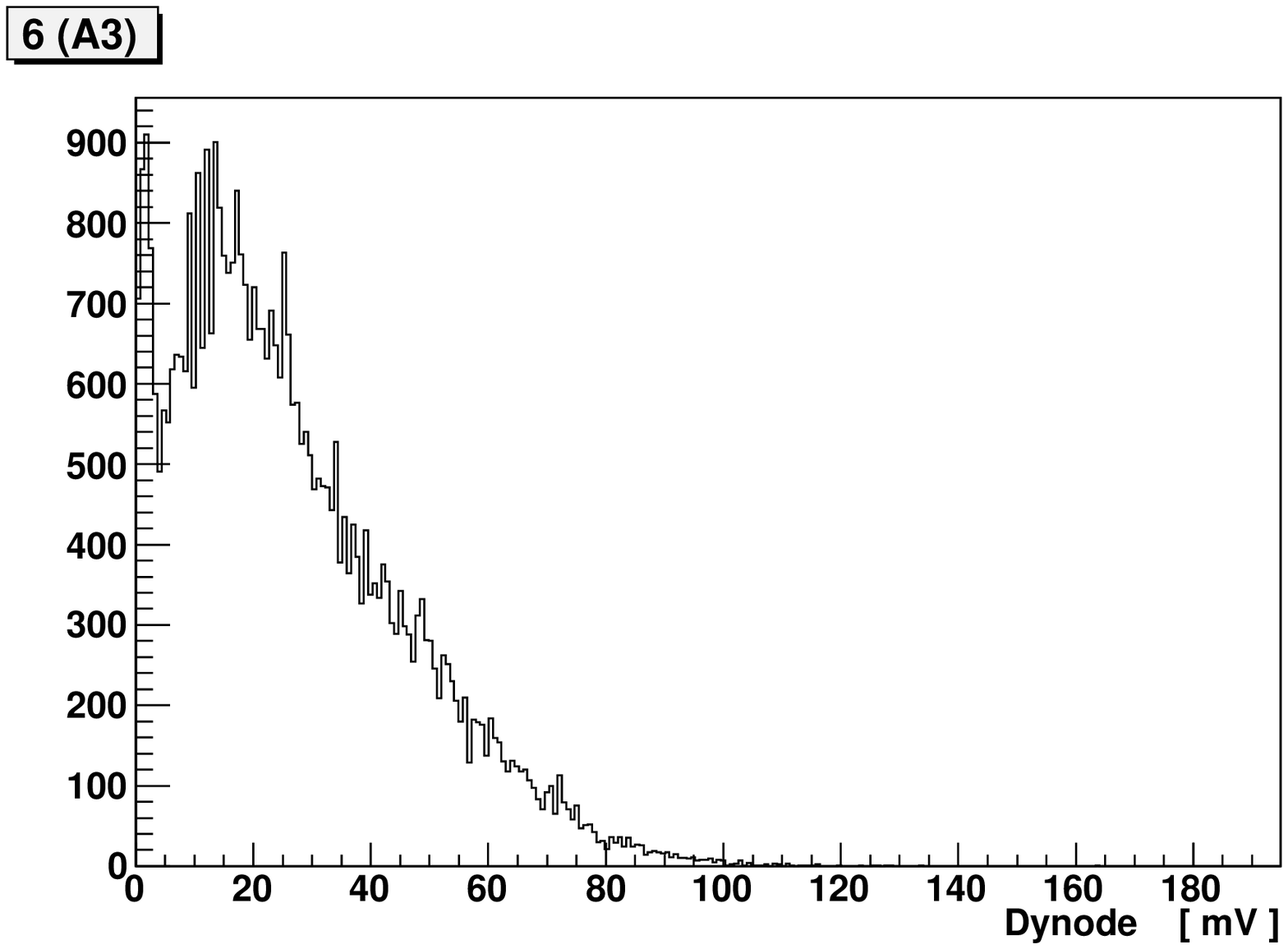}\\
  \includegraphics[width=3cm]{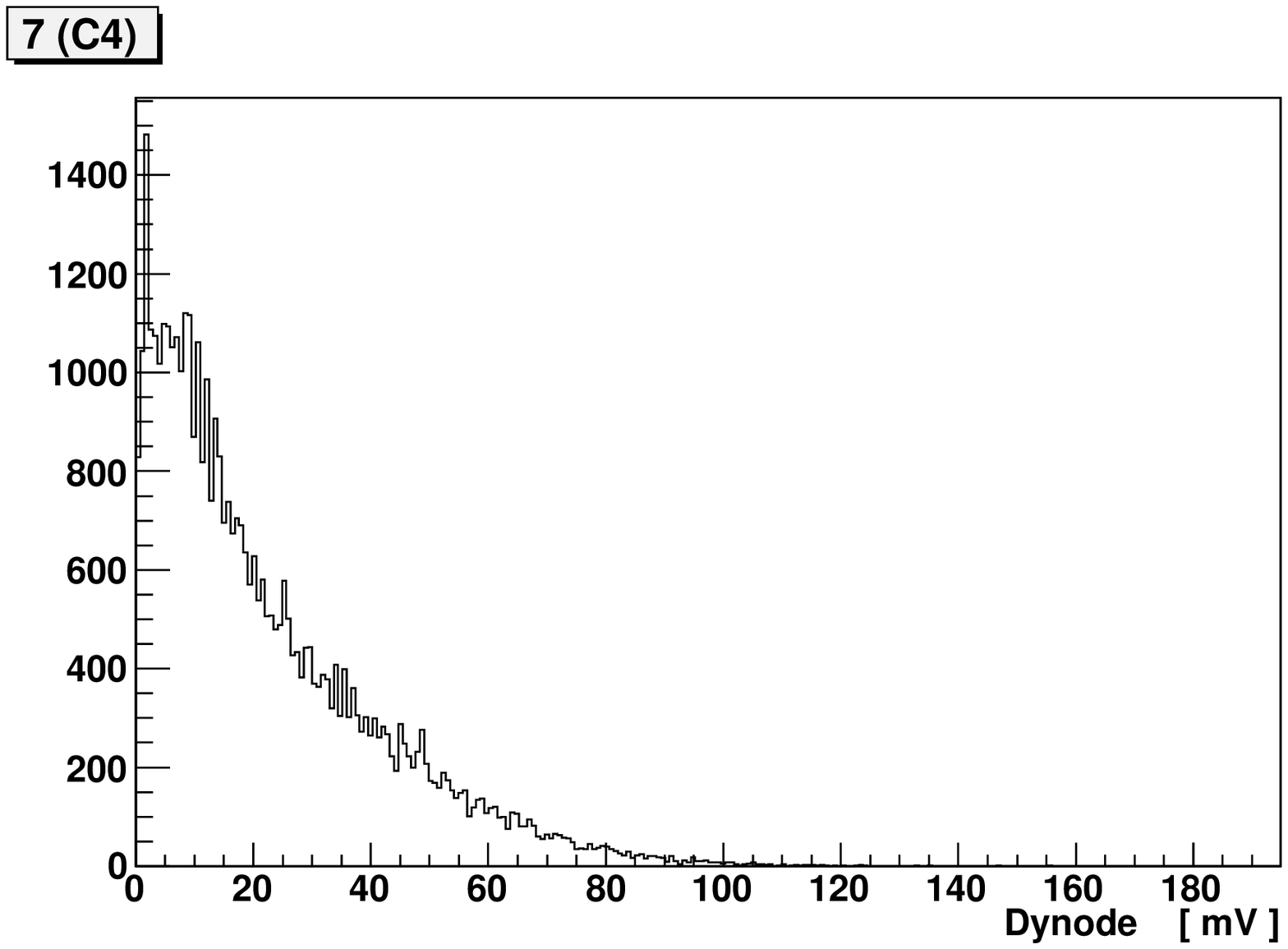}
  \includegraphics[width=3cm]{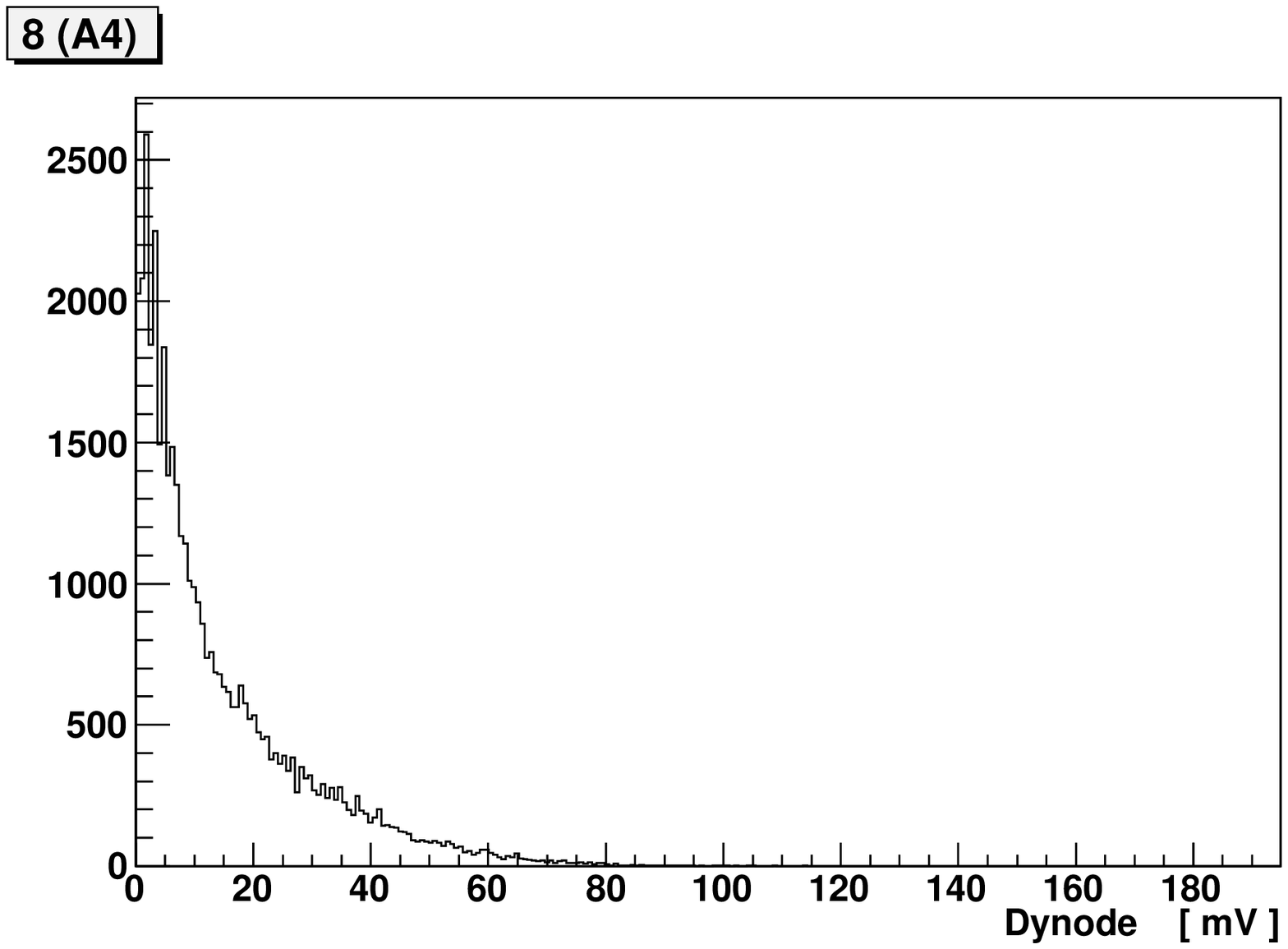}\\
  \includegraphics[width=3cm]{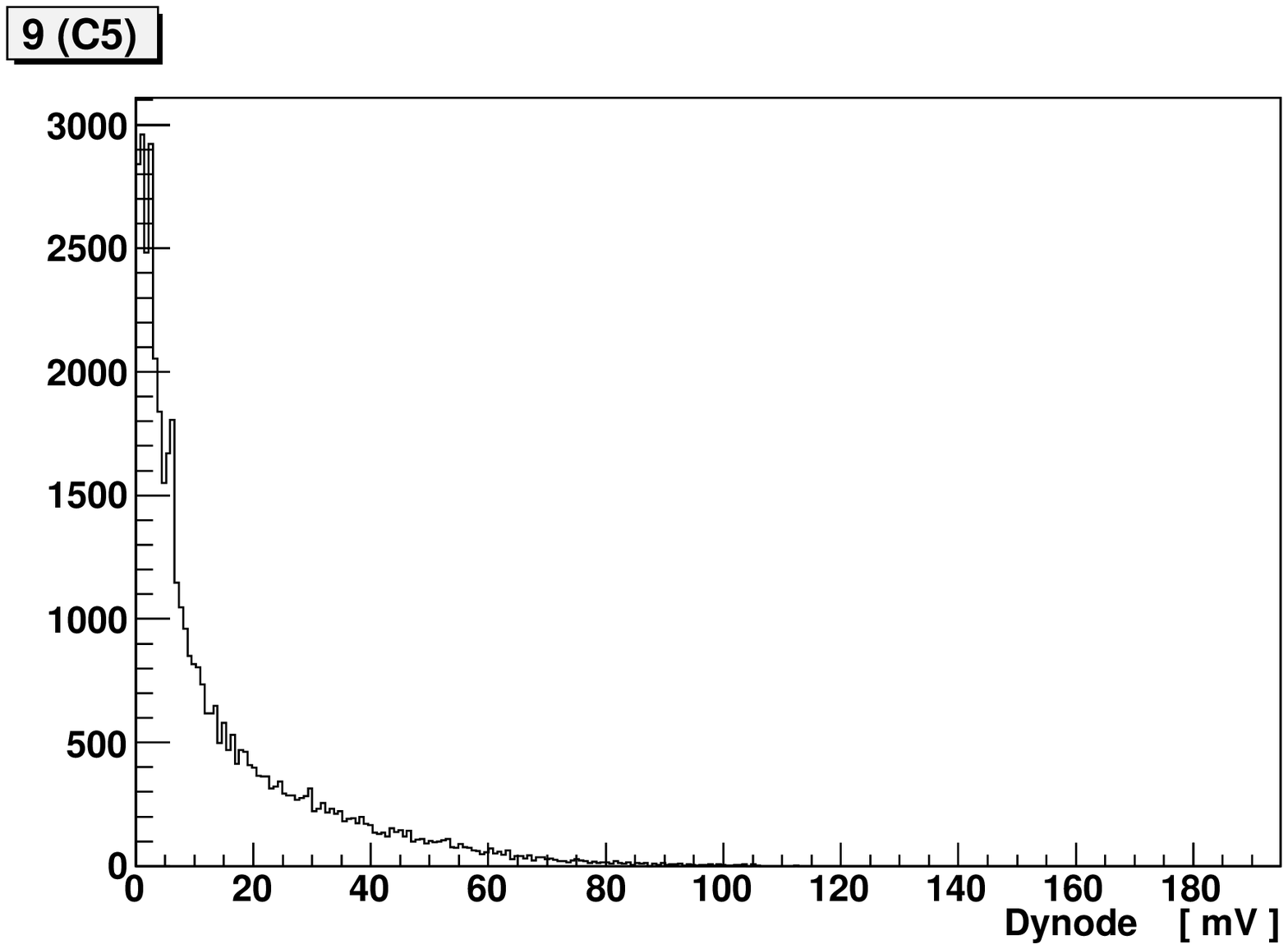}
  \includegraphics[width=3cm]{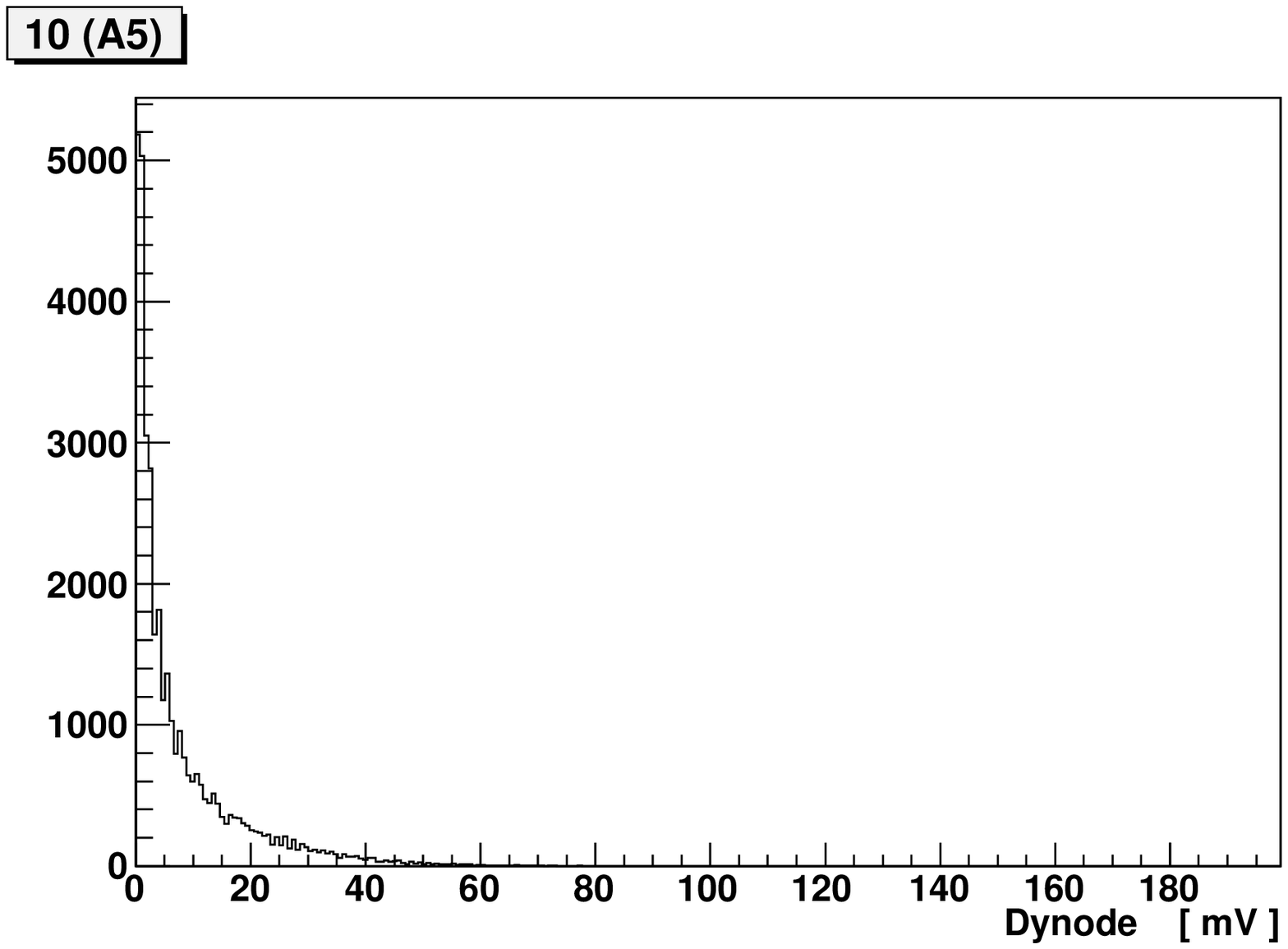}\\
  \includegraphics[width=3cm]{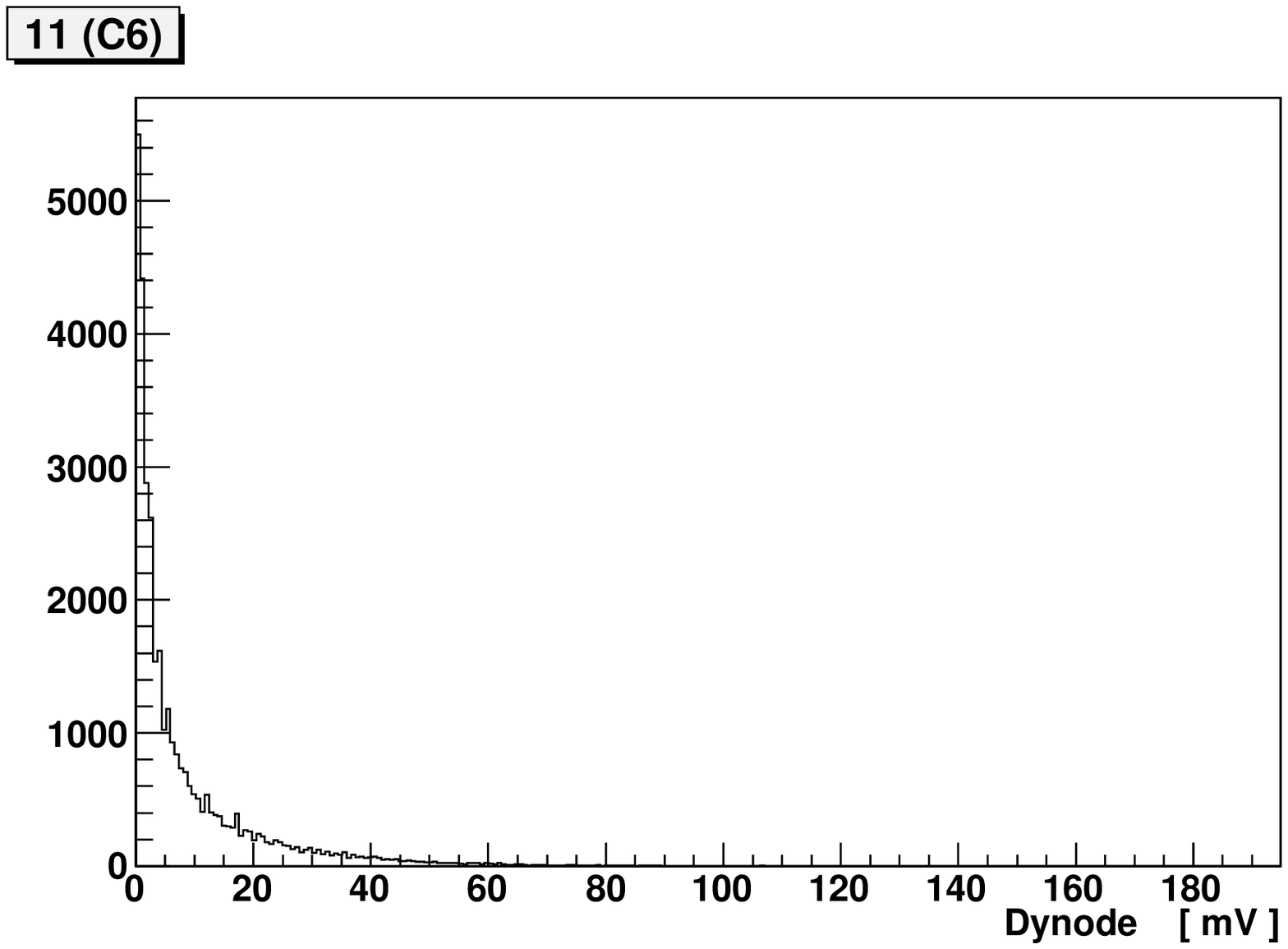}
  \includegraphics[width=3cm]{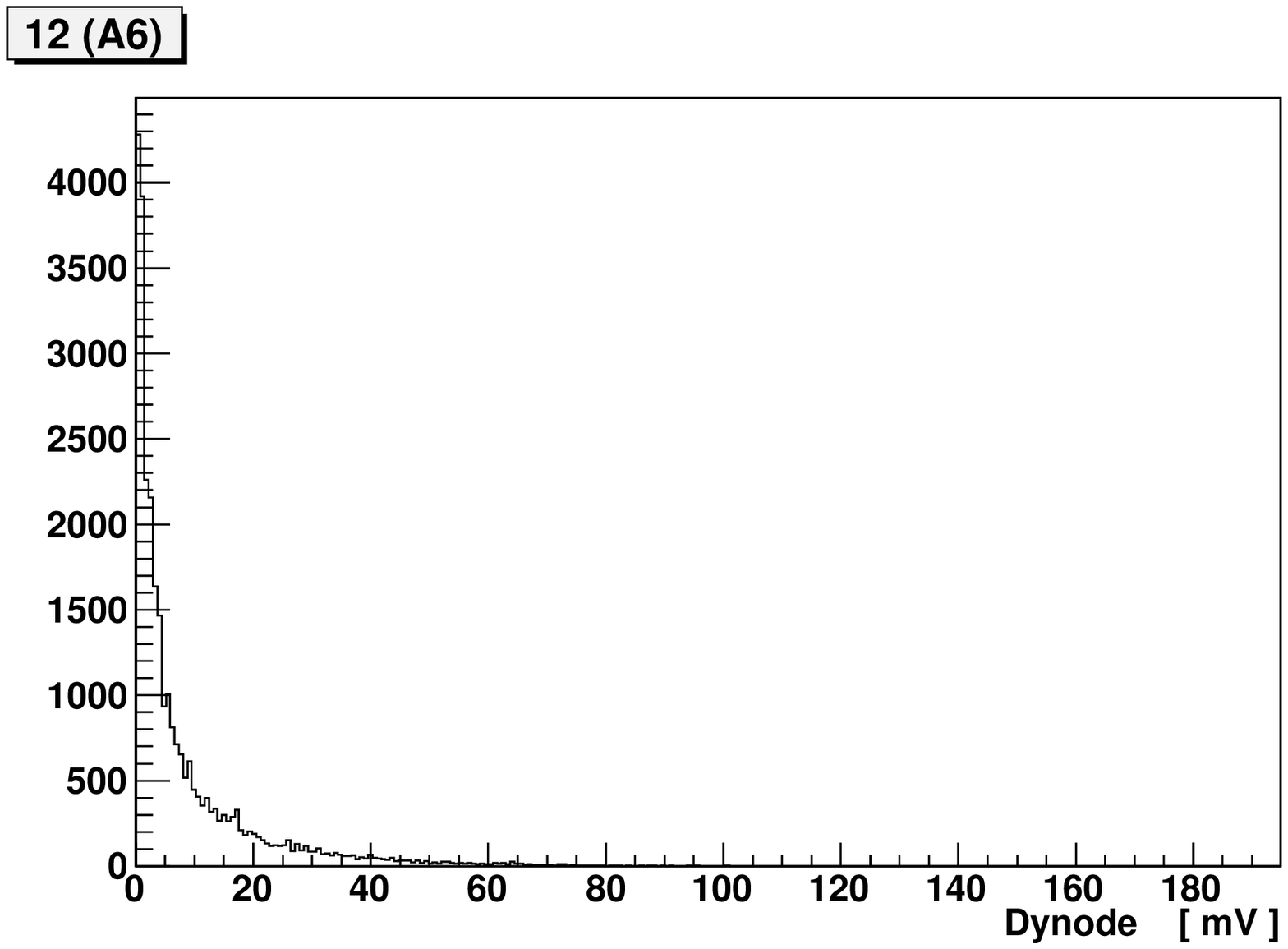}\\
  \includegraphics[width=3cm]{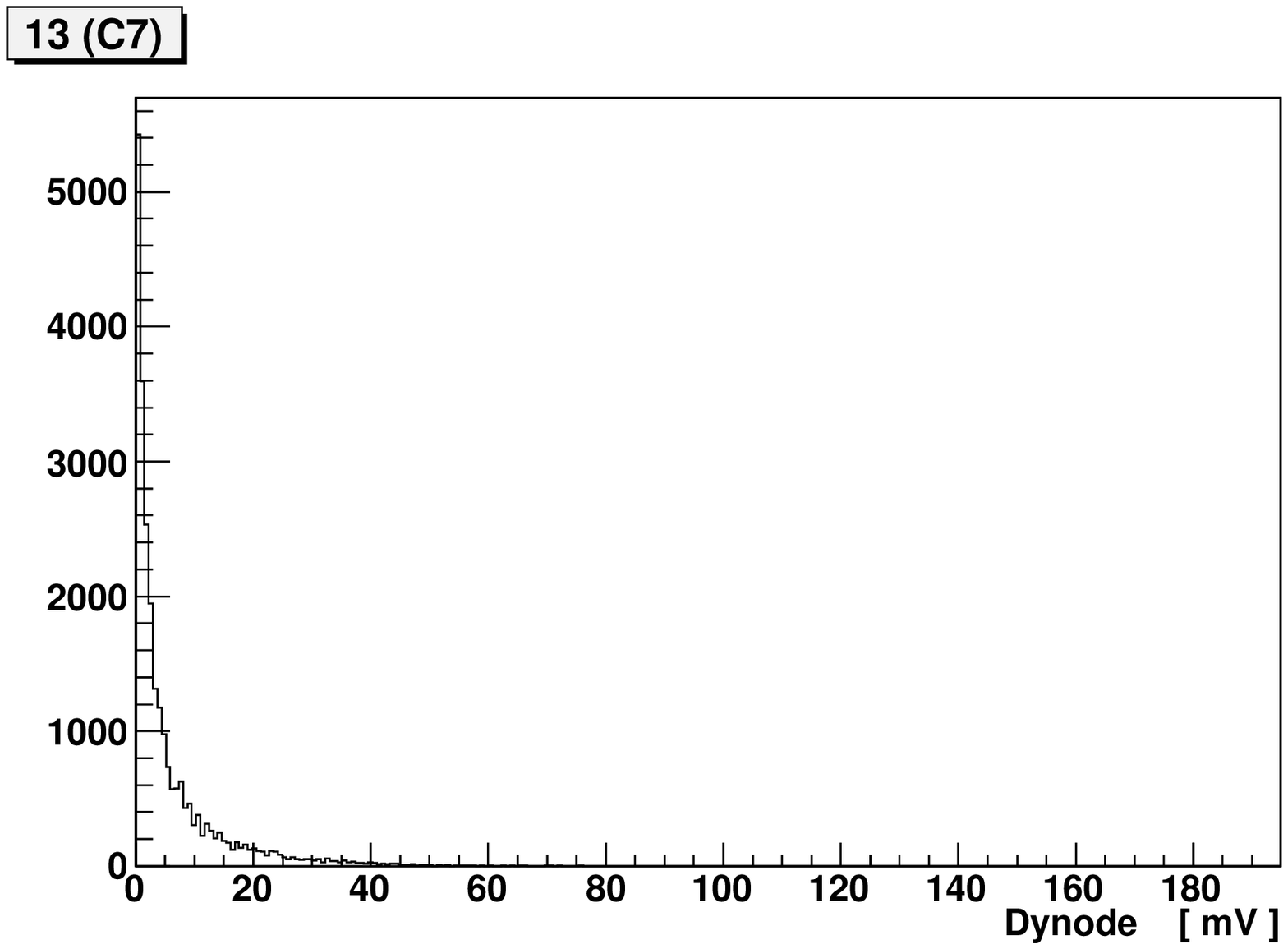}
  \includegraphics[width=3cm]{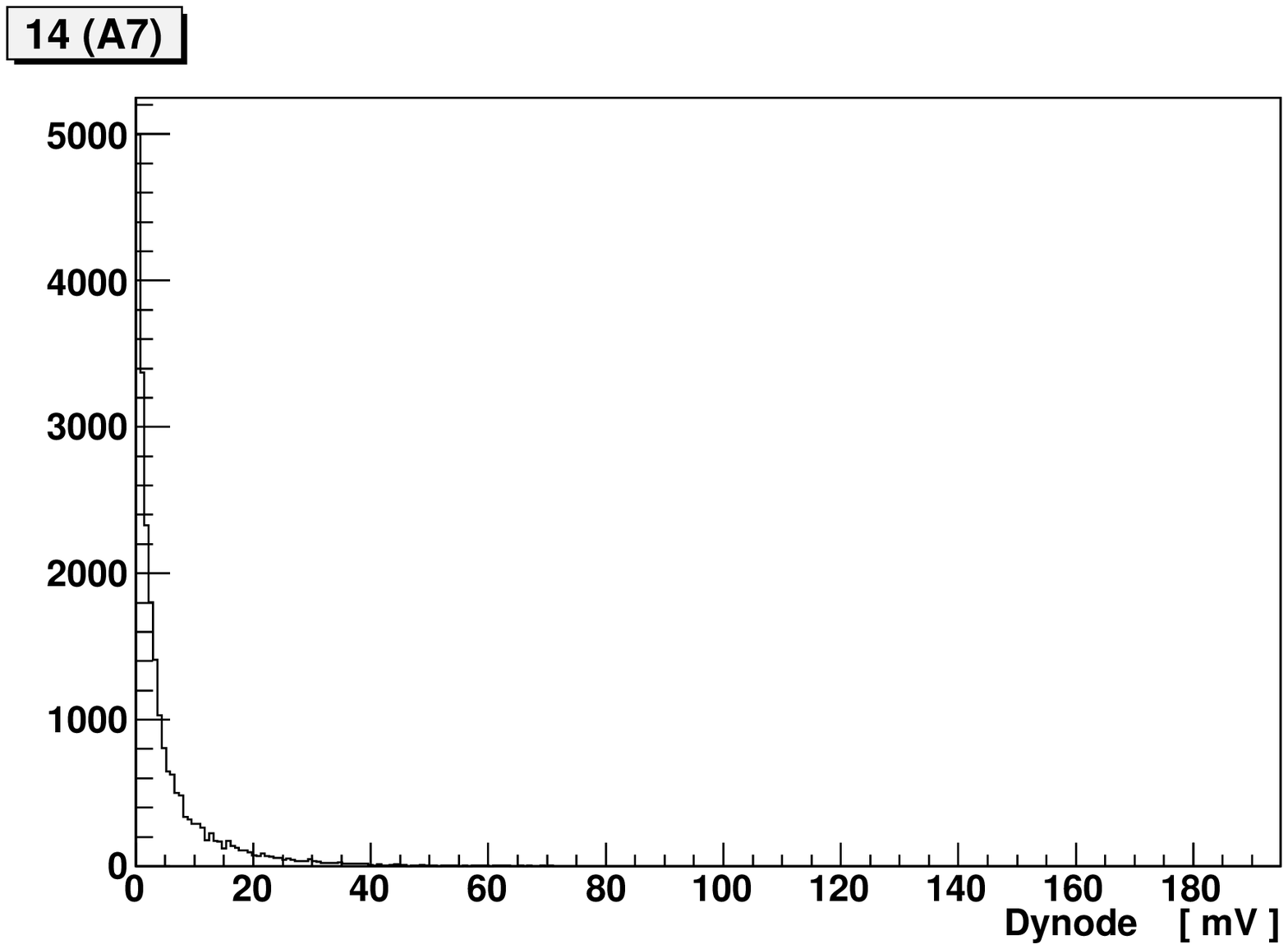}\\
  \includegraphics[width=3cm]{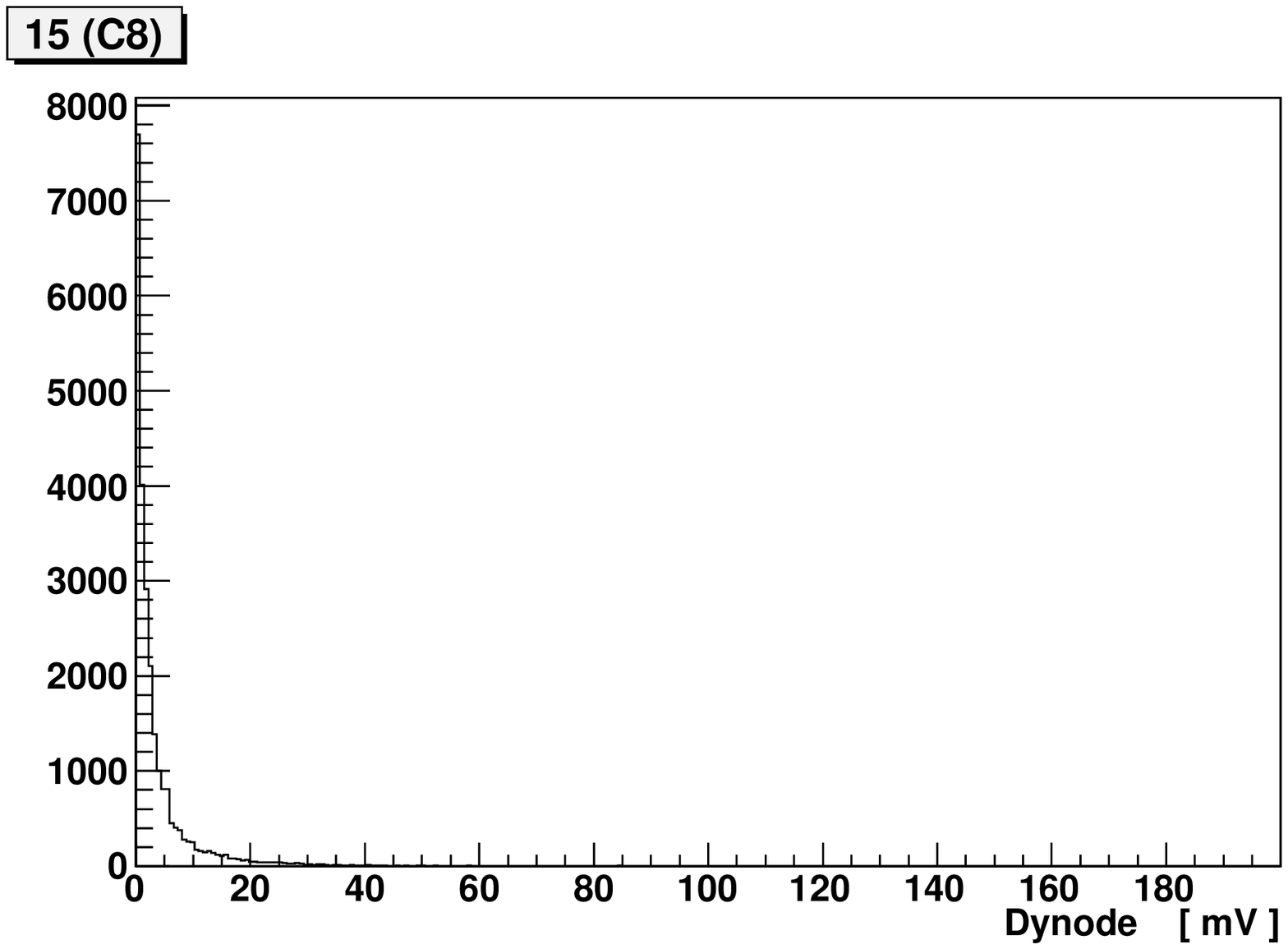}
  \includegraphics[width=3cm]{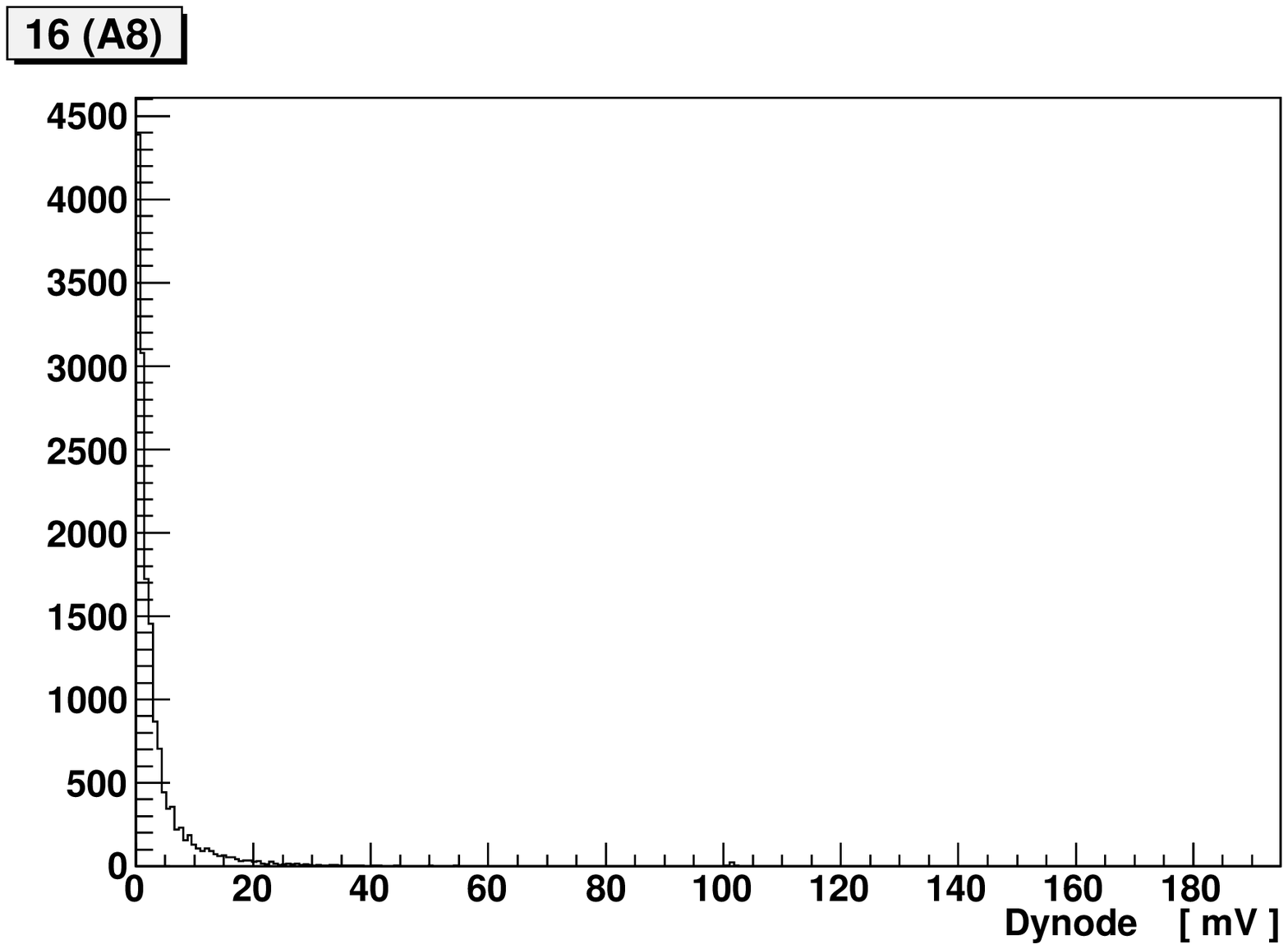}\\
  \includegraphics[width=3cm]{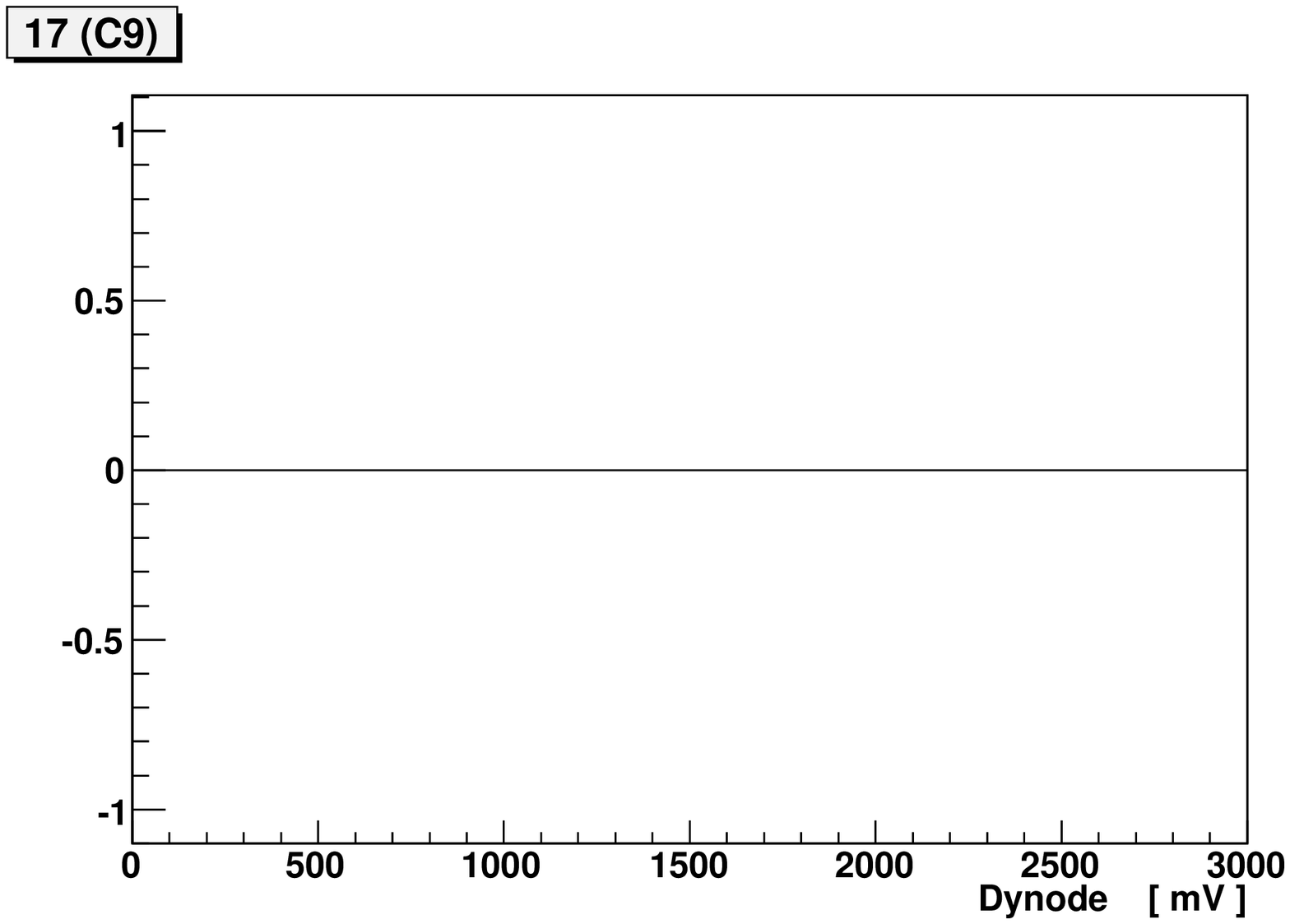}
  \includegraphics[width=3cm]{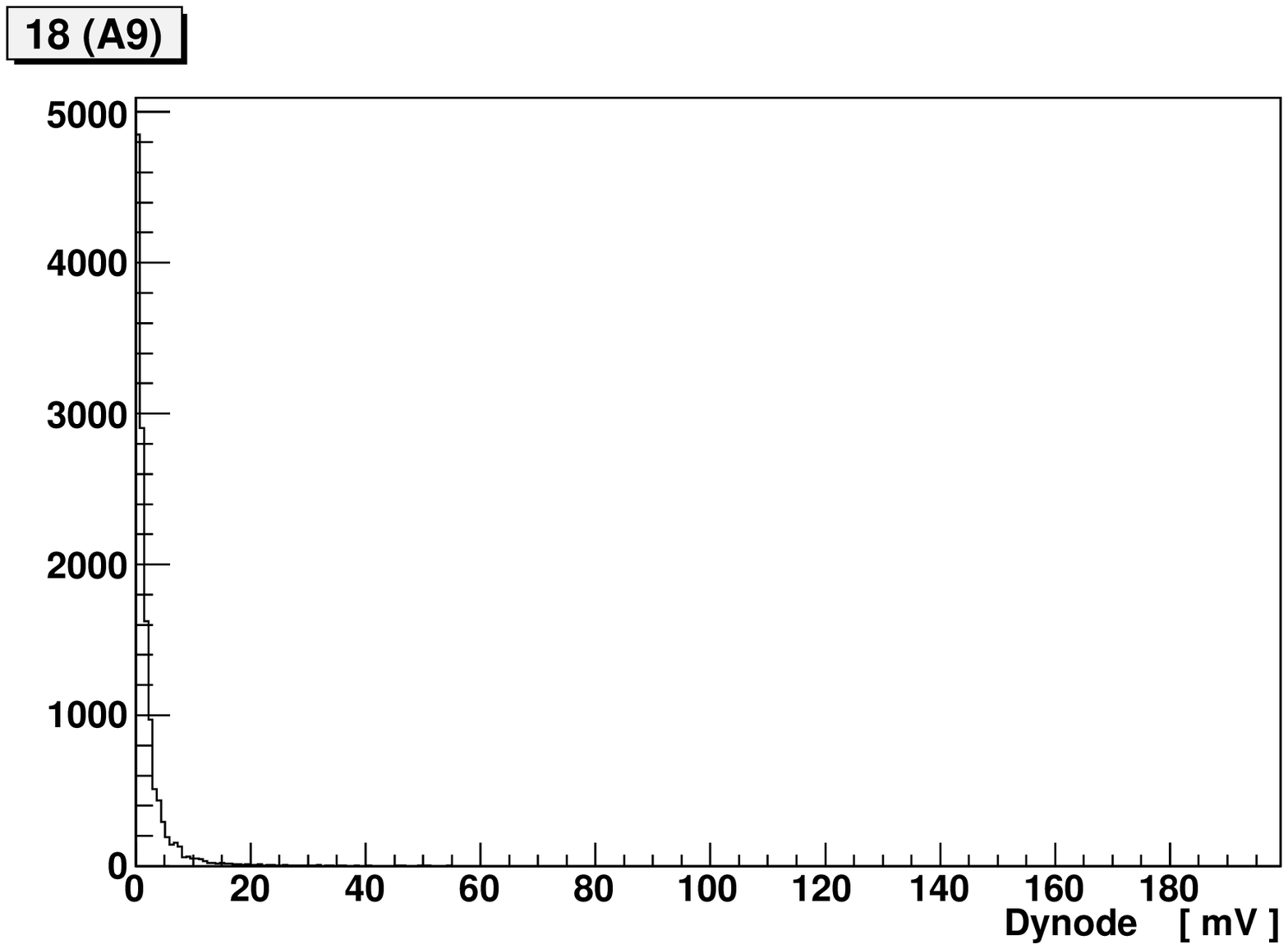}
  \caption{Signal evolution along the particle path for 7 GeV electrons}
  \label{shower}
\end{figure}

Figure \ref{shower} is an illustration to the fact that the shower
is well seen by the Ecal in the present configuration. It shows
the dynode signal of all PMTs during a 7 GeV electrons run. As the
Ecal is horizontal, about 35 radiation lengthes are present. So
the main electromagnetic energy deposition is expected to occur in
the first quarter of the Ecal. Therefore, the higher signals are
shared by PMTs C2, A2 and C3, as seen on figure \ref{shower}.

\begin{figure}[H]
\centering
\includegraphics[width=8cm]{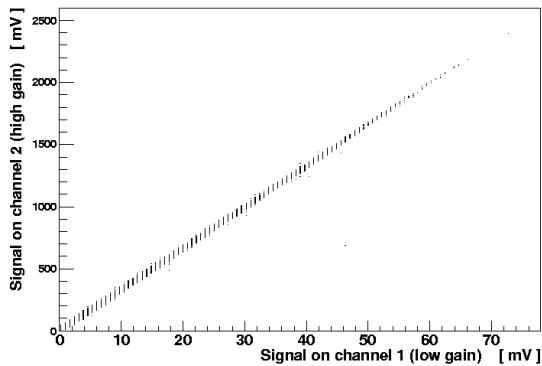}
\caption{Linearity between a high gain channel and a low gain one}
\label{gain}
\end{figure}

This also shows that all the data are correctly transmitted from PMT through the EIBs to the DAQ. Another important check was
the measurement of the gain between two anode channels related to the same pixel. In 2002 test beam, a non linearity had been
observed \cite{loic} and it was important to know if this problem was still there. In this test beam data, we see that the
gain is perfectly linear for all PMTs. Figure \ref{gain} shows this linearity for PMT A2.

The linearity constant is here 33.28. In the following, all PMTs are assumed to have the same gain, that we take equal to
33.3, according to previous test bench measurement made at LAPP \cite{videtherm}.

\subsection{First comparison of A and C amplifiers}

Apart from the study of the bus command and the $\gamma$ trigger, this test beam results were awaited to make a decision
between the two different amplifier types used on A and C side. At this point we should mention that both amplifier types had
been space-qualified after this test beam, so both of them could really be used in AMS02. The study of pedestals and data
taking quality tends to show that the C side amplifier is more stable. The comparison of pedestal fluctuations during stable
periods does not allow to prefer one from the other (rms are always the same order of magnitude). However, the fact that the
first A side EIB had to be replaced may be an indication that this type of amplifier is not very reliable. But this point has
to be compared to the results of the trigger part analysis, since we want to know which amplifier type is the more accurate
for being used on AMS02 board.

\section{Analysis part 2 : $\gamma$ trigger tests}

\subsection{Measurements principle}

\begin{figure}[H]
\centering
\includegraphics[width=15cm]{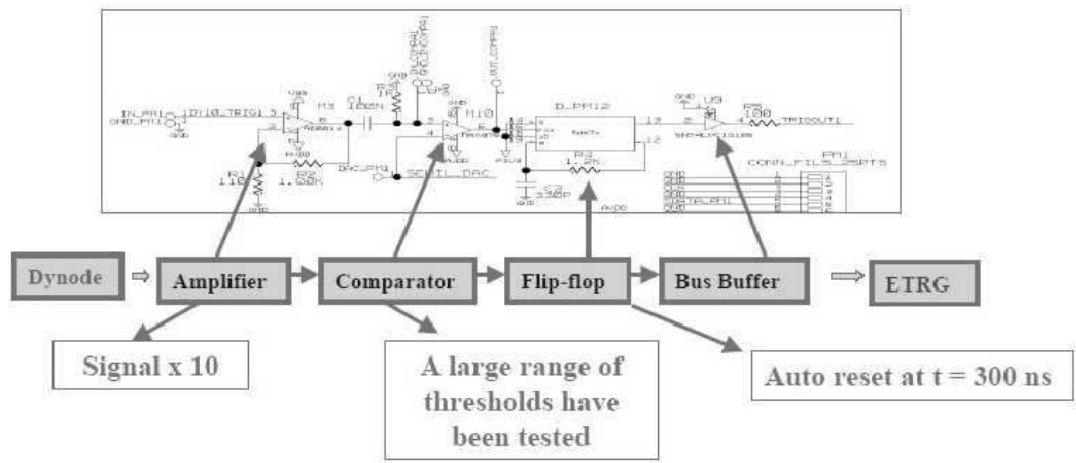}
\caption{Fast dynode signal treatment} \label{elechain}
\end{figure}

The treatment chain applied to the fast dynode signal is summarized on figure \ref{elechain}. First the signal is amplified
($\times 10$), then it is compared to a threshold and a flip-flop returns a binary signal to the EIB. The threshold was
applied from DAC and a large range has been tested during this test beam (from 10 mV to 500 mV). In a first configuration,
the flip-flop was automatically reset after 300 ns, we will see in next section that this has been modified.

\begin{figure}[H]
\centering
\includegraphics[width=15cm]{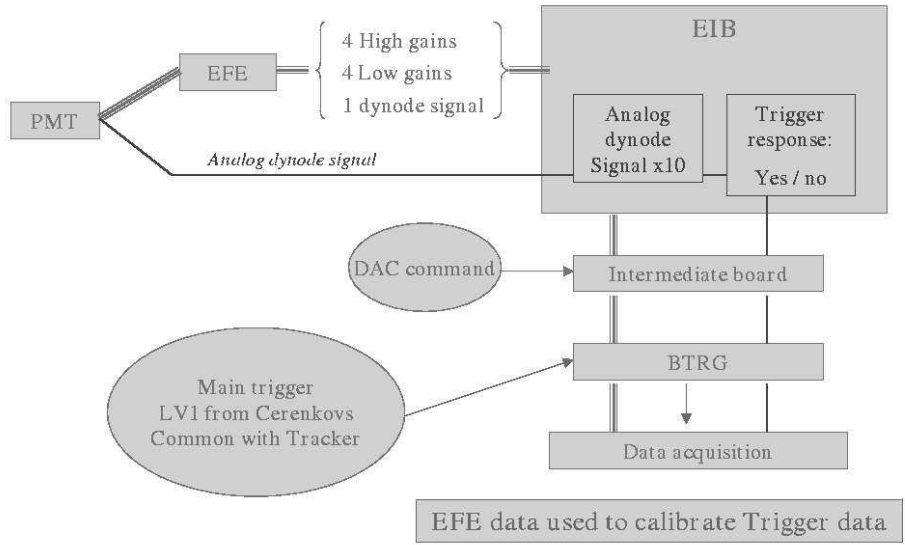}
\caption{Parallel stream of the EFE data and the trigger data} \label{trigecal}
\end{figure}

In order to measure the trigger efficiency, its response had to be recorded whenever the trigger is 0 or 1. But the analog
dynode signal is not available for a measurement since it is used for the trigger, so we use the EFE signal as a
determination of the analog signal. To do so, we can use both the digitized dynode signal or the anode signals. Figure
\ref{trigecal} shows the measurement strategy : it allows to get EFE and trigger data simultaneously.

\subsection{First days adjustments}

\begin{figure}[H]
\centering
\includegraphics[width=10cm]{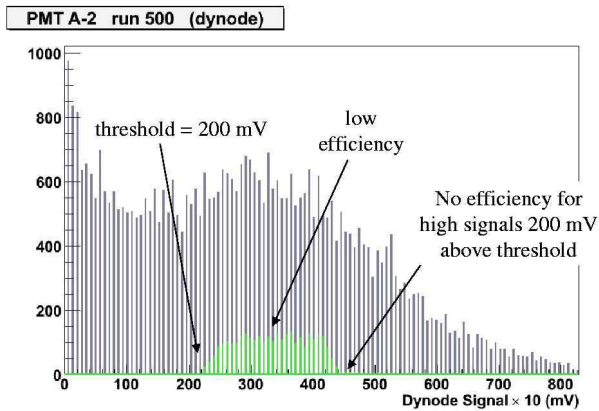}
\caption{Example of a problematic run: superposition of triggered events (green) and all of them (grey)}
\label{noeff}
\end{figure}

In the first days runs, after we checked that data acquisition worked correctly, some problems had been observed. We did
observe that signals with a trigger 1 were present only above the applied threshold. When changing the trigger threshold on
PMTs via the DAC command, the corresponding threshold evolution was observed on signals but there happened to be an
inefficiency for high signals. The ratio of events with trigger at 1 over all the events above the applied threshold is a
measurement of the trigger efficiency. When comparing the amount of data above threshold, it was observed that a lot of
events returned a trigger bit equal to 0, whereas they should have returned 1. The efficiency was only about 20\% and dropped
down to 0\% 200 mV above threshold. Figure \ref{noeff} illustrates this situation: we expect to observe the threshold around
the applied DAC value (here 200 mV), but all events above threshold should have been recorded (i.e. should have returned 1).

\begin{figure}[H]
\centering
\includegraphics[width=15cm]{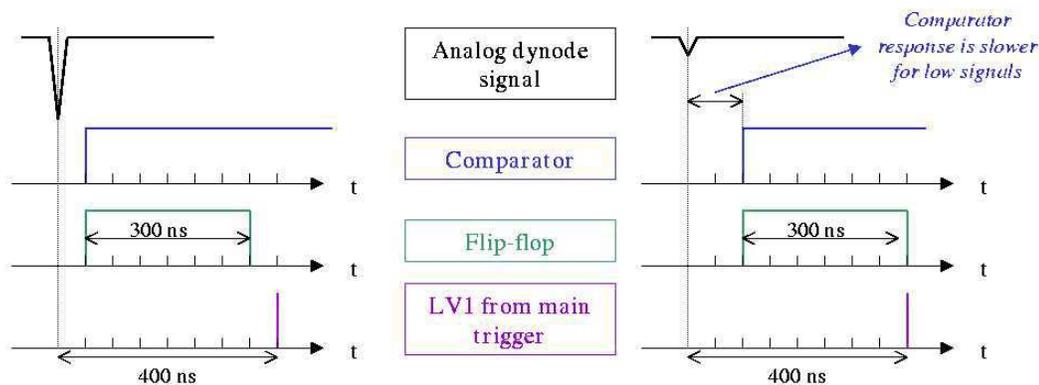}
\caption{Time evolution of signals : the high (fast) signal is erased within the nominal delay,
  unlike the low (slower) one}
\label{autoreset}
\end{figure}

It appeared in fact that the flip-flop auto-reset was too fast, and prompt signals were erased. The acquisition waited for
the main trigger, which unfortunately delivered its signal after the auto-reset has done its job. For low signals, the
comparator response is slower and the probability not to be erased increases. Figure \ref{autoreset} gives a summary of this
situation.

\begin{figure}[H]
\centering
\includegraphics[width=8.2cm]{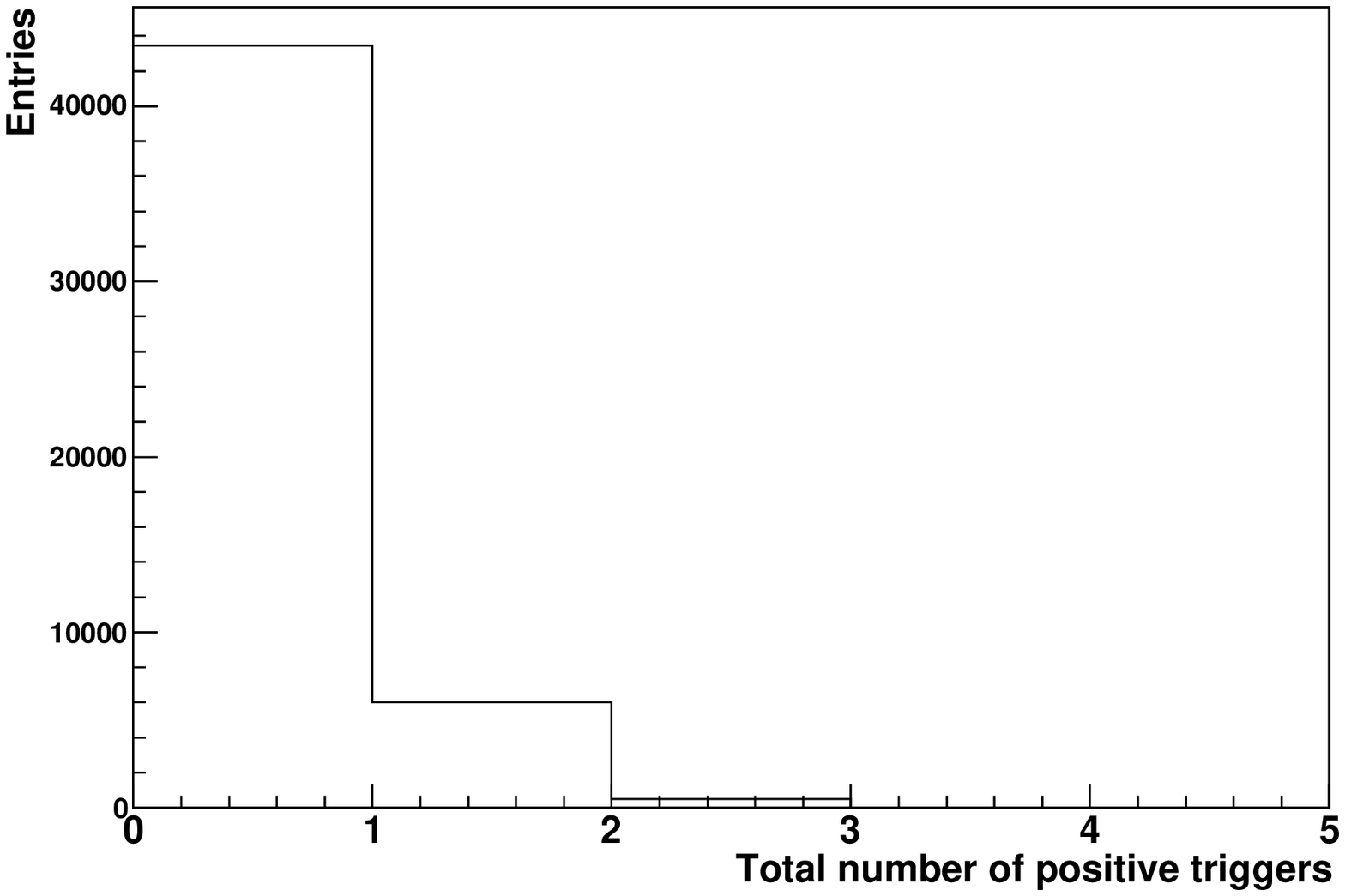}
\includegraphics[width=8.2cm]{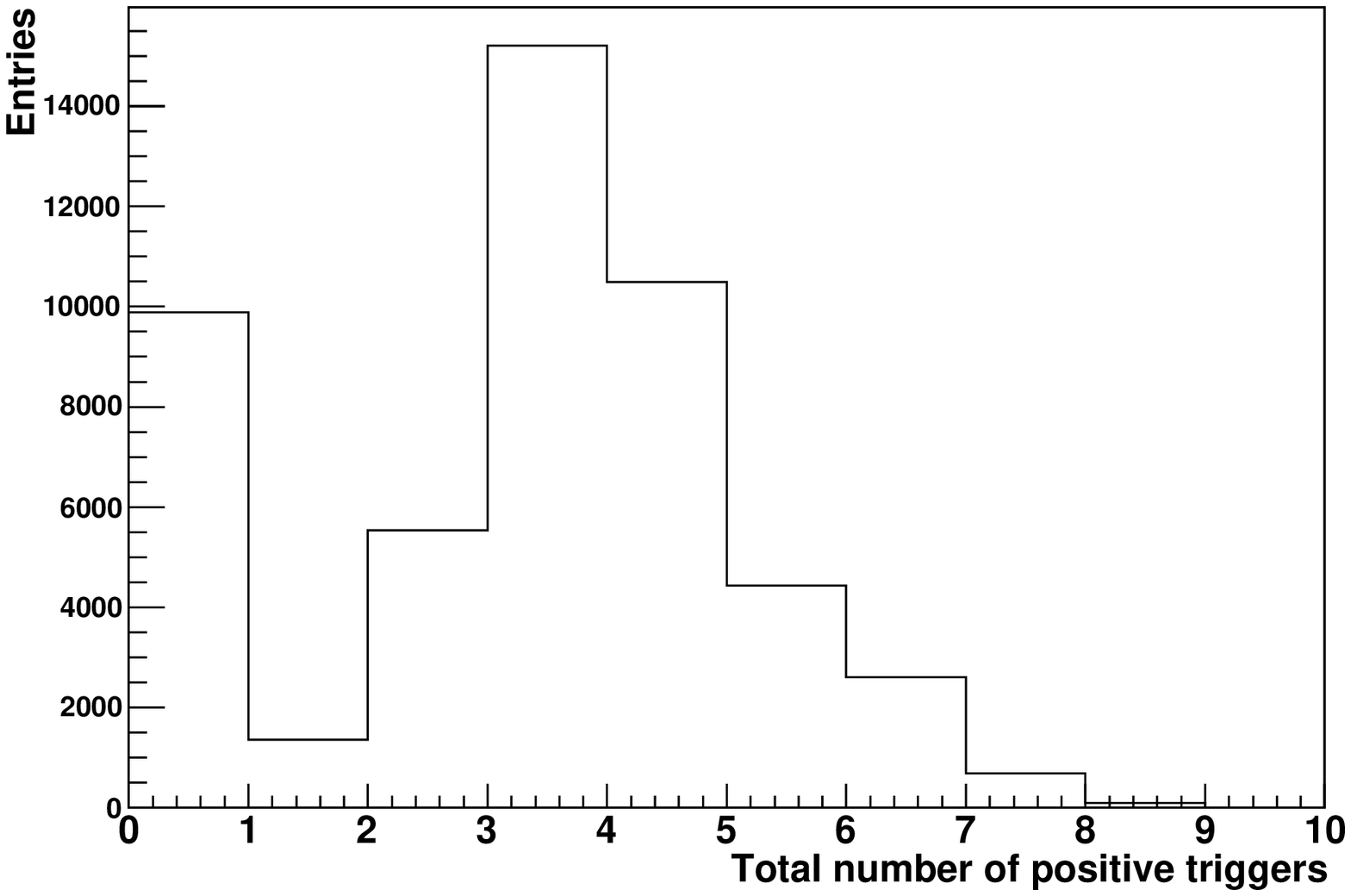}
\caption{Total number of triggers before the auto-reset change
(left) and after (right)} \label{trcount}
\end{figure}

This problem being identified, the auto-reset has been set to 1 $\mu s$, afterward the number of PMT
with a trigger bit equal to 1 largely increased, as shown on figure \ref{trcount}.

After the auto-reset was set to 1 $\mu s$, the typical aspect of plots such as the one of figure \ref{noeff} is shown on
figure \ref{exgood} :

\begin{figure}[H]
\centering
\includegraphics[width=7cm]{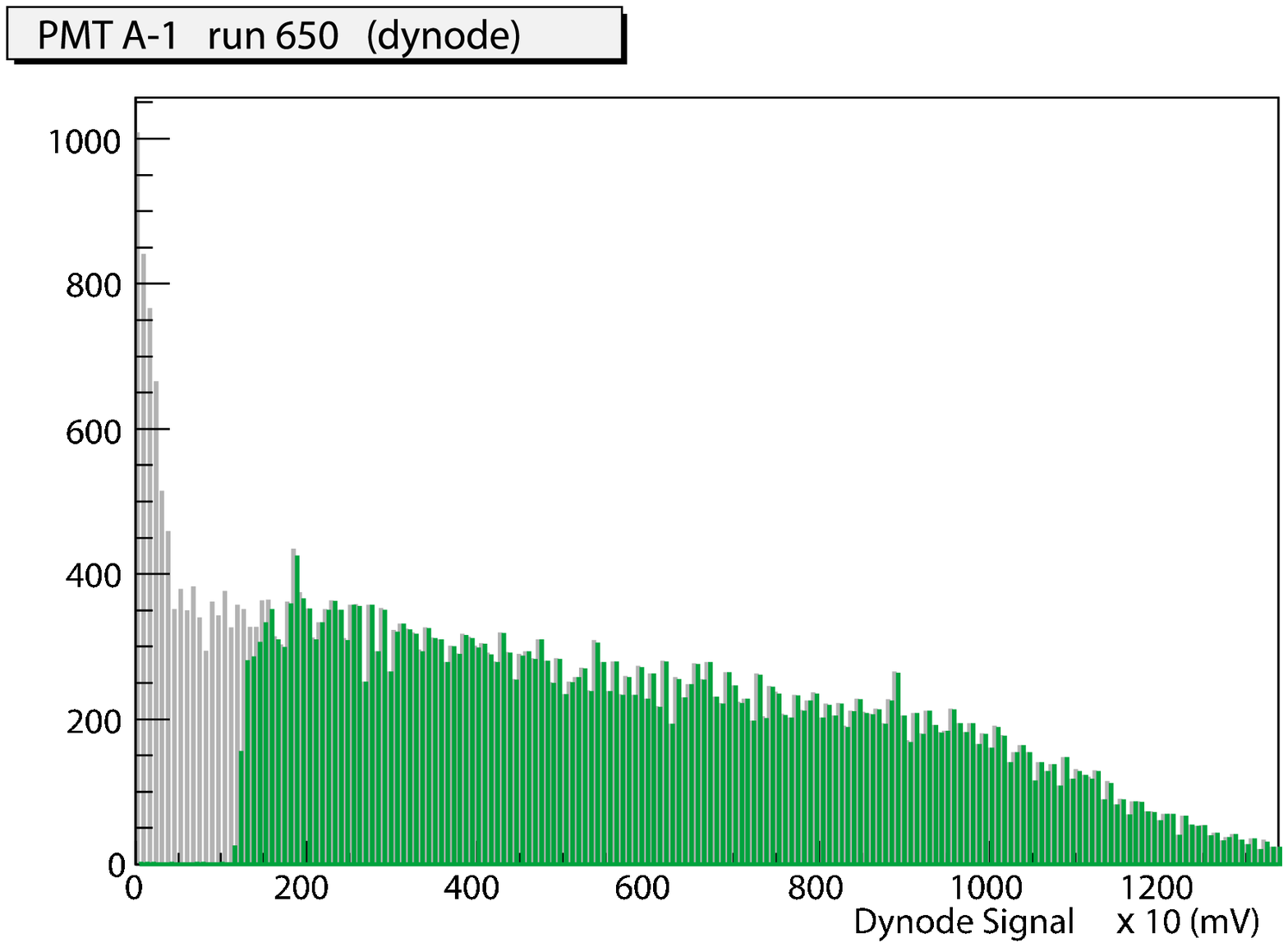}
\includegraphics[width=7cm]{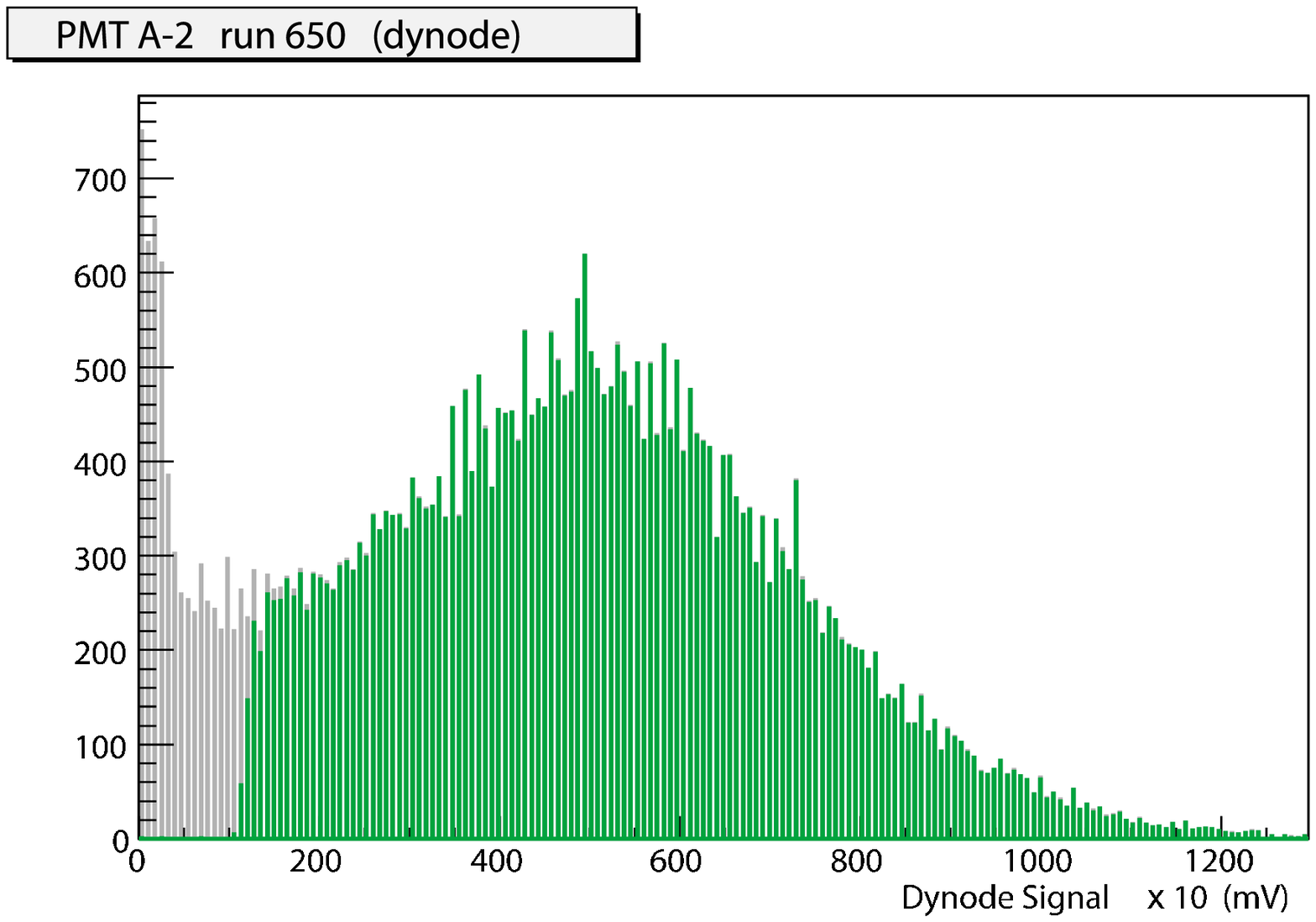}
\includegraphics[width=7cm]{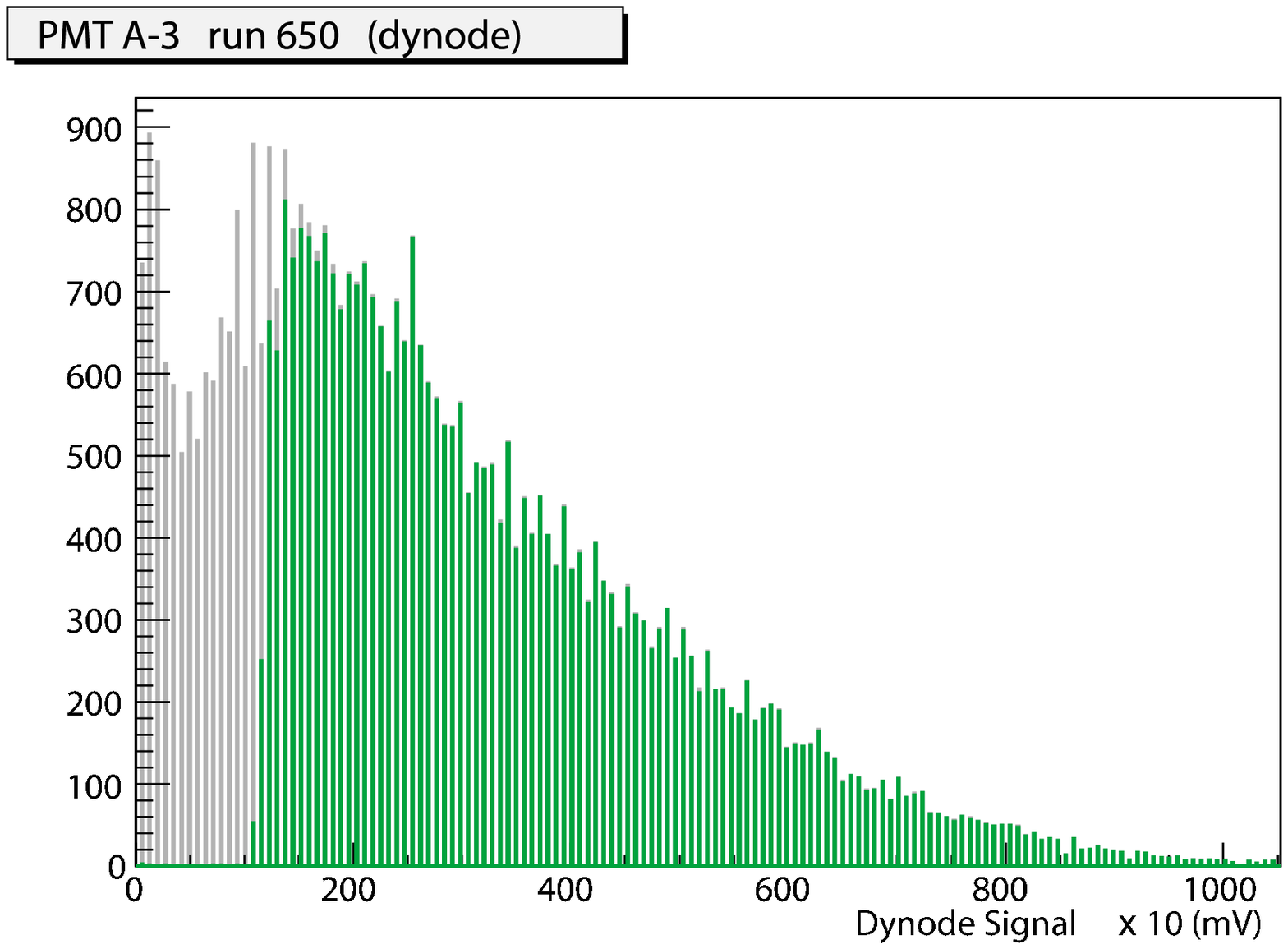}
\includegraphics[width=7cm]{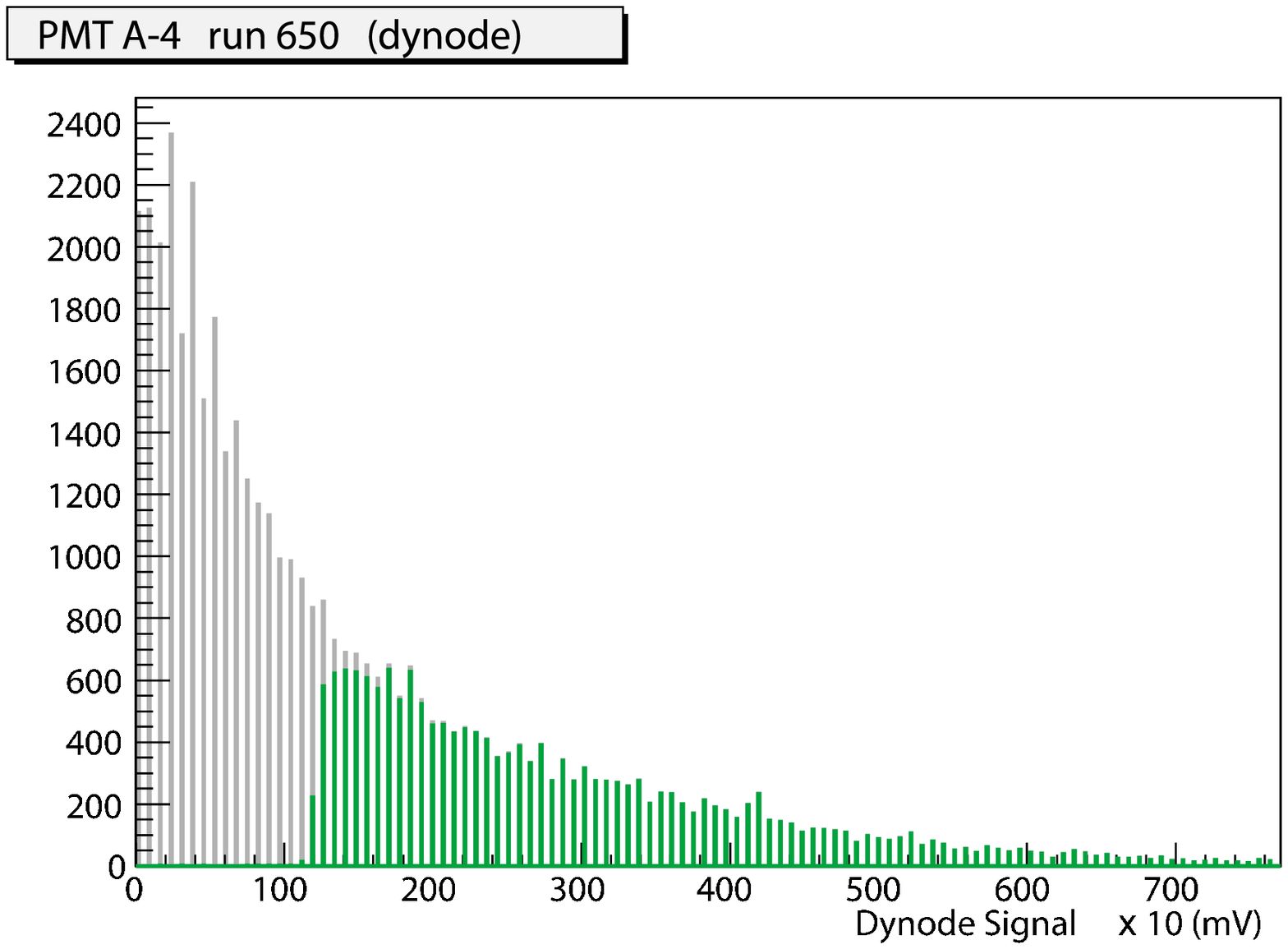}
\caption{Same kind of picture as Fig. \ref{noeff} with auto-reset at 1 $\mu s$ (Run 650 has an applied
  threshold of 100 mV and run 700 an applied threshold of 200 mV)}
\label{exgood}
\end{figure}

One can see that no event is taken below threshold and that above, almost 100\% of them are. Here, the EFE dynode signal is
multiplied by 10 in order to reproduce the amplifier effect. On figure \ref{exgood}, different plots are shown in order to
illustrate the fact that one DAC per EIB was enough since we clearly see that the proper threshold is applied to each of the
PMTs whose efficiency curves are represented. In the following sections, we will see that the effective thresholds were very
homogeneous.

\subsection{Efficiency measurement}

\begin{figure}[H]
\centering
\includegraphics[width=10cm]{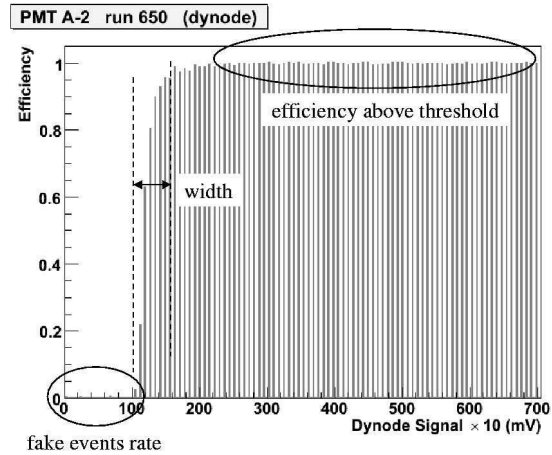}
\caption{Different features of the efficiency curve, and the information that can be extracted
  from it}
\label{eff}
\end{figure}

In order to quantify the trigger efficiency, the ratio of figure \ref{exgood}'s green and grey histograms is performed. Then
the plot of figure \ref{eff} is obtained. It allows to measure :
\begin{itemize}
\item{
  The effective threshold, i.e. the value at which the efficiency curve goes up,
}

\item{
  The number of fake events by counting how many events are recorded below the applied threshold,
}
\item{
  The rise width: how many ADC channels are needed to reach the high efficiency,
}
\item{
  The efficiency above threshold by seeing how many events are missed above the effective threshold.
}
\end{itemize}

\begin{figure}[H]
\centering
\includegraphics[width=8.2cm]{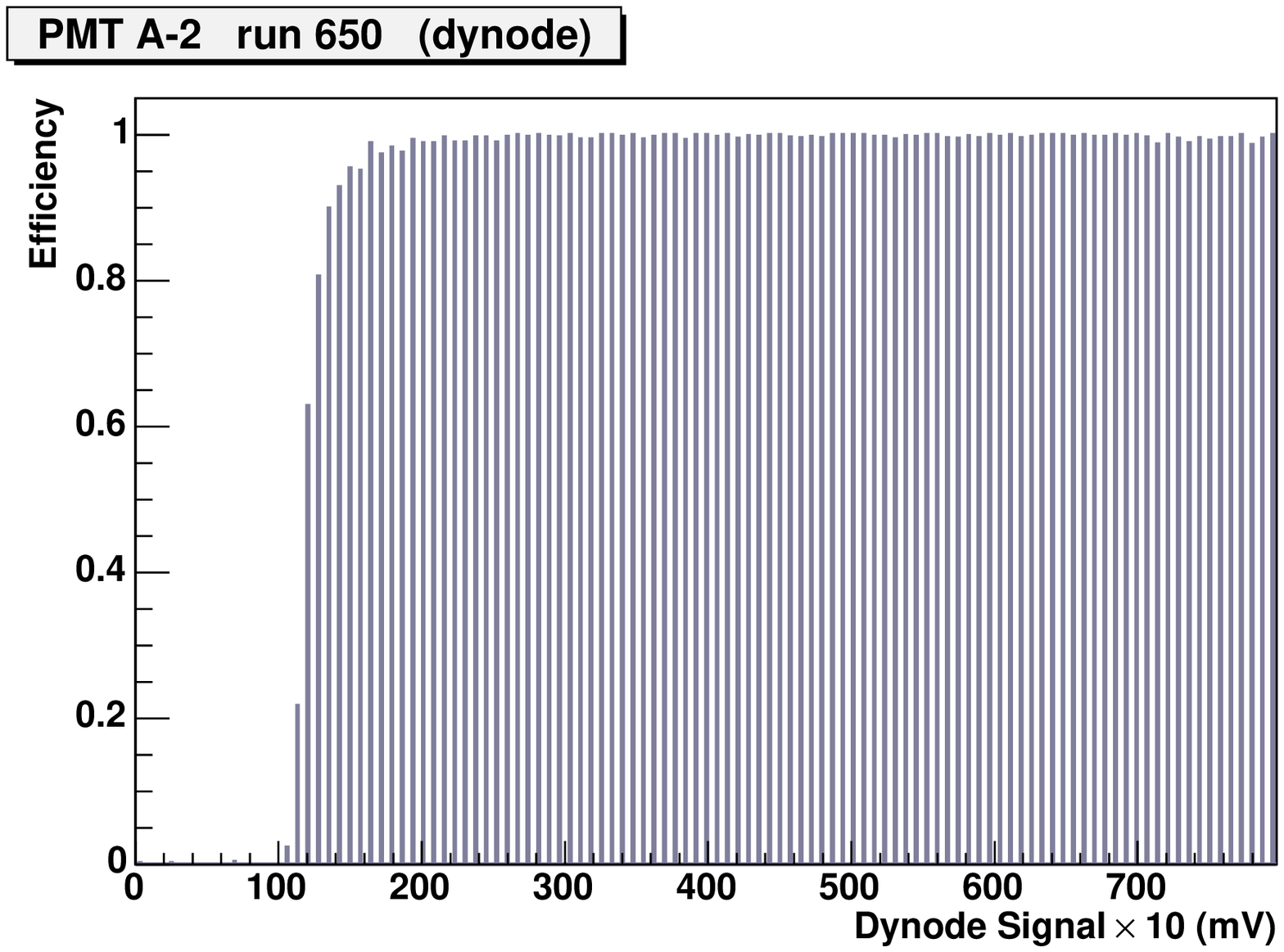}
\includegraphics[width=8.2cm]{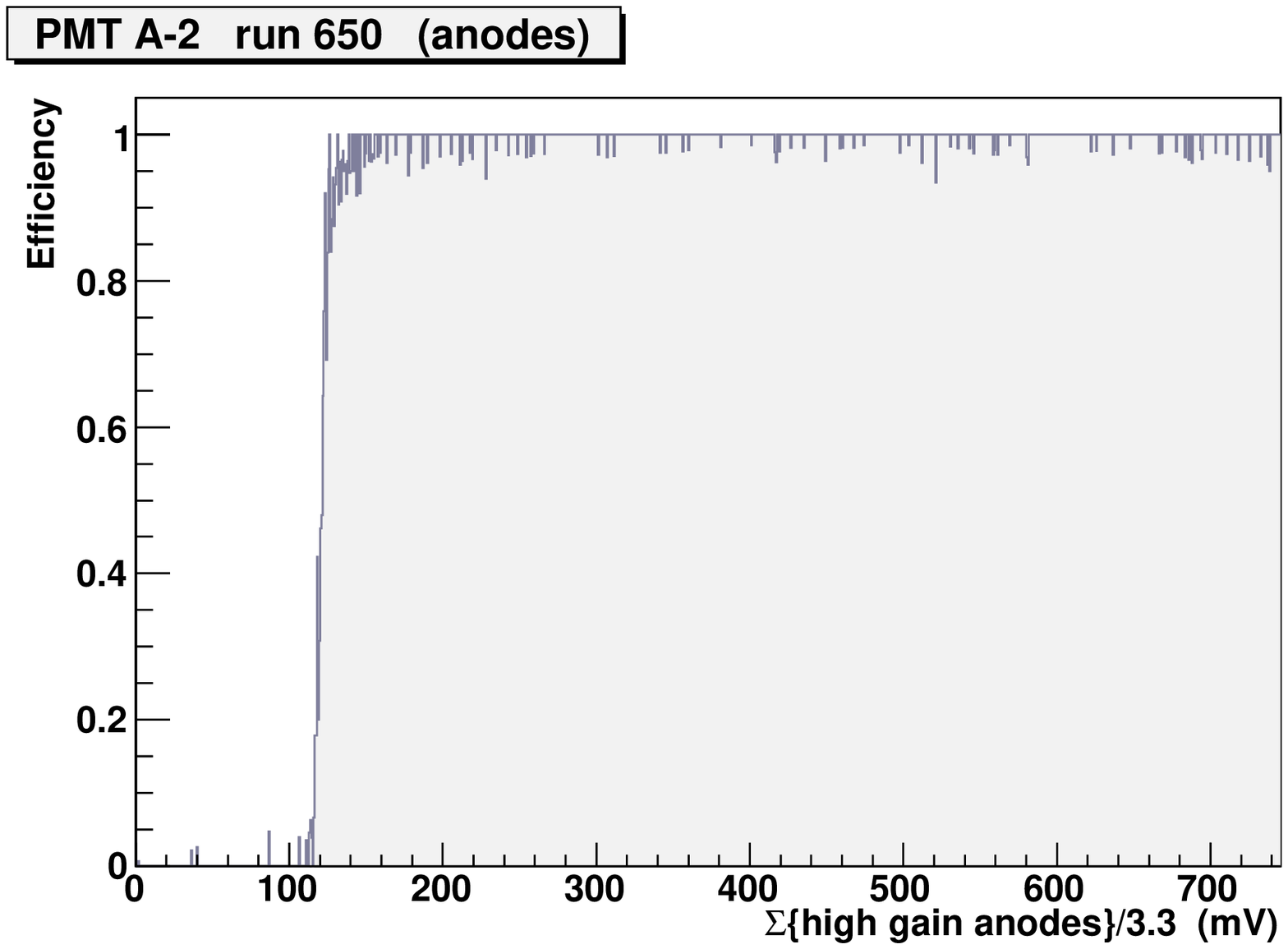}
\caption{Efficiency curve based on the high gain anodes sum}
\label{anodeex}
\end{figure}

\begin{figure}[H]
\centering
\includegraphics[width=8.2cm]{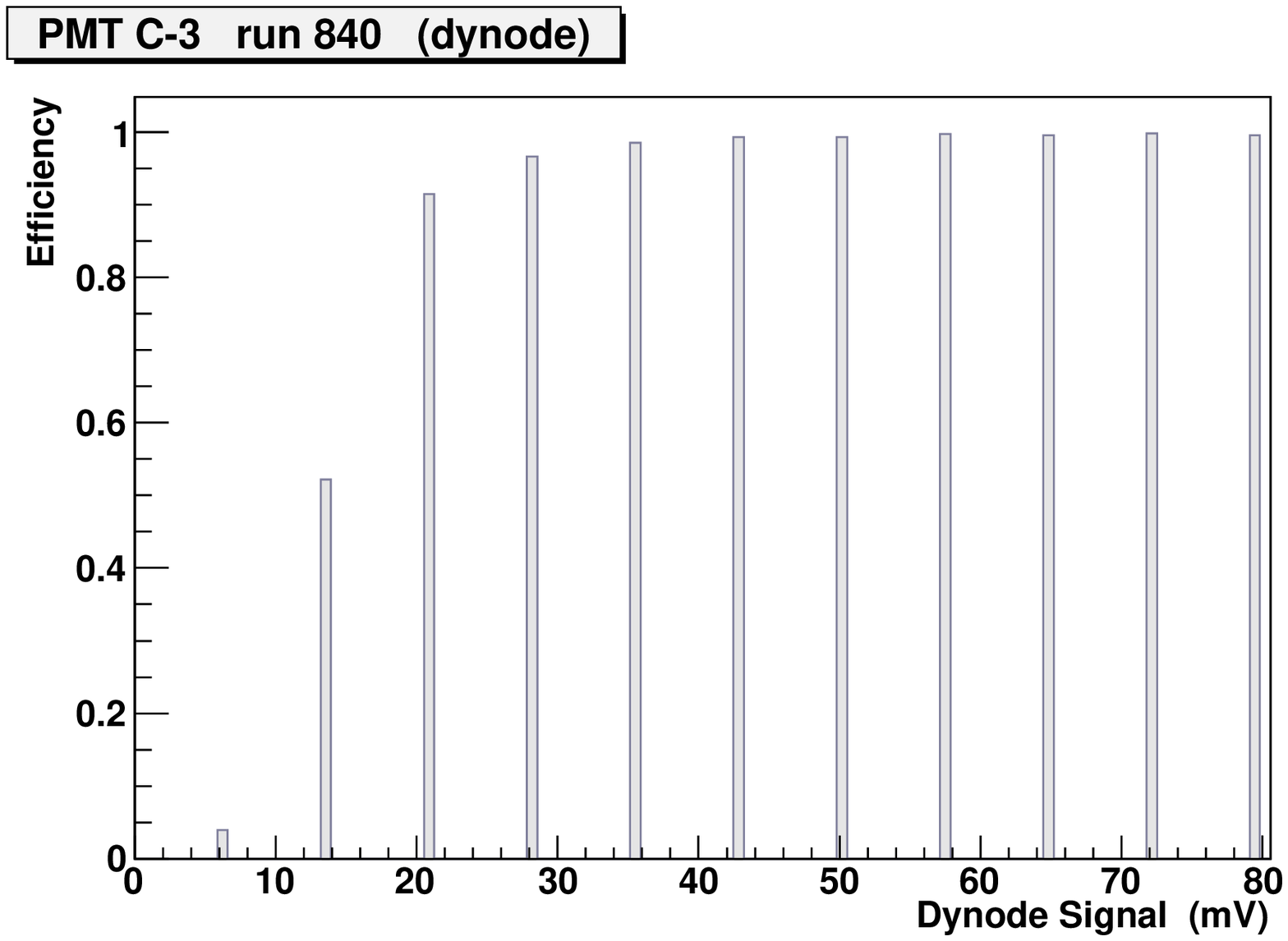}
\includegraphics[width=8.2cm]{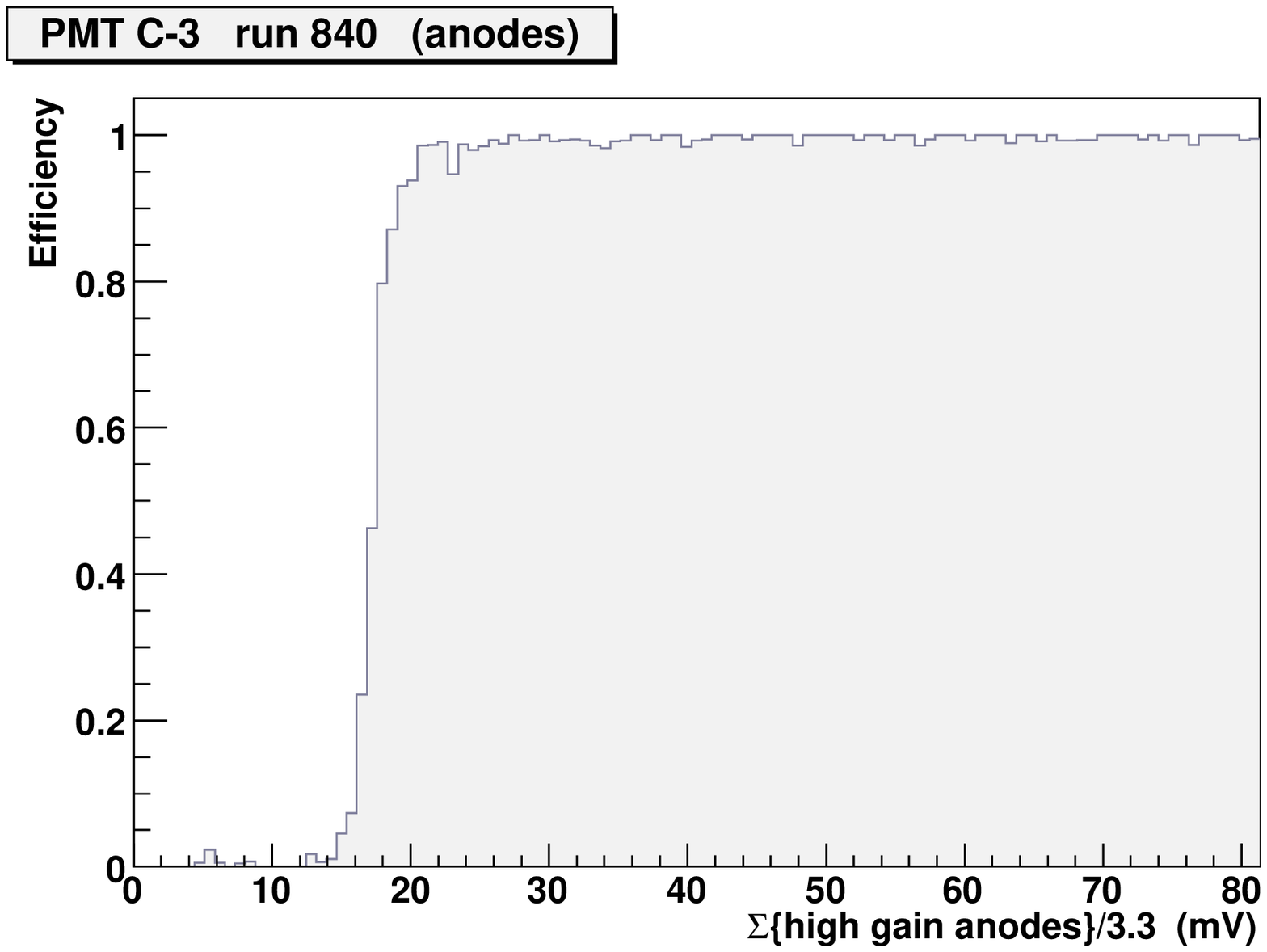}
\caption{Illustration of the importance of the anode measurement (threshold = 20 mV here)}
\label{25mV}
\end{figure}

In order to have more accuracy in the determination of effective thresholds, the anode signals are used instead of the EFE
dynode. Those signals are related one to another by the following expressions:
$$
Dyn=\sum{low\:gain\:anodes}=\sum{\frac{high\:gain\:anodes}{33.3}}
$$
As it was mentionned before, the 33.3 gain between channels is an average value for all PMTs. Finally, the estimation of the
fast dynode signal from the EFE data is summarized in the following formula:
$$
Trigger\:fast\:signal=\sum{\frac{high\:gain\:anodes}{33.3}}\times10=\sum{\frac{high\:gain\:anodes}{3.3}}
$$
This is a key point since the ADC channel width is the same for anodes and dynode, it means that this estimation is 10 times
more precise. In the following, we measure effective thresholds this way. As an illustration, figures \ref{anodeex} and
\ref{25mV} show the comparison between the dynode- and the anode-estimation of the trigger signal. One can see clearly the
accuracy improvement.

It is even more important for low thresholds runs, as shown on next figure.

From this we can deduce the effective threshold, which corresponds to the signal value for which the efficiency goes above a
given value. The ratio of triggered events below threshold over all events provides the fake event probability, and the
increase width is also a quality estimator. Finally, the right part of the histograms provide an efficiency measurement.

\subsection{Dataset}

\begin{figure}[H]
\centering
\includegraphics[width=8.2cm]{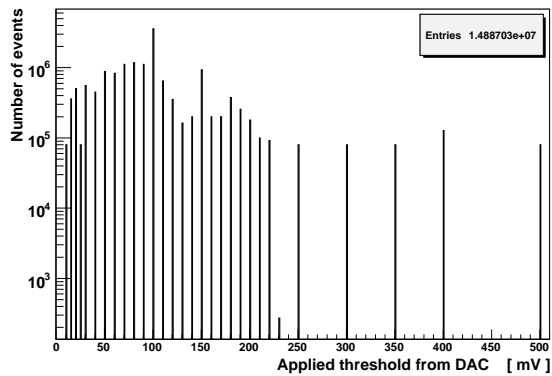}
\caption{Applied threshold statistics for the main period}
\label{threstat}
\end{figure}

In this part of the analysis, we study thresholds. It is important
to test the trigger behaviour on a large range of thresholds
because in the final configuration, different layers of the Ecal
should be applied different threshold. This is because the
standalone $\gamma$ trigger algorithm is based on the longitudinal
development of the shower. So we tried to apply a lot of different
thresholds. The set of data can be divided in 4 categories, each
corresponding to a given period :
\begin{itemize}
\item{
  The starting of data taking, during this period, a few
  applied thresholds have been tested since all starting problems had to be fixed. The part 1 plots
  are from this period and the trigger study does not take these runs into account for two main reasons:
  the A side electronics was not stable and only 2 different thresholds have been applied.
}
\item{
  The main period, in which applied thresholds have been changed in the range from 10 mV to 500 mV.
  Over the $1.7\:10^{7}$ events taken during this test beam, $1.5\:10^{7}$ correspond to this
  period. Figure \ref{threstat} shows the distribution of applied thresholds during this period.
}
\item{
  A short period during which the Ecal acquisition was triggered by its own using the standalone
  trigger. The Ecal was then put in vertical position.
}
\item{
  The last day, still with the Ecal in vertical position. Now the $\gamma$ trigger is used as the main
  trigger and trigs also the Tracker acquisition.
}
\end{itemize}
Most of the events were electrons or electrons+pions. However, for
each of those periods, no particular distinction is made on the
particle type, in order to keep the maximum amount of data. The
two last periods will be detailed in the following. In the next
part, we analyze data from the main period.

\subsection{Results from the main period}

\subsubsection{Run selections}

During this period, the data is sampled in runs of 50.000 events for most of them. Sampling the data is necessary in order to
perform a study of effective thresholds. In this way we can make efficiency plots over one run and follow the evolution of
performances. Some runs present obvious bad data, and they are removed from the analysis. Most of the time, these runs
correspond to an acquisition break because of an access in the zone, or beam instabilities. However, the removed runs
represent a very low fraction of the total amount of events. As it has already been said, the C9 PMT was out of order from
the beginning and must not be taken into account in our discussions. In general, the best results come from the PMTs where
the maximum of the shower is, namely C3-C4 and A2-A3. The last PMTs have very few signal since they are placed
at approximately 35 radiation length and most of the particles were electrons.\\
Each run provide 18 pictures similar to those of figures \ref{anodeex} and \ref{25mV} (one picture per PMT). The study of
those lead to the results presented in the next sections.

\subsubsection{Effective thresholds}

\begin{figure}[H]
\centering
\includegraphics[width=8.2cm]{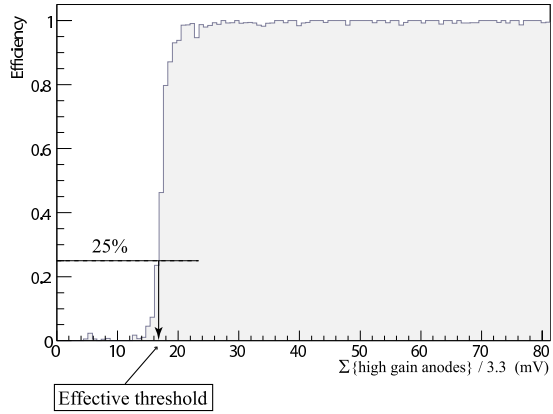}
\caption{Method used here to determine the effective threshold}
\label{25pc}
\end{figure}

\begin{figure}[H]
\centering
\includegraphics[width=8.2cm]{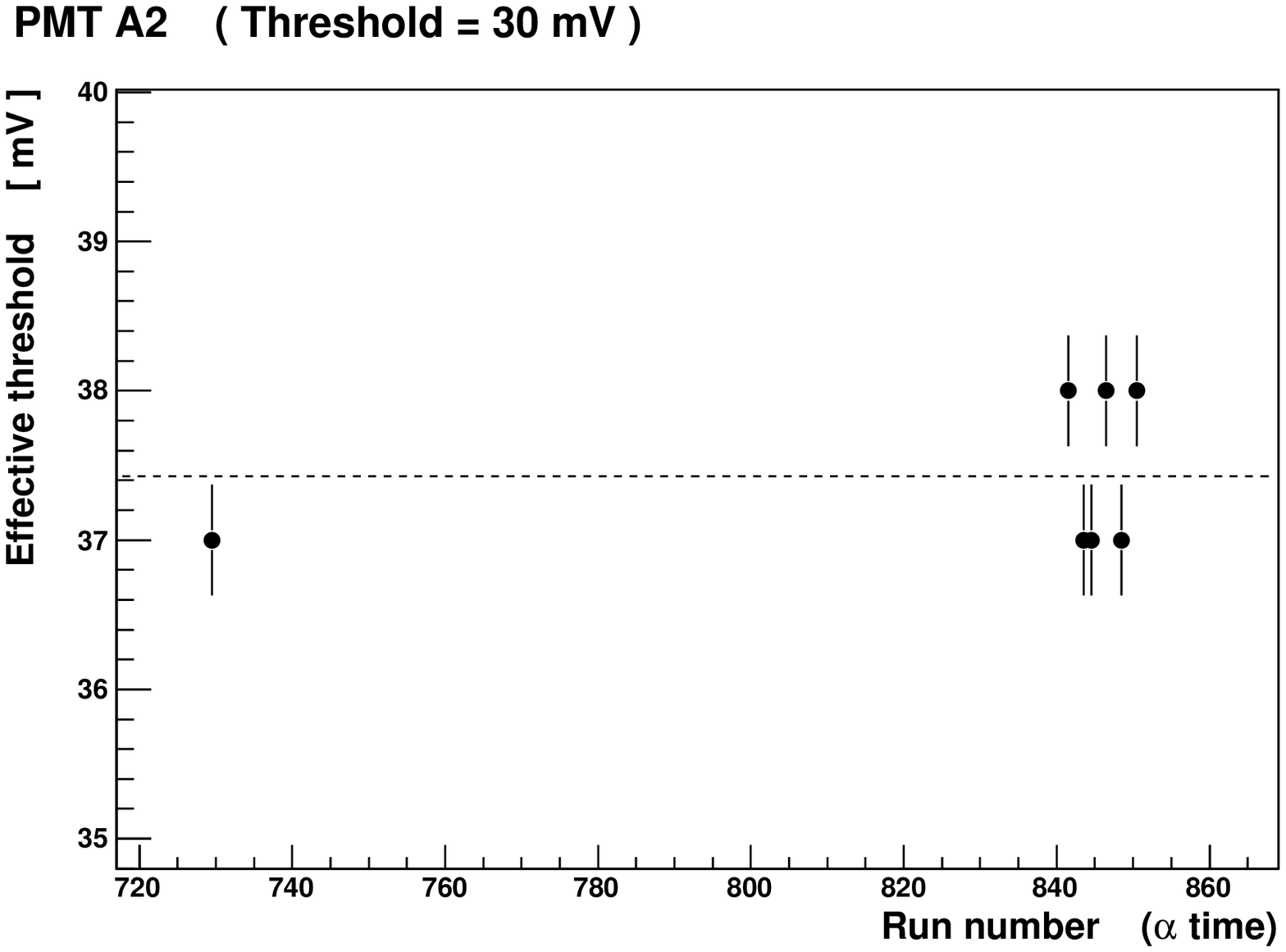}
\includegraphics[width=8.2cm]{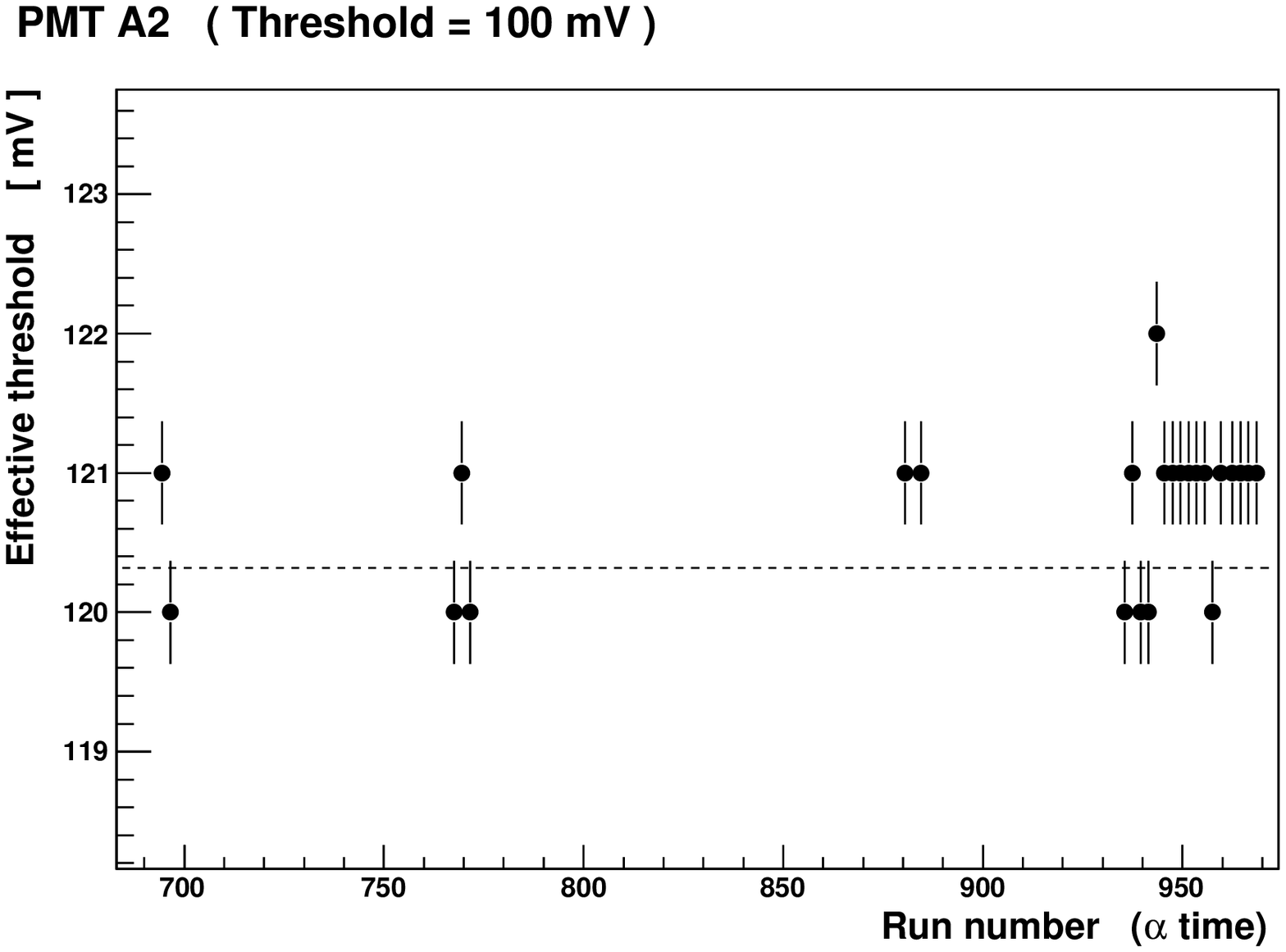}
\caption{Time evolution of 2 effective thresholds on PMT A2} \label{suivi}
\end{figure}

For each run and each PMT, the effective threshold is computed with the following definition : it is the EFE anodes sum value
for which the efficiency goes above 25\%, as it is illustrated on figure \ref{25pc}
Each individual run is analyzed this way. The effective threshold is computed for all PMTs. As an illustration, figure
\ref{suivi} shows the time evolution of 2 effective thresholds (30 mV and 100 mV). On these plots, we have the effective
threshold versus the run number.

\begin{figure}[H]
\centering
\includegraphics[width=8.2cm]{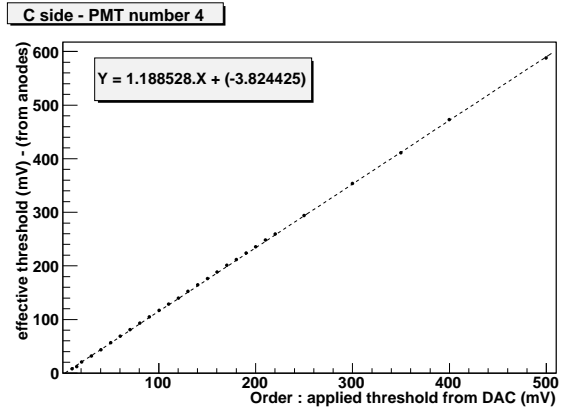}
\caption{Effective threshold versus the order for PMT C4, up to 500 mV}
\label{linex}
\end{figure}

On figure \ref{suivi}, he dashed line is the mean value and the error bars correspond to the width of one ADC channel. They
are not an indication of the rise width (this issue is developed in the next section). For each sample of data having the
same applied threshold, the mean effective threshold is computed as well as the dispertion around this value. At this point
we can plot the effective threshold versus the applied one from DAC. Figure \ref{linex} is an example of such a plot,
obtained for PMT C4. Here the error bars, corresponding to the dispertion around the mean value are smaller than the points.

\begin{figure}[H]
\centering
\includegraphics[width=8.2cm]{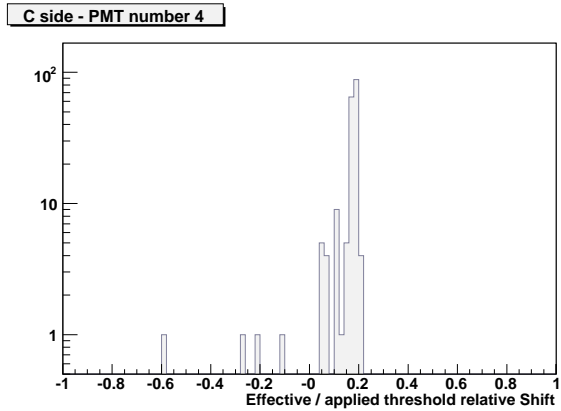}
\caption{Distribution of the effective/applied threshold for PMT C4, during the main period}
\label{distex}
\end{figure}

\begin{figure}[H]
\centering
\includegraphics[width=16cm]{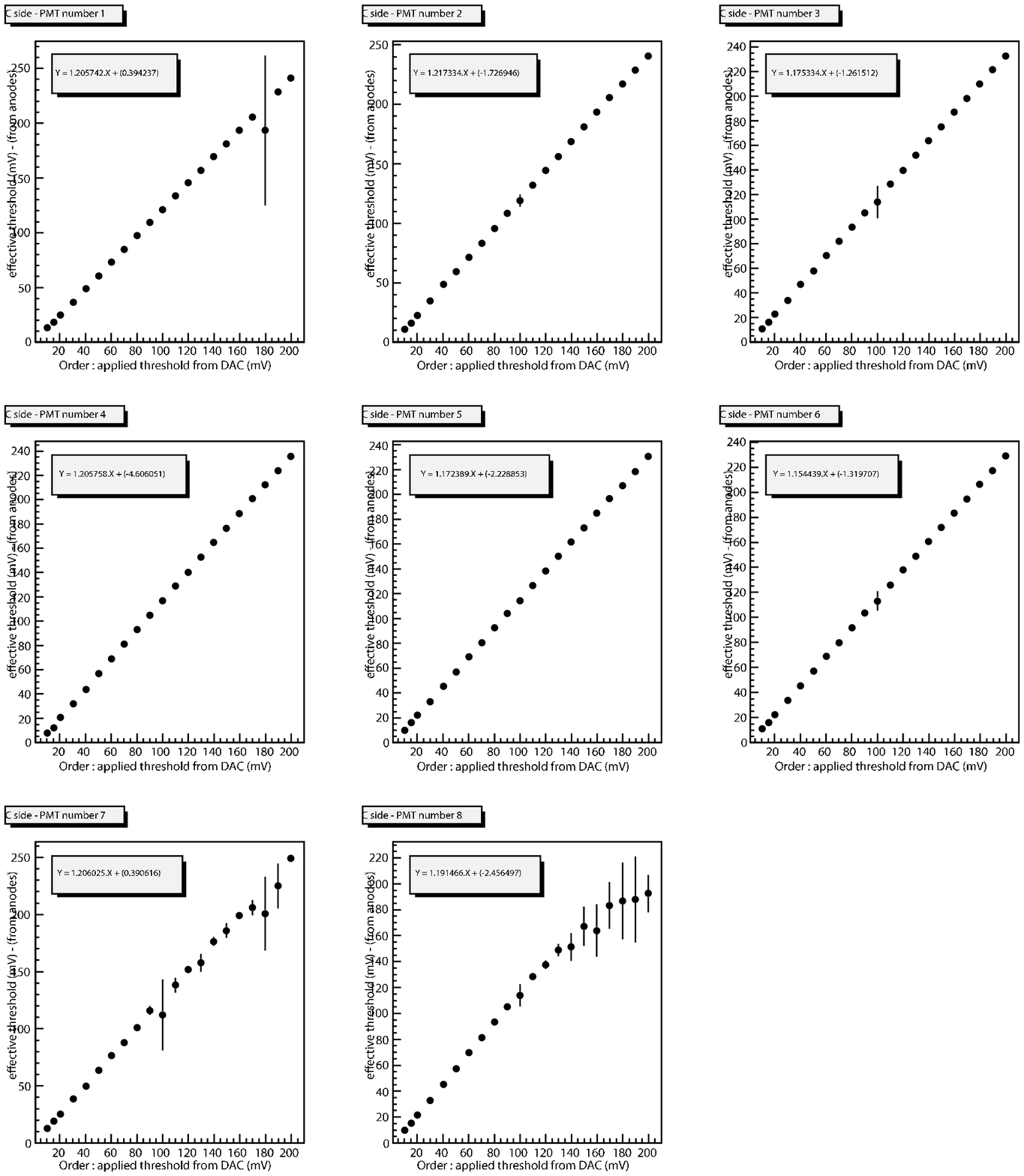}
\caption{Same as previous figure, but for C side}
\label{fitc}
\end{figure}

\begin{figure}[H]
\centering
\includegraphics[width=16cm]{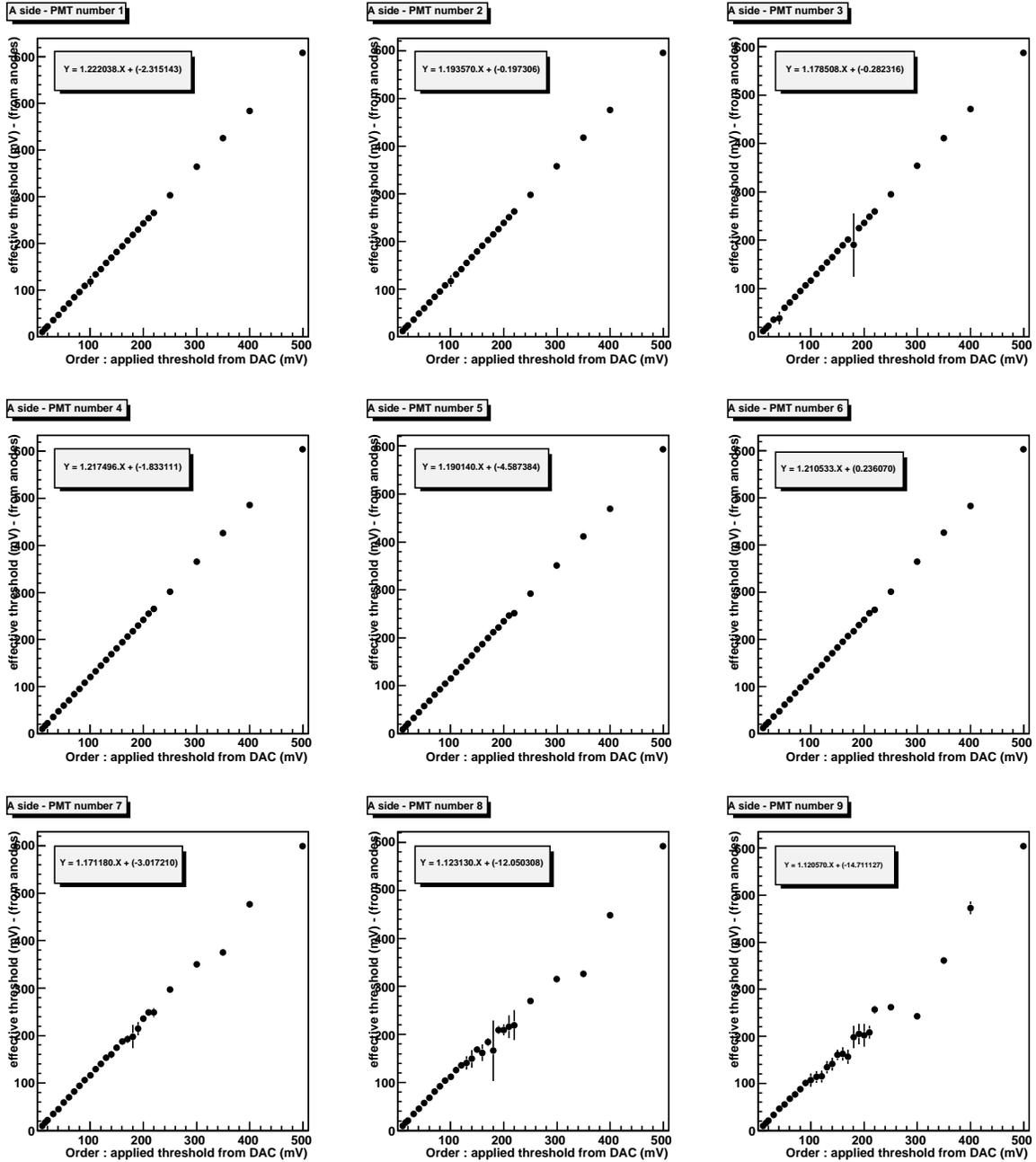}
\caption{Effective thresholds versus applied ones up to 500 mV, for A side}
\label{fita2}
\end{figure}

\begin{figure}[H]
\centering
\includegraphics[width=16cm]{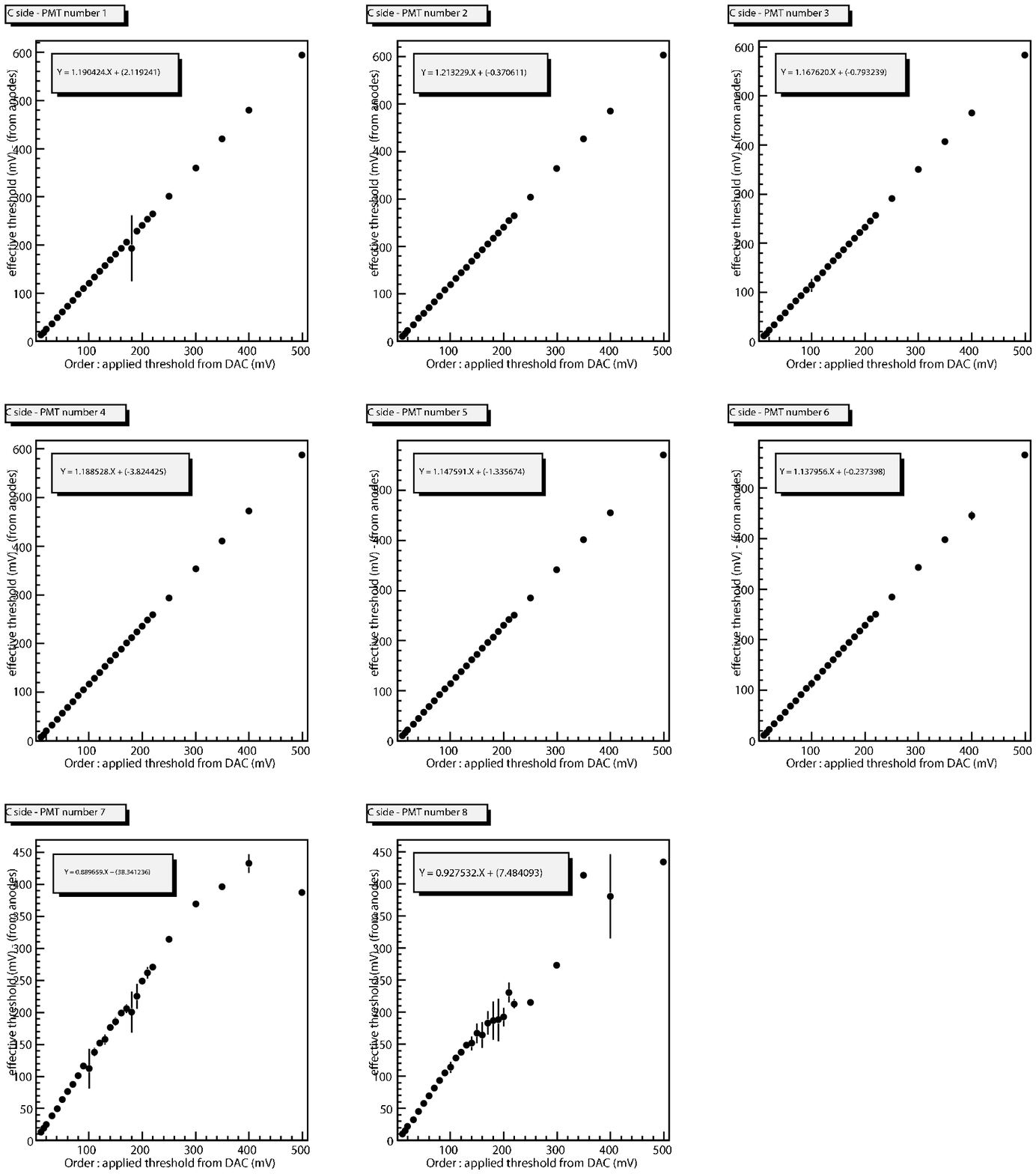}
\caption{Same as previous figure, but for C side}
\label{fitc2}
\end{figure}

On figure \ref{linex} is shown the result of a linear fit to those points. This have been done systematically for all PMTs
and the whole main period. All the results results are shown in the appendix up to a 200 mV applied threshold (figures
\ref{fita} and \ref{fitc}) for easy reading and up to 500 mV (figures \ref{fita2} and \ref{fitc2}) for completeness. One can
also present these results in terms of a relative shift distribution. As an example, figure \ref{distex} shows, for PMT C4,
such a plot. For each run, the relative shift between the effective and applied threshold fills this histogram.

Again, all the distribution are given in the appendix (figures \ref{shifta} and \ref{shiftc}).
Notice that these plots are in log
scale and therefore have a very low dispertion. Next table (figure \ref{tabshift}) summarizes
these results. It gives the mean value of these shifts for each PMT over all the main period,
as well as the dispertion of these distributions.

\begin{figure}[H]
  \centering
  \begin{multicols}{2}
    \begin{tabular}{c|c|c}
      A side & Mean relative shift & RMS \\
      \hline
      1 & 0.23 & 0.012 \\
      \hline
      2 & 0.22 & 0.017 \\
      \hline
      3 & 0.20 & 0.069 \\
      \hline
      4 & 0.22 & 0.027 \\
      \hline
      5 & 0.16 & 0.040 \\
      \hline
      6 & 0.24 & 0.020 \\
      \hline
      7 & 0.19 & 0.036 \\
      \hline
      8 & 0.15 & 0.098 \\
      \hline
      9 & 0.12 & 0.123 \\
    \end{tabular}
    \begin{tabular}{c|c|c}
      C side & Mean relative shift & RMS \\
      \hline
       1 & 0.23 & 0.019 \\
      \hline
       2 & 0.21 & 0.024 \\
      \hline
       3 & 0.17 & 0.023 \\
      \hline
       4 & 0.16 & 0.080 \\
      \hline
       5 & 0.17 & 0.020 \\
      \hline
       6 & 0.16 & 0.015 \\
      \hline
       7 & 0.28 & 0.038 \\
      \hline
       8 & 0.17 & 0.047 \\
       \hline
       9 &-&-
    \end{tabular}
  \end{multicols}
  \caption{Effective/applied thresholds relative shifts : mean value and RMS for all
    PMTs during the main period}
  \label{tabshift}
\end{figure}

We emphasize the fact that the most important result here is the very low dispersion of the shifts and the fact that they are
close from one PMT to another.

\subsubsection{Efficiency rise Width}

We call rise width the number of ADC channels needed to reach almost 100\% efficiency from a given efficiency value. The
results presented here are the number of ADC channels between 25\% and 80\% efficiency.

\begin{figure}[H]
\centering
\includegraphics[width=8.2cm]{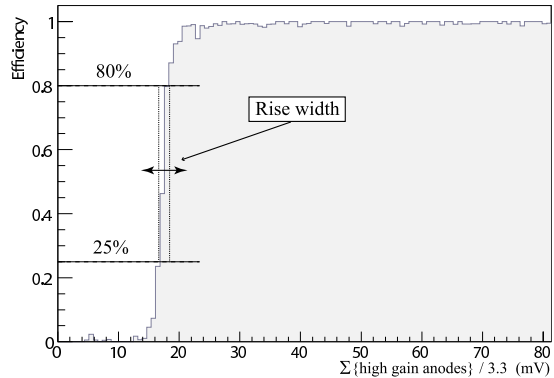}
\caption{Principle of the determination of the efficiency rise width near threshold} \label{}
\end{figure}

\begin{figure}[H]
\centering
\includegraphics[width=16cm]{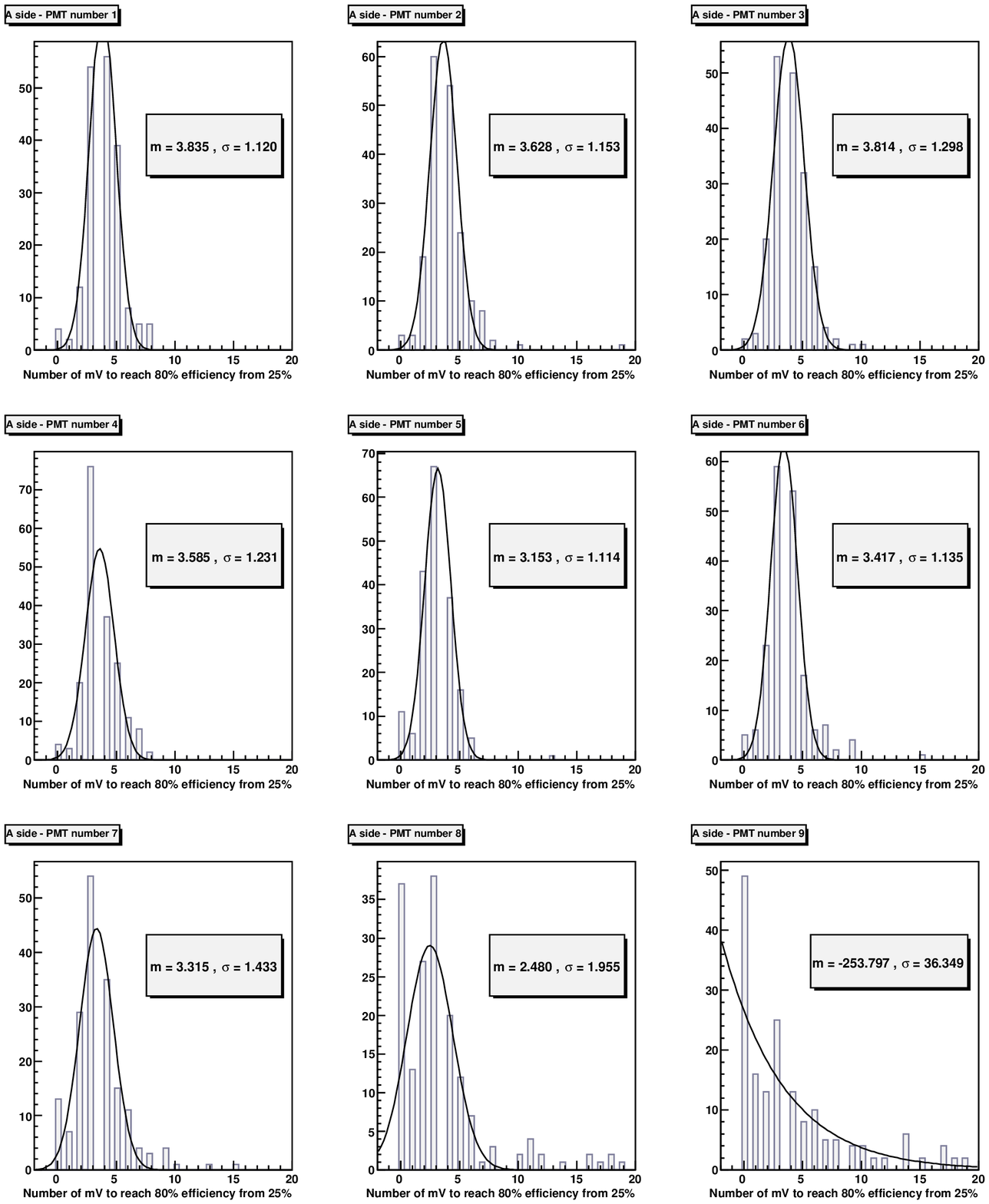}
\caption{Efficiency rise width around threshold for A side PMTs} \label{monta}
\end{figure}

\begin{figure}[H]
\centering
\includegraphics[width=16cm]{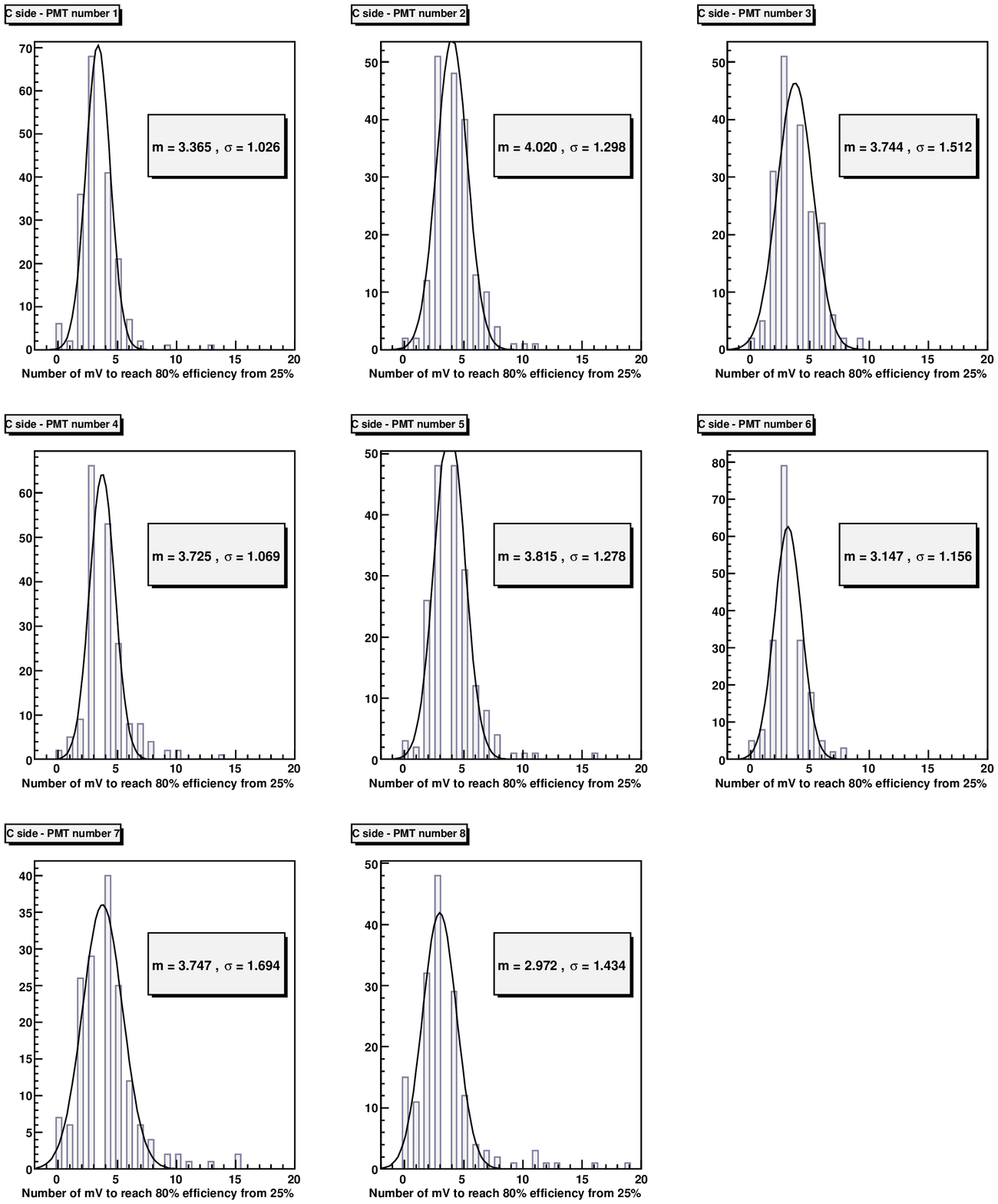}
\caption{Efficiency rise width around threshold for C side PMTs} \label{montc}
\end{figure}

As it has previously been done for the effective thresholds determination, a systematic study involving all sets of data a
has been performed. The results of this analysis appear in figure \ref{tab0}. We learn from figures \ref{monta} and
\ref{montc} that the rise width does not vary too much from one PMT to another. This is an important result since we want an
homogeneous response. The figure \ref{tab0} presents the result of a gaussian fit to the sensitivity distribution (except for
A8 and A9 for which a gaussian does not fit).

\begin{figure}[H]
  \centering
\begin{multicols}{2}  \begin{tabular}{c|c|c}
     A side & Mean ADC channel number & RMS \\
    \hline
    1 & 3.84 & 1.12 \\
    \hline
    2 & 3.62 & 1.15 \\
    \hline
    3 & 3.81 & 1.30 \\
    \hline
    4 & 3.59 & 1.23 \\
    \hline
    5 & 3.15 & 1.11 \\
    \hline
    6 & 3.42 & 1.14 \\
    \hline
    7 & 3.32 & 1.43 \\
    \hline
    8 & 3.45 & 3.79 \\
    \hline
    9 & 4.20 & 4.81 \\
  \end{tabular}

  \begin{tabular}{c|c|c}
    C side & Mean ADC channel  number & RMS \\
    \hline
    1 & 3.37 & 1.03 \\
    \hline
    2 & 4.02 & 1.30 \\
    \hline
    3 & 3.74 & 1.51 \\
    \hline
    4 & 3.73 & 1.07 \\
    \hline
    5 & 3.82 & 1.28 \\
    \hline
    6 & 3.15 & 1.16 \\
    \hline
    7 & 3.75 & 1.69 \\
    \hline
    8 & 2.97 & 1.43 \\
    \hline
    9 & - & - \\
  \end{tabular}\end{multicols}
  \caption{Mean ADC channel numbers needed to reach 80\% from 25\% efficiency}
  \label{tab0}
\end{figure}

\subsubsection{False event rate}

An important feature of the trigger is its noise level. It has to be checked that only a few events are triggered below the
applied threshold. In order to have a determination of the probability for an event that provides a low signal (under
threshold) to be triggered, we perform the ratio of triggered events over all events for signals below the applied threshold.
Table of \ref{tab2} presents the results of this study.

\begin{figure}[H]
\centering
    \begin{tabular}{c|c|c}
      & A side (\%) & C side (\%)\\
      \hline
      1 & 0.057 &  0.065 \\
      \hline
      2 & 0.075 &  0.120 \\
      \hline
      3 & 0.330 &  0.010 \\
      \hline
      4 & 0.050 &  0.110 \\
      \hline
      5 & 0.032 &  0.088 \\
      \hline
      6 & 0.026 &  0.068 \\
      \hline
      7 & 0.026 &  0.063 \\
      \hline
      8 & 0.019 &  0.060 \\
      \hline
      9 & 0.021 &  - \\
    \end{tabular}
    \caption{Mean false event rate (\%) below threshold for all PMTs}
    \label{tab2}
\end{figure}

\begin{figure}[H]
\centering
\includegraphics[width=16cm]{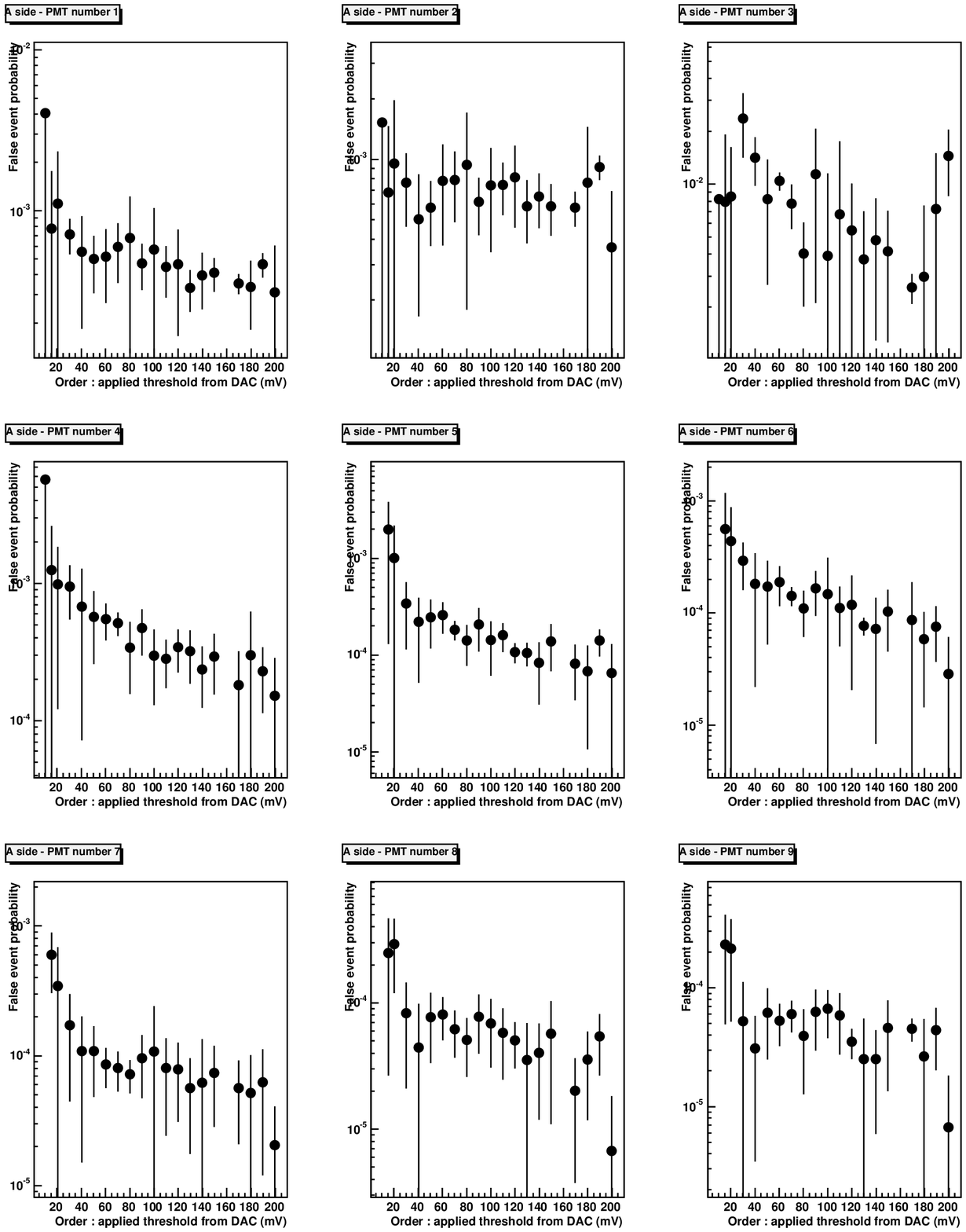}
\caption{False event rate : dependance on threshold for A side} \label{falsedepA}
\end{figure}

\begin{figure}[H]
\centering
\includegraphics[width=16cm]{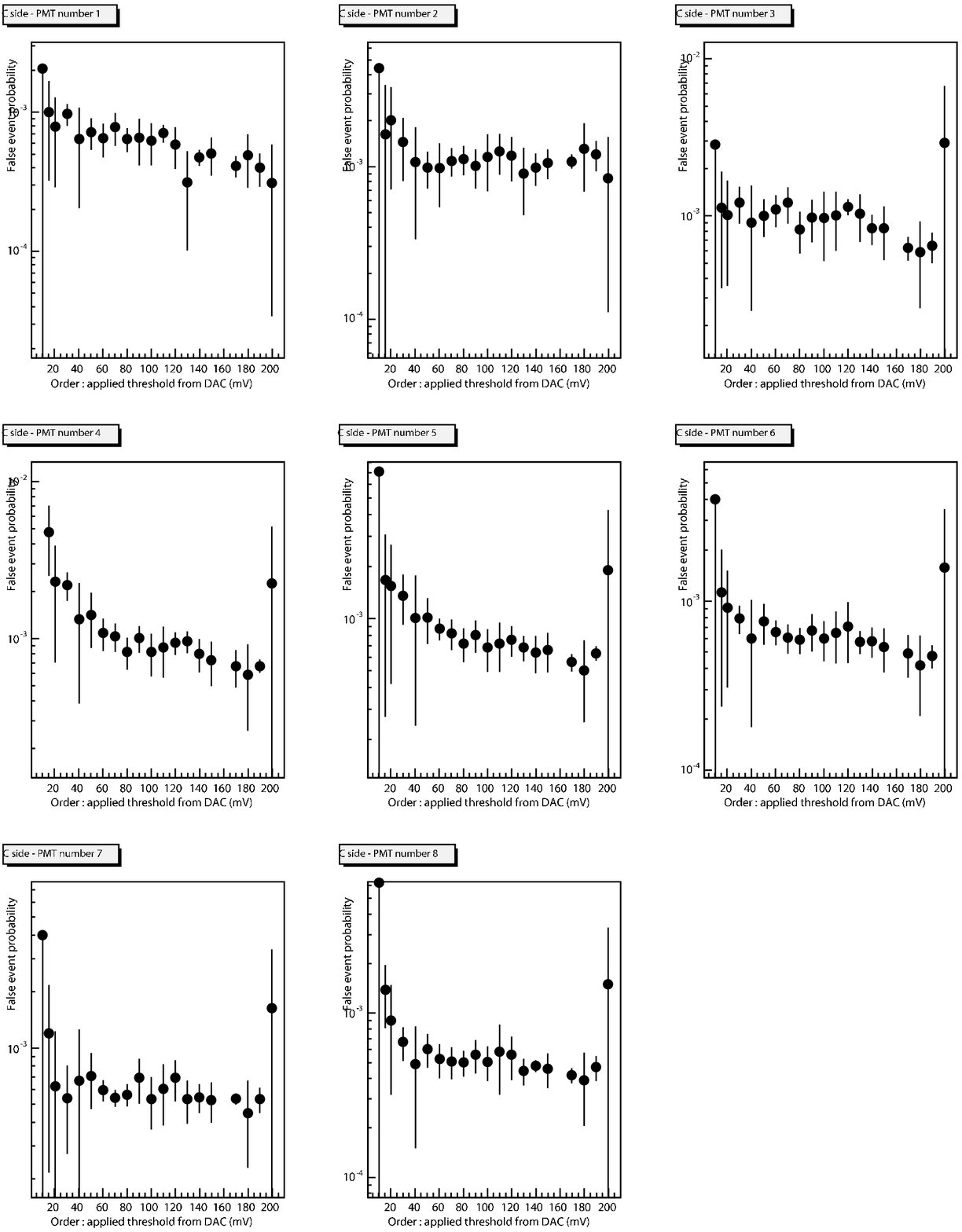}
\caption{False event rate : dependance on threshold for A side} \label{falsedepC}
\end{figure}

Figures \ref{falsedepA} and \ref{falsedepC} show the dependance of this noise level with the applied threshold (the mean
value is computed for each sample of data with the same threshold and the error bars are the RMS of the corresponding
distribution). Clearly the noise level goes down as the threshold goes up and stabilizes around $1\,^{0}\!/_{00}$ or less for
most of the PMTs. As it was mentioned before, we are interested in low thresholds, around 20 mV, in order for AMS02 to trig
on 1 GeV photons, with a local energy deposition of order 100 MeV. It seems that a 15 mV threshold gives more noise because
then the trigger is more sensitive to the electronic noise. However the results for 20 mV are very good, with a false event
rate almost always below $1\,^{0}\!/_{00}$ and never above $2\,^{0}\!/_{00}$ (except for a PMT on the A side).

\subsubsection{Efficiency above threshold}

In order to measure the rate of missed events, \it{i.e. }\rm  above threshold, we perform the ratio of triggered events over
all of them for signals above twice the applied threshold. Then we have the probability to return a 0 for an event that
should return a 1. The more far from the entry point a PMT is, the less it gets signal, for this reason the efficiency above
threshold is only determined on the 5 first PMTs of each side. PMTs that are at the end of the shower have a very low
probability to get a high signal, and subsequently a signal that is twice the applied threshold. Table \ref{tab3} shows the
results of this study.

\begin{figure}[H]
\centering
    \begin{tabular}{c|c|c}
      & A side (\%) & C side (\%)\\
      \hline
      1 & 99.48 & 99.74 \\
      \hline
      2 & 99.68 & 99.72  \\
      \hline
      3 & 99.58 &  99.69 \\
      \hline
      4 & 99.45 &  99.66 \\
      \hline
      5 & 99.37 & 99.42 \\
    \end{tabular}
    \caption{Efficiency above threshold (\%) for the 5 first PMTs of each sides}
    \label{tab3}
\end{figure}
One can notice that the figures decrease as the PMT number goes up. It is very likely due a statistical effect, since the
more far the PMTs are, the less signal they get.

\subsection{Results from the standalone periods}

\subsubsection{Specific setup}

\begin{figure}[H]
    \centering
    \includegraphics[width=9cm]{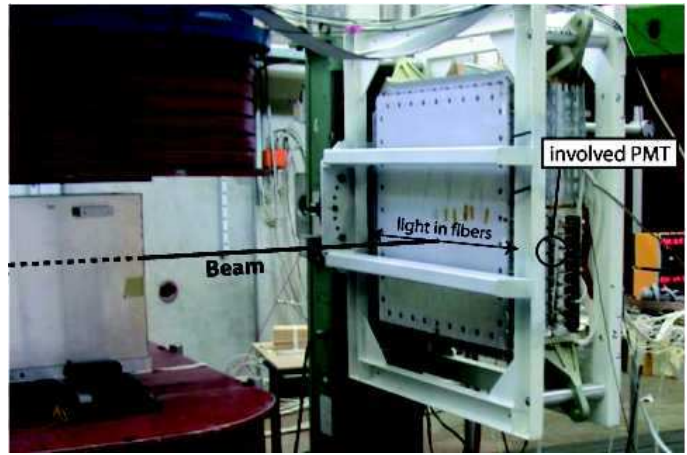}
    \caption{The Ecal in vertical position}
    \label{vertical}
\end{figure}

Although not originally planned, the good results presented above allowed to try another setup, in which the Ecal acquisition
was ruled by its own trigger.During the two last days of data taking, the Ecal was installed in vertical position. From this
moment, only a few PMTs were involved. Figure \ref{vertical} is a picture of the Ecal in vertical position. One can see there
that only 2 PMTs were active on each side. Depending on the exact beam spot position, the touched PMTs were A6, A7, C7, C8.

\begin{figure}[H]
    \centering
    \includegraphics[width=9cm]{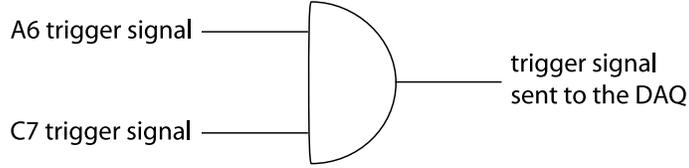}
    \caption{Trigger logic used in the standalone periods}
    \label{or}
\end{figure}

The purpose of this setup was to test the standalone trigger by actually using it to trig the data acquisition. To do so, the
Ecal altitude has been tuned in order for the signal to be shared between PMTs A6 and C7. As this data taking period was not
originally planned, an additional logic box was used to deliver the trigger signal. It returned a "or" of A6 and C7 trigger
signals (see figure \ref{or}).

The standalone period is divided into two subperiods, first the only Ecal was triggered with this system, and then it has
been used as the main trigger for the whole apparatus, including Tracker ladders.

\subsubsection{Analysis and results}

\begin{figure}[H]
\centering
\includegraphics[width=18cm]{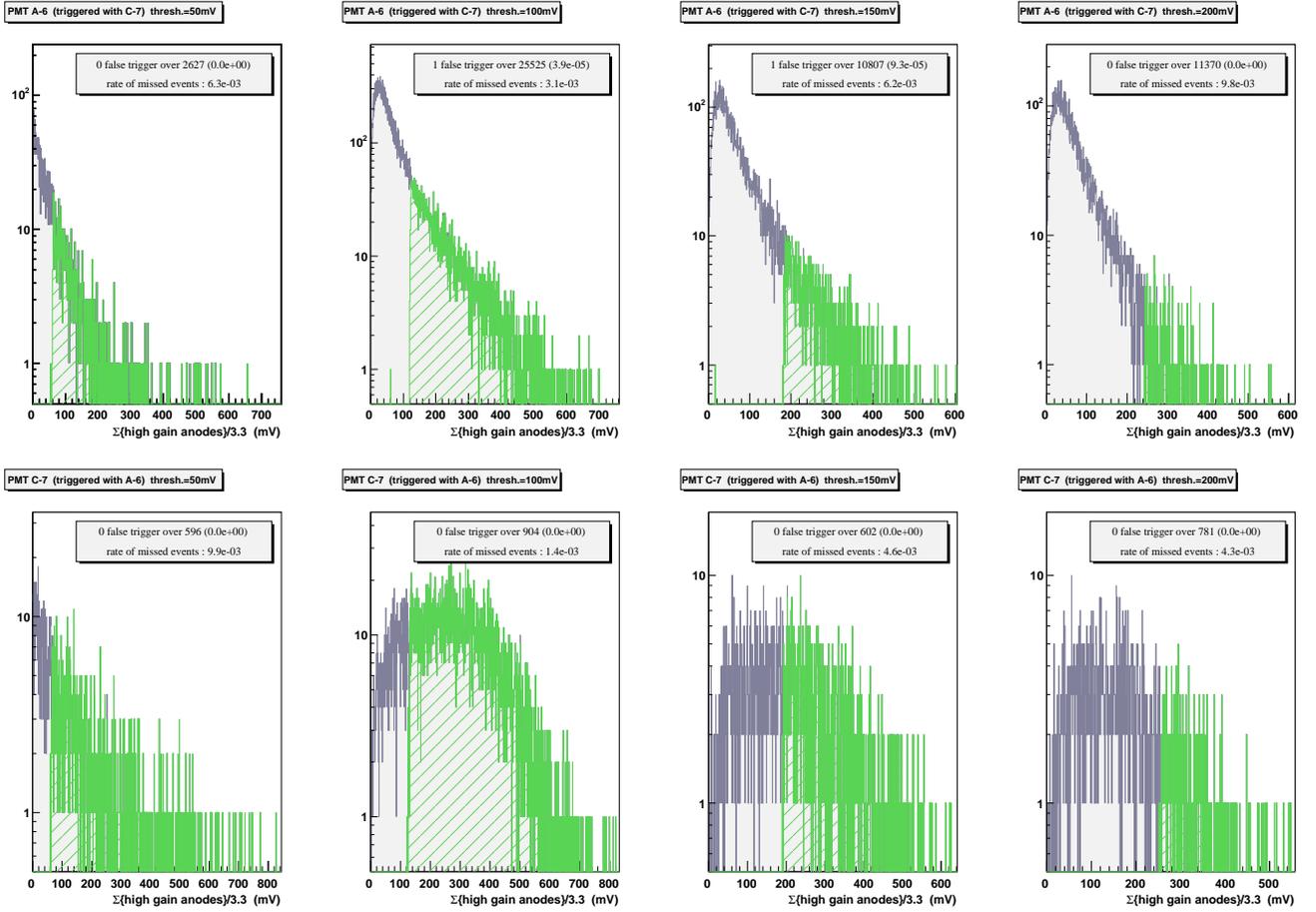}
\caption{Results from the PMTs A6 and C7 during the standalone period} \label{standalone1}
\end{figure}

During the period when the Tracker was triggered on its own, the main trigger still provided the event number. This allowed
to ensure that all the proper events were recorded. The reset of the Ecal trigger was faster than the main one, so that the
Ecal recorded much events than the Tracker during this first period. By counting the redundant event number, we determined
that the Ecal trigger recorded about 3.5 more events than the main trigger. By this and the fact that all the recorded events
are good ones, we validate the system. During this period, the performance measurements can only be determined with one PMT
using the other as the external trigger. All the results of the standalone period are summarized on table \ref{tableSA} and
on figures \ref{standalone1} and \ref{standalone2}. The results presented here are based on the same analysis method than
what was shown before.

\begin{figure}[H]
    \centering
    \begin{tabular}{c|c|c|c|c|c}
    PMT & Applied threshold (mV) & False event rate & Efficiency (\%) & Threshold shift (\%) & Rise width (ADC)\\
    \hline
    A6  & 50 & $<3.8\;10^{-4}$ & 99.37 & 16 & 1 \\
        & 100 & $3.9\;10^{-5}$ & 99.69 & 18 & 2 \\
        & 150 & $9.3\;10^{-5}$ & 99.38 & 20.6 & 2 \\
        & 200 & $< 8\;10^{-5}$ & 99.02 & 21.5 & 1 \\
    \hline
    C7  & 50  & $<1.7\;10^{-3}$ & 99.01 & 22 & 2 \\
        & 100 & $<1.1\;10^{-3}$ & 99.86 & 22 & 2\\
        & 150 & $<1.7\;10^{-3}$ & 99.54 & 26 & 2\\
        & 200 & $<1.3\;10^{-3}$ & 99.57 & 22.5 & 2\\
    \end{tabular}
    \caption{Measured performance during standalone runs}
    \label{tableSA}
\end{figure}

These figures show that the apparatus then worked at the same level of performance as during the main period. It happened
that not enough statistics were available to have a determination of the noise level, we therefore put an upper limit. Figure
\ref{standalone1} shows clearly that the reason why the efficiency is sometime as low as 99.01 \% is the lack of events above
threshold and therefore the implied Poissonnian regime. From these figures we can also infer that the beam spot was closer to
A6's alignment than to C7 one.

\newpage
\section*{Conclusion}

Data analysis from this test beam demonstrate that the Ecal Intermediate Board fulfils the physics requirements. The
electronic chain is validated, logic signals and data are perfectly handled. The $\gamma$ trigger performance, for what
concerns the analog part, is in good agreement with awaited specifications, with good efficiency, threshold accuracy, and low
noise level. The C side amplifier has been space qualified and chosen for equipping AMS02 Ecal flight model, for its
performance and reliability in comparison to the A side one. The presented analysis methods are used in the algorithms of
production tests. The Ecal is currently being fully equipped and should undergo a final beam test at the end of 2006.

\newpage

\section*{Additional figures mentioned in the note}

\begin{figure}[H]
\centering
\includegraphics[width=16cm]{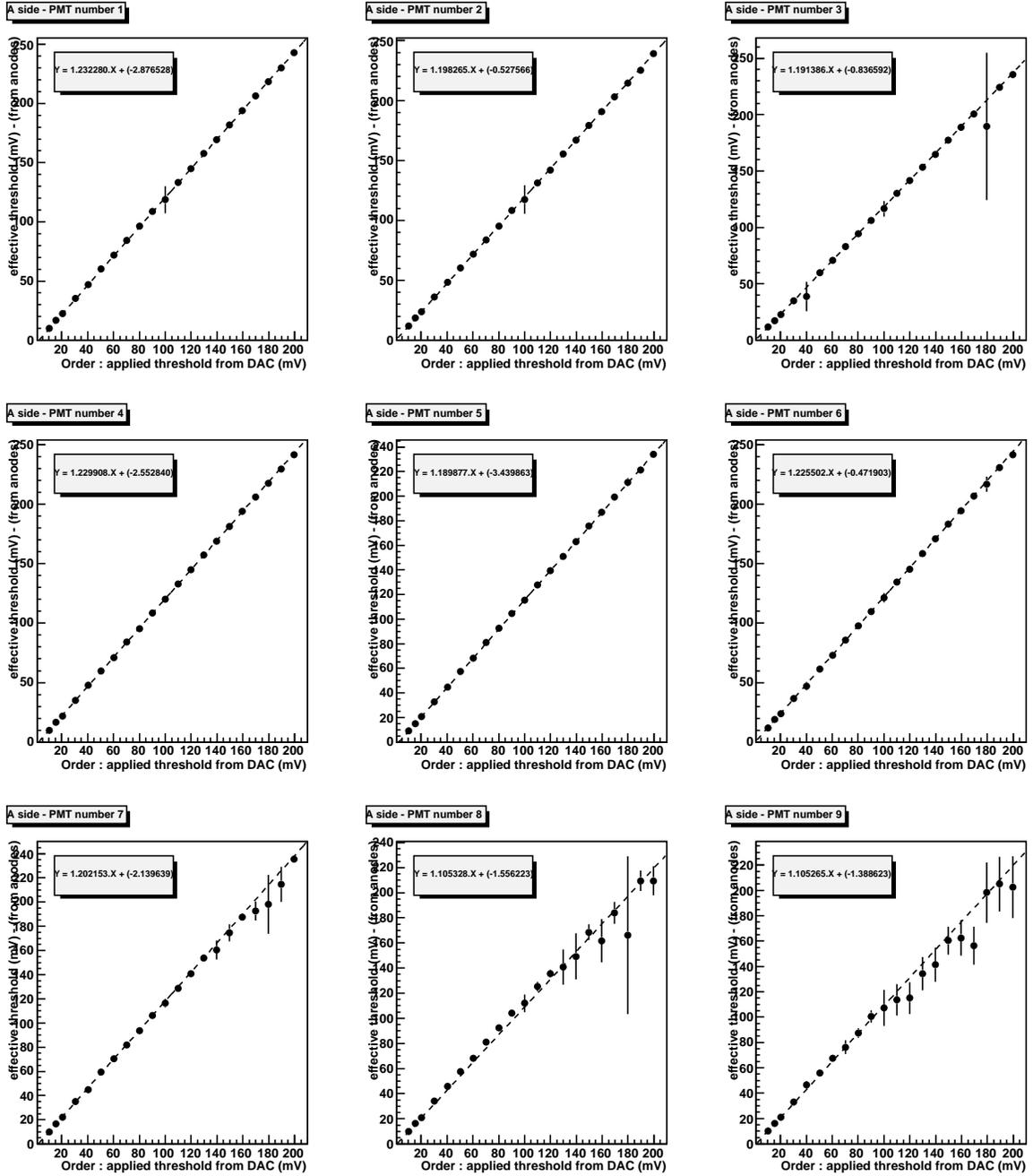}
\caption{Effective threshold versus applied ones, for all A side PMTs}
\label{fita}
\end{figure}


\begin{figure}[H]
\centering
\includegraphics[width=16cm]{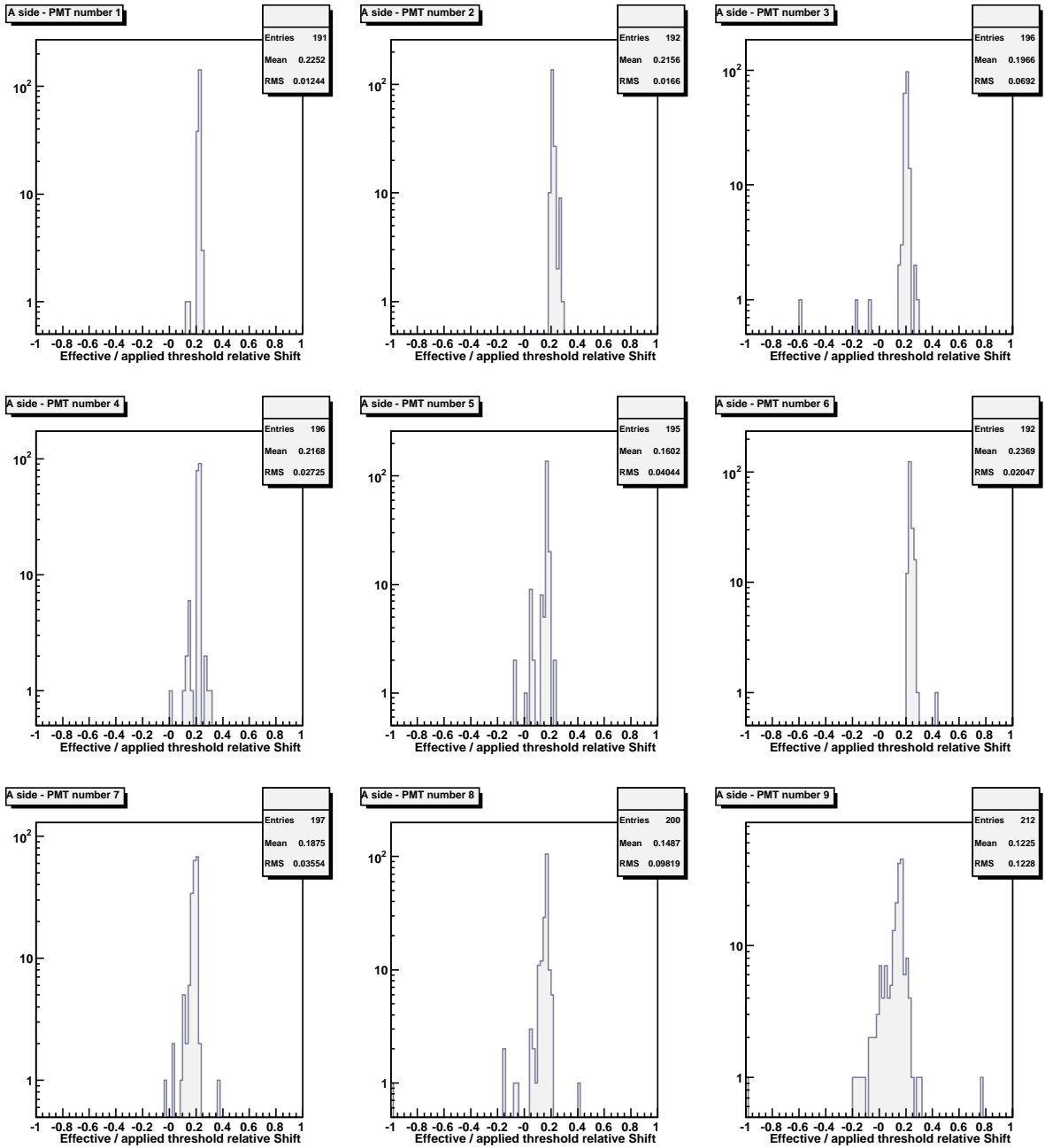}
\caption{Effective / applied threshold relative shifts on A side}
\label{shifta}
\end{figure}
\begin{figure}[H]
\centering
\includegraphics[width=16cm]{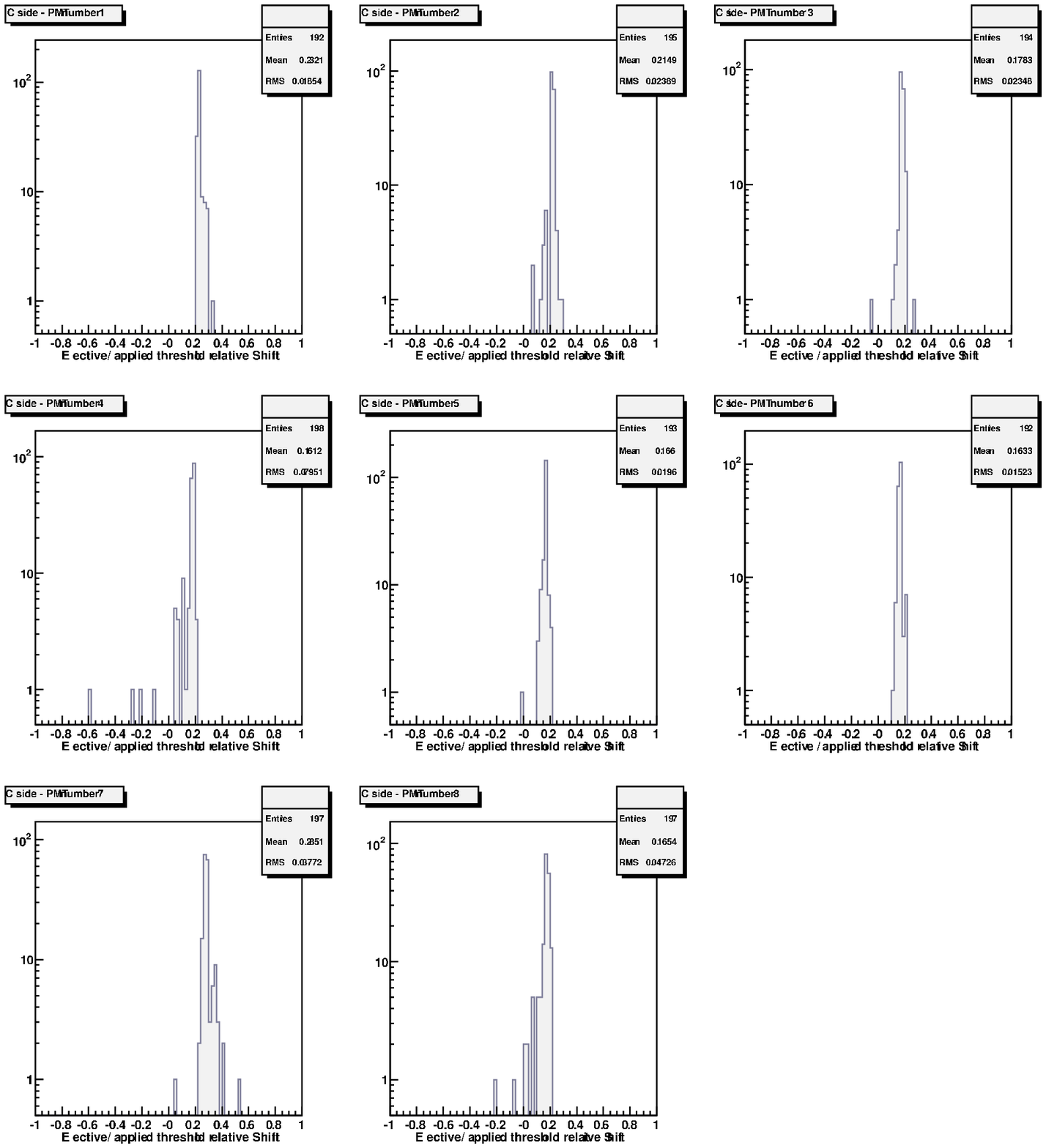}
\caption{Same as previous figure, but for C side}
\label{shiftc}
\end{figure}



\newpage

\section*{Acknowledgements}
We would like to thank all the members of LAPP mechanics, electronics and physics teams. We gratefully thank Divic Rapin and
the whole Tracker team.


\begin{thebibliography}{0}
\bibitem{note0} C. Goy, S. Rosier-Lees, AMS note 2001-06-04
\bibitem{note1} S. Di Falco, $et\;al.$, AMS note 2003-08-01
\bibitem{note2} C. Adloff, D. Fougeron, R. Hermel, S. Rosier-Lees, AMS note 2003-11-01
\bibitem{loic} L. Girard, PhD thesis LAPP-T-2004-04, 2004
\bibitem{dani} D. Grosjean, master thesis, June 2004
\bibitem{divic} D. Rapin, presentation at Tracker-Ecal meeting, CERN, June 2004
\bibitem{franck} F. Cadoux, presentation at Tracker-Ecal meeting, CERN, July 2004
\bibitem{videtherm} J.P. Vialle, Z. Li, TIM presentation, CERN, April 2005

\end{thebibliography}
\end{document}